\documentclass[prd,twocolumn,nofootinbib,showpacs,superscriptaddress]{revtex4}

\usepackage{amsfonts}
\usepackage{amsmath}
\usepackage{bm}
\usepackage{graphicx}
\usepackage[latin1]{inputenc}
\usepackage{latexsym}
\usepackage{rotating}

\newcommand{\mb}[1]{\mbox{\boldmath $#1$}}

\def \met  {\mbox{g}}

\def \sch  {\mbox{\tiny Sch}}
\newcommand{\GW}{{\mbox{\tiny GW}}}
\newcommand{\ZM}{{\mbox{\tiny ZM}}}
\newcommand{\RW}{{\mbox{\tiny RW}}}
\newcommand{\CPM}{{\mbox{\tiny CPM}}}
\newcommand{\ADM}{{\mbox{\tiny ADM}}}
\newcommand{\CLA}{{\mbox{\tiny CLA}}}
\newcommand{\CUT}{{\mbox{\tiny cut}}}
\newcommand{\MIN}{{\mbox{\tiny min}}}
\newcommand{\FIT}{{\mbox{\tiny fit}}}
\newcommand{\LOW}{{\mbox{\tiny low}}}
\newcommand{\UP}{{\mbox{\tiny up}}}
\newcommand{\recoil}{{\mbox{\tiny recoil}}}
\newcommand{\PN}{{\mbox{\tiny PN}}}
\newcommand{\UPUNO}{{\mbox{\tiny up,1}}}
\newcommand{\UPDOS}{{\mbox{\tiny up,2}}}

\begin{document}

\def\jnl@style{\it}
\def\aaref@jnl#1{{\jnl@style#1}}

\def\aaref@jnl#1{{\jnl@style#1}}

\def\aj{\aaref@jnl{AJ}}                   
\def\apj{\aaref@jnl{ApJ}}                 
\def\apjl{\aaref@jnl{ApJ}}                
\def\apjs{\aaref@jnl{ApJS}}               
\def\apss{\aaref@jnl{Ap\&SS}}             
\def\aap{\aaref@jnl{A\&A}}                
\def\aapr{\aaref@jnl{A\&A~Rev.}}          
\def\aaps{\aaref@jnl{A\&AS}}              
\def\mnras{\aaref@jnl{MNRAS}}             
\def\prd{\aaref@jnl{Phys.~Rev.~D}}        
\def\prl{\aaref@jnl{Phys.~Rev.~Lett.}}    
\def\qjras{\aaref@jnl{QJRAS}}             
\def\skytel{\aaref@jnl{S\&T}}             
\def\ssr{\aaref@jnl{Space~Sci.~Rev.}}     
\def\zap{\aaref@jnl{ZAp}}                 
\def\nat{\aaref@jnl{Nature}}              
\def\aplett{\aaref@jnl{Astrophys.~Lett.}} 
\def\apspr{\aaref@jnl{Astrophys.~Space~Phys.~Res.}} 
\def\physrep{\aaref@jnl{Phys.~Rep.}}      
\def\physscr{\aaref@jnl{Phys.~Scr}}       

\let\astap=\aap
\let\apjlett=\apjl
\let\apjsupp=\apjs
\let\applopt=\ao

\title[Gravitational Recoil from Binary Black Hole Mergers:the Close-Limit Approximation]
{Gravitational Recoil from Binary Black Hole Mergers: the Close-Limit Approximation}

\author{Carlos F. Sopuerta}
\affiliation{Institute for Gravitational Physics and Geometry and
Center for Gravitational Wave Physics, \\
The Pennsylvania State University, University Park, PA 16802, USA}
\affiliation{Department of Astronomy \& Astrophysics,
The Pennsylvania State University, University Park, PA 16802, USA}
\author{Nicol\'as Yunes}
\affiliation{Institute for Gravitational Physics and Geometry and
Center for Gravitational Wave Physics, \\
The Pennsylvania State University, University Park, PA 16802, USA}
\affiliation{Department of Physics, The Pennsylvania State University,
University Park, PA 16802, USA}
\author{Pablo Laguna}
\affiliation{Institute for Gravitational Physics and Geometry and
Center for Gravitational Wave Physics, \\
The Pennsylvania State University, University Park, PA 16802, USA}
\affiliation{Department of Astronomy \& Astrophysics, 
The Pennsylvania State University, University Park, PA 16802, USA}
\affiliation{Department of Physics, The Pennsylvania State University,
University Park, PA 16802, USA}

\date{\today}

\begin{abstract}
The coalescence of a binary black hole system is one of the main sources of
gravitational waves that present and future detectors will study.   Apart from
the energy and angular momentum that these waves carry, for unequal-mass
binaries there is also a net flux of linear momentum that implies a recoil
velocity of the resulting final black hole in the opposite direction. Due to the
relevance of this phenomenon in astrophysics, in particular for galaxy merger
scenarios, there have been several attempts to estimate the magnitude of this
velocity.  Since the main contribution to the recoil comes from the last orbit
and plunge, an approximation valid at the last stage of coalescence is well
motivated for this type of calculation.  In this paper, we present a computation
of the recoil velocity based on the close-limit approximation scheme, which
gives excellent results for head-on and grazing collisions of black holes when
compared to full numerical relativistic calculations. We obtain a maximum recoil
velocity of $\sim 64\,$ km/s for a symmetric mass ratio $\eta =
M^{}_1M^{}_2/(M^{}_1+M^{}_2)^2 \sim 0.19$ and an initial proper separation of
$4\,M$, where $M$ is the total ADM mass of the system. This separation is the
maximum at which the close-limit approximation is expected to provide accurate
results.  Therefore, it cannot account for the contributions due to inspiral and
initial merger.  If we supplement this estimate with PN calculations up to the
innermost stable circular orbit, we obtain a lower bound for the recoil
velocity, with a maximum around $84$ km/s. This is a lower bound because it
neglects the initial merger phase.  We can however obtain a rough estimate by
using PN methods or the close-limit approximation. Since both methods are known
to overestimate the amount of radiation, we obtain in this way an upper bound for
the recoil with maxima in the range of $220-265$ km/s. We also provide non-linear
fits to these estimated upper and lower bounds. These estimates are subject to
uncertainties related to issues such as the choice of initial data and higher
effects in perturbation theory. Nonetheless, our estimates are consistent with
previous results in the literature and suggest a narrower range of possible
recoil velocities.
\end{abstract}

\pacs{04.30.Db, 95.30.Sf, 98.35.Jk, 98.62.Js}

\preprint{IGPG-0618-2}

\maketitle

\section{Introduction}\label{intro}
The inspiral and merger of binary black holes systems is one of the most
interesting sources of gravitational waves that both earth-based interferometric
antennas (LIGO~\cite{ligo}, VIRGO~\cite{virgo}, GEO600~\cite{geo} and
TAMA~\cite{tama}) and space-based ones (LISA~\cite{lisa}) will detect. These
waves carry both energy and momentum away from the system, leading to the
adiabatic shrinking of the orbit, due to the former, and a recoil of the merged
object by conservation of momentum, due to the latter. The magnitude of this
recoil is of astrophysical importance because it determines whether the merged
hole will be ejected from its host galaxy.

Possible observational evidence for such a recoil may be the observations of
faint galaxies~\cite{Binggeli:1998bj,Binggeli:2000bb} where the lack of a dense
nucleus has been associated with the central black hole being ejected after
merger~\cite{Merritt:2004xa}. There is also evidence of an ejection of a
supermassive black hole in ongoing galaxy mergers, either because of recoil or
because of slingshot due to the presence of 3 or more supermassive black holes
in the merger~\cite{Haehnelt:2006hd}. The gravitational recoil has also been
shown to have important consequences in hierarchical merging scenarios and the
observable structure of galaxy nuclei. Recoil velocities of a few hundred km/s
could be large when compared to escape velocities of dwarf galaxies, globular
clusters and dark matter halos~\cite{Haiman:2004ve,Merritt:2004xa}. For
super-massive black holes at the centers of galaxies, a kick of this magnitude
could potentially transfer energy to the stars in the
nucleus~\cite{Merritt:2004xa}. There are thus very important astrophysical
aspects that can be refined or clarified with a better understanding of the
black hole kicking process.

Mass distributions without symmetries that undergo gravitational collapse of any
sort will exhibit momentum ejection and recoil of the center of mass of the
remnant due to the strong emission of gravitational
radiation~\cite{Bonnor:1961br,Papapetrou:1962ap,Peres:1962ap,Bekenstein:1973jd,Redmount:1989rr}.
Of particular interest is the case of unequal-mass black-hole binary systems. An
intuitive picture of how the system gets a {\em kick} after the merger is the
following~\cite{Wiseman:1992dv,Favata:2004wz}:  From the {\em center of mass}
point of view, the lighter black hole will move faster than the heavier one, and
hence it will beam forward gravitational radiation stronger. Then, there will be
a net flux of linear momentum carried by the gravitational radiation in the
direction of the lighter black hole, and this will cause a recoil of the
{\em center of mass} in the opposite direction. The first analytic studies of
this subject were carried out by Fitchett and
Detweiler~\cite{Fitchett:1983fc,Fitchett:1984fd}; Oohara and
Nakamura~\cite{Oohara:1983on}; Nakamura and Haugan~\cite{Nakamura:1983nh};
and Wiseman~\cite{Wiseman:1992dv}. Due to the strong non-linearity of the merger
phase, analytic studies have difficulties in obtaining an accurate estimate of
the recoil velocity. The first quasi-Newtonian analytic calculations were
presented in~\cite{Fitchett:1983fc,Fitchett:1984fd}, while a post-Newtonian (PN)
analysis have been carried out
in~\cite{Wiseman:1992dv,Blanchet:2005rj,Damour:2006tr}.  Estimates using
black-hole perturbation theory have been given
in~\cite{Favata:2004wz,Hughes:2004ck}, and a estimate that combines full
numerical relativity and perturbation theory, the {\em Lazarus} approach, is
given in~\cite{Campanelli:2004zw}.

Full numerical relativistic simulations are a natural approach to this problem
since they can in principle handle the non-linearities of the gravitational
field during the merger.  The challenge is the resolution that the computational
resources impose.   Some calculations have already been carried out in different
scenarios to estimate recoil velocities.  The first one was done by Anninos and
Brandt~\cite{Anninos:1998wt} for the case of the head-on collision of two
unequal-mass black holes.  Their numerical calculations were effectively
2-dimensional since they made use of the axisymmetry of the configuration. Using
the same type of numerical calculations they also estimated the gravitational
radiation recoil from highly distorted black holes~\cite{Brandt:1999zh}. More
recently, and due to the significant advances in 3-dimensional numerical
relativity in the binary black hole
problem~\cite{Pretorius:2005gq,Campanelli:2005dd,Baker:2005vv},
estimates of the radiation recoil velocity have also
appeared~\cite{Herrmann:2006ks,Baker:2006vn}.

Each of the approaches described above has its own limitations. Analytic
approaches are able to provide accurate estimates in their region of validity.
However, the largest contribution to the recoil velocity occurs during merger,
precisely where the approximation methods break down. Numerical simulations, in
principle, have the opportunity of producing estimates with a minimal number of
assumptions.  However, as we have mentioned, these calculations have also
limitations and use initial data that is only an approximation to the actual
astrophysical configurations. Therefore, the error bars on the computed
distribution of recoil velocities relative to the distribution present in nature
are believed to be large and, even worse, are difficult to estimate.  It is then
not surprising to find disagreements on the estimated recoil velocity as
calculated with different methods.

In this paper, we present estimates of the recoil velocity using an approach not
used before, the close-limit approximation (CLA), which combines both analytical 
and numerical techniques.  The CLA was introduced by Price and
Pullin~\cite{Price:1994pm}, who showed that this approximation method provides
accurate results compared to those obtained from numerical
relativity~\cite{Anninos:1993zj} for head-on collisions of two black holes (see
also~\cite{Anninos:1995vf}). The CLA scheme is based on the assumption that in
the last stage of coalescence, when the black holes are sufficiently {\em close}
to each other, the system behaves, up to a certain degree of approximation, as a 
single distorted hole. Then, the CLA scheme consists of establishing an
appropriate correspondence between the binary black hole system and a single
perturbed hole.  Once this correspondence is made, one can extract initial data
that can be evolved by the perturbative relativistic equations. From the outcome
of the evolution, one can estimate the fluxes of energy, angular momentum, and
linear momentum carried away to spatial infinity by the gravitational radiation
emitted. The CLA scheme has been developed and applied by a number of
authors~\cite{Gleiser:1996yc,Gleiser:1998rw,Krivan:1998er,Nicasio:1998aj,Gleiser:2001in,Khanna:2001ch,Sarbach:2001tj,Khanna:2002qp}.
In particular, Andrade and Price~\cite{Andrade:1996pc} used the CLA to estimate
the recoil velocity of a head-on collision of unequal-mass black holes starting
from rest.

Since the CLA applies to the last stage of the merger of two black holes, it is
very appealing to use it to estimate the recoil velocity of the merger of an
unequal-mass black hole binary system. With this scheme, we obtain a maximum
recoil velocity of $\sim \{17,33,64\}\,$ km/s for a symmetric mass ratio
$\eta = M^{}_1M^{}_2/(M^{}_1+M^{}_2)^2 \sim 0.19$ and initial proper separations
of $\{3,3.5,4\}\,M$, with $M$ the total ADM mass. Beyond a proper separation of
$4\,M$ the CLA is not expected to provide accurate
results~\cite{Andrade:1996pc}. Therefore, this method cannot account for the
contributions during the inspiral and initial merger phase. Supplementing this
estimate with PN calculations up to the innermost stable circular orbit (ISCO),
we obtain a lower bound for the recoil velocity, with a maximum of $\sim 84\,$
km/s. This lower bound neglects the initial merger phase, for which we can
obtain an approximate estimate by using either PN methods or the CLA. Since both
methods are known to overestimate the amount of radiation during the early
merger phase, we obtain, thus, an upper limit for the recoil with maxima in the
range of $220-265\,$ km/s.  We also perform non-linear fits to these bounds and
obtain
\begin{equation}
v_{\FIT}  = a \eta^2 \sqrt{ 1 - 4 \eta} \left(1 + b \eta + c \eta^2 \right),
\label{v-fit}
\end{equation}
where $a=7782\,$ km/s, $b=-2.507$ and $c=2.727$ for the lower bound and
$a=14802\,$ km/s, $b=-1.1339$ and $c=1.4766$ for the upper bound.

The plan of this paper is as follows: Sec.~\ref{calculation} describes the main
procedure of our calculation; Sec.~\ref{bbhid} constructs initial data for a
quasicircular binary black hole system in the $3+1$-formalism;
Sec.~\ref{close-limit} maps these initial data to a single perturbed black hole
spacetime, such that it is suitable for a CLA evolution; Sec.~\ref{resultscla}
describes the numerical implementation and presents results from the evolution
within the CLA scheme; Sec.~\ref{totalrecoil} estimates the lower and upper
bounds, as well as constructing the non-linear fits to these bounds; we finish
in Sec.~\ref{discussion} with a summary and discussion of the main results and
points to future research to obtain improved estimates.

The conventions that we use throughout this work are the following: For the
4-dimensional spacetime, we use Greek letters for the indices and a semicolon
for the covariant derivative.  The Schwarzschild metric admits a 2+2
decomposition consisting of the warped product of a Lorentzian 2-dimensional
manifold with the 2-sphere (see~\cite{Gerlach:1979rw,Gerlach:1980tx}). On the
2-dimensional Lorentzian manifold indices are denoted by capital Latin letters,
the covariant derivative associated with the 2-dimensional metric is represented
by a vertical bar, and the Levi-Civita antisymmetric tensor by
$\epsilon^{}_{AB}\,$.  On the 2-sphere indices are denoted by the lower-case
Latin letters $a,b,\ldots,h$, the covariant derivate by a colon, and the
Levi-Civita antisymmetric tensor by $\epsilon^{}_{ab}\,.$ When using the $3+1$
decomposition of spacetime quantities, spatial indices are denoted by the
lower-case Latin letters $i,j,k,\ldots\,$. Uncontrolled remainders are denoted
with ${\cal{O}}(A)$ or ${\cal{O}}(A,B)$, which stands for terms of order $A$ and
terms of order $A$ or $B$ respectively. Although usually, when dealing with
order symbols, $A$ and $B$ must be dimensionless, here they will not be, but can
be made to be dimensionless through division by the appropriate factor. We also
use physical units in which $G = c = 1$.

\section{Description of our Calculation}\label{calculation}

In this paper, we use the CLA scheme to calculate the recoil velocities after
the merger of an unequal-mass binary black hole system. Due to the complexity of
the calculation, we discuss here the different steps involved, while getting a
glimpse of the general scheme. First, we need to construct initial data
corresponding to a non-spinning binary black hole system in quasicircular
orbital motion. The method employed to construct the data is the standard one:
we solve the constraints on an initial slice using the York-Lichnerowicz
conformal decomposition. Then, the solution needs to be expanded in two
parameters: the separation of the two black holes, based on the main assumption
of the CLA, that is, small separation between the holes; and their linear
momenta, rooted in an additional {\em slow motion}
approximation~\cite{Khanna:2000dg}.

The second step is to establish a map between this initial data and the generic
initial data corresponding to a perturbed single black hole. In this work we
only consider the case in which the single black hole is a non-rotating
Schwarzschild hole. There is also the possibility of considering a Kerr black
hole (see~\cite{Khanna:2000dg} for details), but the CLA machinery in that case
is more intricate. After expanding the initial data in the separation and
linear momenta, it is straightforward, after some coordinate changes, to
identify a Schwarzschild background and its perturbations.

Once the perturbations have been identified, we need to calculate initial data
suitable for evolving the linearized (around the Schwarzschild background)
Einstein equations. The spherical symmetry of the background allows us to
separate the linearized equations. Then, by decomposing the perturbations in
spherical harmonics we obtain decoupled equations for each mode. Moreover, by
appropriately reparameterizing the perturbations, we can decouple the equations
for each individual mode, so that the problem reduces to solving a master
equation for a complex combination of the metric perturbations. These master
equations (usually known as the Regge-Wheeler and Zerilli-Moncrief equations)
are 1-dimensional wave-type equations containing a potential that accounts for
the effect of the background spacetime curvature. Therefore, the problem of
providing initial data reduces to finding initial conditions for these master
functions.

The initial data contains three parameters that we need to specify.  These
parameters are associated with the initial distance between the holes, the mass
ratio, and the initial linear momentum. The mass ratio is an independent
parameter that will be used to study the functional behavior of the recoil
velocity. The distance and linear momentum parameters determine the dynamical
character of the binary and, therefore, they must be chosen carefully. To that
end, we use the standard method of minimizing the binding energy of the system,
so that the binary is in a quasi-circular orbit. The expressions that we obtain
are formally the same as in Newtonian theory, although they cannot be assigned
the same interpretation, since they are expressed in terms of {\em bare}
parameters. In order to relate these parameters to meaningful physical ones, we
must introduce a proper separation and a physical mass ratio. The proper
separation can be calculated by evaluating the minimum proper distance between
the marginally trapped surfaces surrounding each individual hole.

The final step is to solve the master equations and evaluate the different
physical quantities of interest. The metric waveforms $h^{}_+$ and
$h^{}_\times$, together with the fluxes of energy, angular and linear momentum
carried away by gravitational waves can be computed in terms of the master
functions and their first time derivatives. In this paper, we include the
general formulae for the linear momentum fluxes in terms of the perturbation
master functions. We present several plots of these quantities, together with
plots of the recoil velocities for different initial separations.

\section{Initial Data}\label{bbhid}

In this section, we begin the initial data construction for an unequal-mass 
binary black hole system suitable to the CLA scheme. To that end, we extend the
results of Andrade and Price~\cite{Andrade:1996pc}, who carried out the
calculation for unboosted head-on collisions, and also extend the results of
Khanna {\em et al}~\cite{Khanna:2000dg}, who constructed data for equal-mass
black holes in a quasicircular orbit. Our calculation not only allows for
arbitrary mass ratios, but it also includes higher-order terms in the expansion
of the initial data, which are essential in the calculation of the recoil.

In order to solve the Hamiltonian and momentum constraints, we use the conformal
transverse-traceless method of Lichnerowicz, York and
others~\cite{Lichnerowicz:1944al,York:1971hw,York:1972sj,York:1973jw,Cook:2000vr}.
The 3-metric $\gamma^{}_{ij}$ is decomposed in terms of a conformal factor
$\Phi$ and an auxiliary metric $\hat\gamma^{}_{ij}$, $\gamma^{}_{ij} =
\Phi^4\hat\gamma^{}_{ij}\,,$ which here we assume to be conformally flat:
\begin{equation}
ds^2 = \gamma^{}_{ij}dx^idx^j = \Phi^4 (dR^2 + R^2d\Omega^2)\,, \label{bl_metric}
\end{equation}
where $d\Omega^2 = d\theta^2 + \sin^2\theta d\varphi^2\,$ is the line element of
the $2$-sphere. For the extrinsic curvature $K^{}_{ij}\,,$ we choose a
{\em maximal} initial slice, that is, $K^{}_{ij}$ is trace free:
$\gamma^{ij}K^{}_{ij}=0$.  Then, we also conformally decompose the trace-free
extrinsic curvature $K^{}_{ij}$ as
\begin{equation}
K^{}_{ij} = \Phi^{-2}\hat{K}^{}_{ij}\,, \label{pextrinsic}
\end{equation}
and we further make the choice that the longitudinal part of $\hat{K}^{}_{ij}$ 
vanishes, so that $\hat{K}^{}_{ij}$ is a symmetric transverse traceless tensor.
Then, the momentum and Hamiltonian constraints reduce to
\begin{equation}
\hat{\nabla}^j\hat{K}^{}_{ji} = 0\,, \label{momentumc}
\end{equation}
\begin{equation}
\hat{\nabla}^2\Phi = -\frac{1}{8}\Phi^{-7}\hat{K}^{}_{ij}\hat{K}^{ij}\,,
\label{hamiltonianc}
\end{equation}
where $\hat{\nabla}^{}_i$ and $\hat{\nabla}^2$ denote the covariant derivative
and Laplacian associated with the flat 3-metric $\hat{\gamma}^{}_{ij}\,.$  The
momentum constraint [Eq.~(\ref{momentumc})] can be exactly solved using the
method of Bowen and York~\cite{Bowen:1980yu}. For a single black hole located at
$\mb{R}=\mb{R}^{}_o$ with linear momentum $\mb{P}$ it can be written as follows:
\begin{equation}
\hat{K}^{one}_{ij} = \frac{3}{2|\mb{R}-\mb{R}^{}_o|^2}\left[
  2P^{}_{(i}n^{}_{j)} - (\hat{\gamma}^{}_{ij}-n^{}_i n^{}_j) P^k
  n^{}_k \right]\,, \label{cextcur}
\end{equation}
$P^i$ is the ADM momentum of a single hole, while $n^i_{}$ is a unit vector in
flat three-dimensional space directed from the location of the single hole to an
arbitrary point, namely
\begin{equation}
n^i = \frac{R^i - R^i_o}{|\mb{R}-\mb{R}^{}_o|}\,,
\label{unit-vec}
\end{equation}
and the vertical bars, $|\cdot|$, denote the norm of vector in the flat
3-dimensional space. In order to construct a solution for two holes, we can
simply superpose two solutions of the type of Eq.~(\ref{cextcur}).

\begin{figure}
\includegraphics[scale=0.6,clip=true]{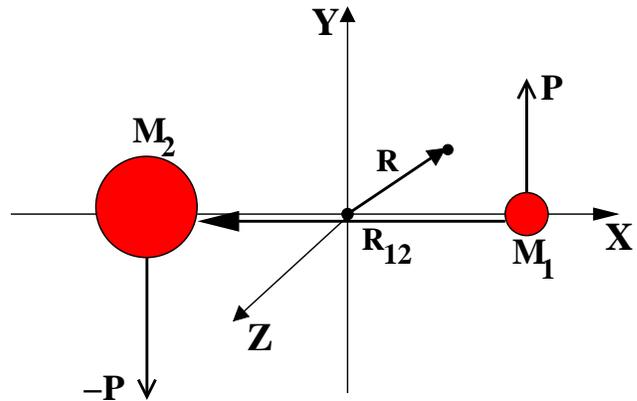}
\caption{\label{phys_scen} Schematic diagram of the initial physical configuration.
The linear momenta are parallel to the $Y$-axis and span the $X$-$Y$ plane, 
so that the angular momentum is aligned with the $Z$-axis.}
\end{figure}

Before constructing the extrinsic curvature, it will be usefull to first
describe the initial physical configuration. The system we are modeling consists
of two black holes with masses $M^{}_1$ and $M^{}_2$ located on the $X$-axis, a
coordinate distance $d$ apart, as shown in Figure~\ref{phys_scen}. In this
figure, $\mb{R}^{}_1\,$, $\mb{R}^{}_2\,,$ and $\mb{R}$ are radial vectors that
point from the origin to hole $1$, hole $2\,,$ and an arbitrary point,
respectively. Moreover, $\mb{R}^{}_{12} = \mb{R}^{}_2 - \mb{R}^{}_1$ is also a
vector that points from hole $1$ to $2$, and $\mb{P}^{}$ and $-\mb{P}^{}$ are
the linear momenta associated with holes 1 and 2 respectively. Since the linear
momenta are parallel to the $Y$-axis, the orbital angular momentum is directed
along the $Z$-axis.
With such physical scenario, the solution of Eq.~(\ref{momentumc}) can be
written as~(see also~\cite{Khanna:2000dg}):
\begin{equation}
\hat{K}^{}_{ij} = \hat{K}^{ one}_{ij}[\mb{R}^{}_o=\mb{R}^{}_1,\mb{P}] +
\hat{K}^{ one}_{ij}[\mb{R}^{}_o=\mb{R}^{}_2,-\mb{P}]\,,
\label{cextrinsic}
\end{equation}
where we have defined $\mb{P} = P\,\mb{\hat{y}}$. The ADM momentum corresponding
to $\hat{K}^{one}_{ij}$ is simply $\mb{P}$ and the one associated with
$\hat{K}^{}_{ij}$ is zero.

Let us now concentrate on the solution to the Hamiltonian constraint.  Using
Eq.~(\ref{cextrinsic}) in the Hamiltonian constraint leads to a source term
quadratic in $P$. We now introduce a {\em slow-motion} approximation, where we
assume that the linear momentum $P$ is {\em small}, in the sense that terms of
${\cal{O}}(v^2)$ are much smaller than terms of ${\cal{O}}(v)$, where $v$ is a
measure of the orbital velocity. We, thus, neglect terms of ${\cal O}(P^2)$, so
that the Hamiltonian constraint reduces to the Laplace equation
\begin{equation}
\nabla^2 \Phi = 0\,.
\end{equation}
The solution of this equation depends on our choice of topology. If we choose
the initial slice to have a single asymptotically flat region, the solution to
the conformal factor is the Misner solution~\cite{Misner:1960cw}, but if one
chooses the slice to have three asymptotically-flat regions, the solution is the
Brill-Lindquist~\cite{Brill:1963bl} one. In this paper, we adopt the latter and
the conformal factor takes the form of the Newtonian potential:
\begin{equation}
\Phi = 1 + \frac{m^{}_1}{2|\mb{R}-\mb{R}^{}_1|} +
\frac{m^{}_2}{2|\mb{R}-\mb{R}^{}_{2}|} \,,
\label{cfactor}
\end{equation}
where $m^{}_1$ and $m^{}_2$ denote the {\em bare} masses of each individual
hole. One reason for choosing the Brill-Lindquist (BL) solution is that it is
simpler to manipulate, while it has also been shown~\cite{Andrade:1996pc} to
lead to essentially the same results when calculating recoil velocities for
head-on collisions. We remark that, although terms of ${\cal O}(P^2)$ have been
neglected, they can be straightforwardly added in a perturbative fashion, but
this will be studied elsewhere.

Let us comment further on the topology of the initial slice associated with the
BL solution, as it is important in some calculations. As we have already 
mentioned, this solution has three asymptotically-flat regions: one of them,
$\Sigma^{}_0$, corresponds to the region far from the two holes, $R = |\mb{R}|
\gg |\mb{R}^{}_1| = R^{}_1$ and $R \gg |\mb{R}^{}_2| = R^{}_2$; the other two,
$\Sigma^{}_1$ and $\Sigma^{}_2$, are associated with hole $1$ and $2$
respectively. By simple inspection of the conformal factor
[Eq.~(\ref{cfactor})], the solution seems ill-behaved at the location of the
holes, $\mb{R}=\mb{R}^{}_1$ and $\mb{R}=\mb{R}^{}_2$, although it is actually
not. Near each hole, the geometry is invariant under the transformation:
$|\mb{R}-\mb{R}^{}_\Lambda|\,\rightarrow\,m^{2}_\Lambda/(4R'^{}_\Lambda)$
($\Lambda=1,2$). The value $m^{}_\Lambda/2$ coincides with the intersection of
the event horizon with the initial slice for a single hole and it is a fixed
point in the transformation. This value is sometimes referred to as the
{\em throat}, joining two asymptotically-flat regions. Therefore, the points
$\mb{R} = \mb{R}^{}_1$ and $\mb{R} = \mb{R}^{}_2$ are just an {\em image} of the
infinities of $\Sigma^{}_1$ and $\Sigma^{}_2$. For a single hole, there are two
asymptotically-flat regions, and its mass, equal to $m$, is the same independent
of which region we evaluate it on. In the case of a binary system, the
gravitational interaction between the holes will change the value of the
individual masses. Actually, there is not an invariant measure of them in
$\Sigma^{}_0$, but such a measure does exist on $\Sigma^{}_1$ and $\Sigma^{}_2$.
Doing the calculation yields the following result~\cite{Brill:1963bl}
\begin{equation}
M^{}_1 = m^{}_1\left(1 + \frac{m^{}_2}{2 d} \right) \,,~~
M^{}_2 = m^{}_2\left(1 + \frac{m^{}_1}{2 d} \right) \,, \label{physicalmasses}
\end{equation}
where $d = |\mb{R}^{}_{12}|\,.$ In $\Sigma^{}_0$, we can compute the total mass
of the binary system, the ADM mass of the system.  We call the result $M$ and it
is given by
\begin{equation}
M = m^{}_1 + m^{}_2 \,. \label{themass}
\end{equation}
Eqs.~(\ref{cextrinsic}), (\ref{cfactor}), and~(\ref{pextrinsic}) are the initial
data. We should note that, apart from our choice of initial data (BL conformal
factor and Bowen-York extrinsic curvature), there are other possible choices
that can be used in the CLA scheme. Some examples of other possible data sets
are the following: a Misner conformal factor with a Bowen-York extrinsic
curvature with inversion symmetry through the throats; Kerr-Schild initial
data~\cite{Sarbach:2001tj}.

The next step in the construction of the initial data is to put it in a form
suitable for the CLA scheme. Before doing so, however, it is convenient to study
the parameters that determine the configuration described by the data. To begin
with, let us introduce the {\em bare} mass ratio
\begin{equation}
q = \frac{m^{}_2}{m^{}_1}\,.
\label{q}
\end{equation}
The initial configuration can then be fully specified in terms of the parameters
$(q,d,P)$. Since $q$ and $d$ are {\em bare} parameters, in the sense that we
cannot give them a physical meaning, we are going to introduce analogous
parameters, which can be given a physical interpretation. First, we introduce
the mass ratio between the individual masses of the holes as computed in their
respective asymptotically-flat regions:
\begin{equation}
Q = \frac{M^{}_2}{M^{}_1}\,. \label{Q}
\end{equation}
The quantities $q$ and $Q$ are related through $d$ via~\cite{Andrade:1996pc}:
\begin{eqnarray}
Q =  q\frac{1 + [(1+q)/2](M/d)}{1 + [(1+q)q/2](M/d)} \,, \label{mass-ratio1}
\end{eqnarray} 
\begin{eqnarray}
q = \frac{Q-1}{2}\left(1+\frac{M}{2d}\right) + \sqrt{Q +
\left[\frac{Q-1}{2}\left(1+\frac{M}{2d}\right) \right]^2}\,. \label{mass-ratio2}
\end{eqnarray}
Defining a useful distance is a more difficult matter. In this paper, we use
the same distance as Andrade and Price~\cite{Andrade:1996pc}, which is the
proper distance between the points where the marginally trapped surfaces,
surrounding each individual hole, cross the $X$-axis (we obviously refer to
crossing points closer to the opposite hole). When the initial configuration
does not present a common apparent horizon, these marginally trapped surfaces
are the individual components of the apparent horizon. If $x^{}_1$ and $x^{}_2$
stand for these crossing points, the distance we have just defined is given by
$D = \int^{x^{}_1}_{x^{}_2}\Phi^2\,dx\,,$ which yields
\begin{widetext}
\begin{equation}
D = \Delta x\left\{1 + \frac{m^2_1}{4(X^{}_1-x^{}_1)(X^{}_1-x^{}_2)} 
+ \frac{m^2_2}{4(x^{}_1-X^{}_2)(x^{}_2-X^{}_2)}  \right\} - 
M^{}_1\ln\left(\frac{X^{}_1-x^{}_1}{X^{}_1-x^{}_2}\right) +
M^{}_2\ln\left(\frac{x^{}_1-X^{}_2}{x^{}_2-X^{}_2} \right)\,, \label{D}
\end{equation}
\end{widetext}
where $\Delta x = x^{}_1 - x^{}_2 > 0\,,$ and $X^{}_1$ ($=R^{}_1$) and $X^{}_2$
($=-R^{}_2$) are the $x$-coordinate locations of holes 1 and 2 respectively in
the conformal space. In summary, we can determine our initial configuration
either by specifying the set $(q,d,P)$ or the set $(Q,D,P)\,.$

The CLA scheme assumes that the black holes are sufficiently close enough, which
allows us to expand the initial data in $R \gg R^{}_1$ and $R \gg R^{}_2$.  In
this sense, it is useful to choose the coordinate origin, $\mb{R} = 0\,,$ in
such a way that it coincides with the {\em bare} center of mass of the binary
black hole, that is:
\begin{equation}
m^{}_1\mb{R}^{}_1 + m^{}_2\mb{R}^{}_2 = 0\,. \label{barecm}
\end{equation}
We can write then $\mb{R}^{}_1$ and $\mb{R}^{}_2$ in terms of the separation
vector $\mb{R}^{}_{12}$ via
\begin{equation}
\mb{R}^{}_1 = \chi^{}_1 \, \mb{R}^{}_{12}\,, \qquad
\mb{R}^{}_2 = \chi^{}_2 \, \mb{R}^{}_{12}\,,
\end{equation}
where we have defined 
\begin{equation}
\chi^{}_{1} = -\frac{q}{1 + q}\,, \qquad 
\chi^{}_2 = \frac{1}{1 + q}\,.
\end{equation}
Then, it is natural to expand the initial data in the following dimensionless
parameter $\epsilon$
\begin{equation}
\epsilon = |\mb{\epsilon}| = \frac{d}{R} \,, \qquad
{\mb{\epsilon}} = \frac{{\bf{R}_{12}}}{R} \,.
\end{equation}
A key formula for performing these expansions is the following:
\begin{equation}
\frac{1}{|\mb{R}-\mb{R}^{}_A|^N} = \frac{1}{R^N}\sum_{\ell=0}^{\infty}
C^{(N/2)}_{\ell}(\mb{\hat{\epsilon}} \cdot \mb{\hat{R}}) \; 
\left( \chi^{}_A\,\epsilon\right)^\ell\,, \label{keyformula}
\end{equation}
where we have introduced the following unit vectors
\begin{eqnarray}
\mb{\hat{R}} & = & \frac{\mb{R}}{R}\,, \qquad
\mb{\hat{\epsilon}} = \frac{\mb{\epsilon}}{\epsilon}\,, 
\end{eqnarray}
and where $C^{(N/2)}_{\ell}$ denote the Gegenbauer polynomials. These
polynomials, also known as ultra-spherical polynomials, are a generalization of
the Legendre polynomials to $(N/2 + 2)$-dimensional spaces, which are common in
angular momentum theory~\cite{Condon:1970cs,Arfken}. For the special cases of
$N=0\,,\,1\,,2\,,$ these polynomials reduce to Legendre polynomials and
Chebyshev polynomials of type~1 and~2 respectively. We refer to
Appendix~\ref{spec-func} for more details on these polynomials.

We now use all these definitions and results to expand the conformal extrinsic
curvature given by Eq.~(\ref{cextrinsic}) in $\epsilon$ to arbitrary order. The
result we obtain can be written as follows:
\begin{widetext}
\begin{eqnarray}
\hat{K}^{}_{ij} &=& \frac{3}{2 R^2} \sum_{\ell=0}^{\infty}\left(\frac{d}{R}
\right)^{\ell +1} \left( \chi_1^{\ell + 1} - \chi_2^{\ell+1} \right)
\left\{ C_{\ell +1}^{(3/2)} \left[ 2 P^{}_{(i}\hat{R}^{}_{j)} -
(\mb{P}\cdot\mb{\hat{R}})\, \hat{\gamma}^{}_{ij}\right] + C_{\ell +1}^{(5/2)}
(\mb{P} \cdot \mb{\hat{R}})\, \hat{R}^{}_{i} \hat{R}^{}_{j} \right. \nonumber \\
&+& \left.  C_{\ell}^{(3/2)} \left[ (\mb{P} \cdot \mb{\hat{\epsilon}}) \,
\hat{\gamma}^{}_{ij} - 2 P^{}_{(i} \hat{\epsilon}^{}_{j)} \right]
- C_{\ell}^{(5/2)} \left[ (\mb{P} \cdot \mb{\hat{\epsilon}}) \,
\hat{R}^{}_i\hat{R}^{}_j + 2 (\mb{P} \cdot \mb{\hat{R}}) \,
\hat{R}^{}_{(i}\hat{\epsilon}^{}_{j)} \right] \right\} \nonumber \\
&+&  \left(\frac{d}{R}\right)^{\ell+2}  \left( \chi_1^{\ell + 2} -
\chi_2^{\ell+2} \right) C_{\ell}^{(5/2)} \left[ (\mb{P} \cdot \mb{\hat{R}}) \,
\hat{\epsilon}^{}_i \hat{\epsilon}^{}_j + 2 (\mb{P} \cdot \mb{\hat{\epsilon}})\,
\hat{R}^{}_{(i} \hat{\epsilon}^{}_{j)} \right] \nonumber \\
&-& \left(\frac{d}{R}\right)^{\ell+3} \left( \chi_1^{\ell + 3} -
\chi_2^{\ell+3}\right) C_{\ell}^{(5/2)} (\mb{P} \cdot \mb{\hat{\epsilon}})\,
\hat{\epsilon}^{}_i \hat{\epsilon}^{}_j \,,    \label{ultraK}
\end{eqnarray}
\end{widetext}
where for simplicity we have omitted the argument of the Gegenbauer polynomials,
which still is $\mb{\hat{\epsilon}} \cdot \mb{\hat{R}}$. It is worth noting that
the $\epsilon^0$ term has identically vanished due to the choice of coordinate
origin, which coincides with the {\em bare} center of mass. Another interesting
fact is that only combinations of the $(3/2)$- and $(5/2)$-Gegenbauer
polynomials appear due to the combination of odd powers in the denominators of
the extrinsic curvature.

In this paper we are going to consider terms up to ${\cal O}(\epsilon^3)$, which
is enough to get a gravitational recoil effect and actually the dominant part of
it (see, e.g.~\cite{Andrade:1996pc}). Extensions of our calculations to higher
order are in principle straightforward, but we are not going to present them
here.  Then, up to this order of approximation we can rewrite Eq.~(\ref{ultraK})
as follows:
\begin{widetext}
\begin{eqnarray}
\hat{K}^{}_{ij} & =  & \frac{9}{2} \frac{d}{R^3} \left\{
  \mb{\hat{R}}\cdot\mb{\hat\epsilon} + \frac{1}{2}
  \frac{d}{R}\frac{1-q}{1+q}[5(\mb{\hat{R}}\cdot\mb{\hat\epsilon})^2 -
  1] \right\}
  \left[(\mb{P}\cdot\mb{\hat{R}})\hat{\gamma}_{ij}-2P^{}_{(i}\hat{R}^{}_{j)}
  \right] 
\nonumber \\ 
& - & \frac{3}{2} \frac{d}{R^3} \left\{ 1 + 3 \frac{d}{R}
  \frac{1-q}{1+q} \mb{\hat{R}}\cdot\mb{\hat\epsilon} \right\}
  \left[(\mb{P}\cdot\mb{\hat\epsilon})\hat{\gamma}_{ij}-
  2P^{}_{(i}\hat\epsilon^{}_{j)}\right]
\nonumber \\ 
& - & \frac{15}{2} \frac{d}{R^3} \left\{
  \mb{\hat{R}}\cdot\mb{\hat\epsilon} + \frac{1}{2} \frac{d}{R}
  \frac{1-q}{1+q}[7(\mb{\hat{R}}\cdot\mb{\hat\epsilon})^2- 1]
  \right\}(\mb{P}\cdot\mb{\hat{R}}) \hat{R}^{}_i\hat{R}^{}_j 
\nonumber \\ 
& + & \frac{3}{2} \frac{d}{R^3} \left\{ 1 + 5 \frac{d}{R}
  \frac{1-q}{1+q}\mb{\hat{R}}\cdot\mb{\hat\epsilon} \right\} \left[
  (\mb{P}\cdot\mb{\hat\epsilon})\hat{R}^{}_i\hat{R}^{}_j +
  2(\mb{P}\cdot\mb{\hat{R}})\hat{R}^{}_{(i}
  \hat\epsilon^{}_{j)}\right] 
\nonumber \\ 
& - & \frac{3}{2} \frac{d^2}{R^4} \frac{1-q}{1+q} \left[
  (\mb{P}\cdot\mb{\hat{R}})\hat\epsilon^{}_i\hat\epsilon^{}_j +
  2(\mb{P}\cdot\mb{\hat\epsilon})\hat{R}^{}_{(i}
  \hat\epsilon^{}_{j)}\right] + {\cal{O}}(P d^3) \,.
\label{K-first}
\end{eqnarray}
\end{widetext}
The lowest-order contribution is of ${\cal{O}}(P d)$ and it is the only
contribution used by Khanna {\em et al}~\cite{Khanna:2000dg} for grazing
collisions of equal-mass black holes. The next contribution is of
${\cal{O}}(P d^2)$ and, as far as we know, this is the first time it has been
considered.

Let us now look at the conformal factor [Eq.~(\ref{cfactor})]. Using
Eq.~(\ref{keyformula}), we can also expand $\Phi$ in Gegenbauer polynomials
to obtain
\begin{eqnarray}
\Phi = 1 + \frac{M}{2R} + \sum_{\ell \geq 2}^{\infty} 
C^{(1/2)}_\ell(\mb{\hat{R}} \cdot \mb{\hat{\epsilon}}) \;
\epsilon^{\ell} \left( m_1 \chi^{\ell}_1 + m_2 \chi^{\ell}_2 \right)\,.
\end{eqnarray}
It is important to recall that in solving the Hamiltonian constraint we have
used a {\em slow-motion} approximation, neglecting terms of ${\cal O}(P^2)$.
The terms we are, thus, neglecting are of ${\cal{O}}(P^2 d^2)$. This expression
can also be written in terms of the parameters $(q,d)$ and $M$, and in terms of
Legendre polynomials.  In this way we obtain
\begin{equation}
\Phi = 1 + \frac{M}{2R} + \sum_{\ell \geq 2}^{\infty} \phi^{}_\ell
\left(\frac{M}{R}\right)^{\ell+1} P^{}_\ell(\mb{\hat{R}} \cdot
\mb{\hat{\epsilon}}) + {\cal{O}}(P^2 d^2)\,,
\end{equation}
where $P_{\ell}$ denotes the Legendre polynomials and where the coefficients
$\phi^{}_\ell$ are given by
\begin{equation}
\phi^{}_\ell = \frac{1}{2} \left\{(-1)^\ell +
  q^{\ell-1}\right\}\frac{q}{(1+q)^{\ell + 1}}
  \left(\frac{d}{M}\right)^\ell \,.  
\end{equation}
The $\ell = 1$-term vanishes due to the choice of the origin of coordinates in
Eq.~(\ref{barecm}). Finally, the expansion of the conformal factor up to third
order in $d$ is given by
\begin{eqnarray}
\Phi &=& 1 + \frac{M}{2 R} + \frac{1}{2} \frac{q}{(1 + q)^2}
\frac{M d^2}{R^3} P_2(\mb{\hat{R}} \cdot
\mb{\hat{\epsilon}}) 
\nonumber \\
&-& \frac{1}{2} \frac{q(1 - q)}{(1 + q)^3}
\frac{M d^3}{R^4} P_3(\mb{\hat{R}} \cdot
\mb{\hat{\epsilon}}) +  {\cal{O}}(P^2 d^2,d^4)\,. \label{cfactord3}
\end{eqnarray}
With this, we finish the construction of initial data to be used in the CLA
scheme. To summarize, we remark that this construction is based on expansions on
two different parameters: $P$ (related to the {\em slow-motion} approximation)
and $d$ (related to the assumption that the holes are close to each other). 
Since $P$ and $d$ have dimensions, the meaning of these expansions is that 
terms of order $d^N$ and/or $P^M$ are smaller than terms of order $d^{N-1}$ 
and/or $P^{M-1}$. As we are going to see later, these expansions will provide
the leading contribution of the multipoles $\ell=2$ and $\ell=3$.

\section{The Close Limit Approximation} \label{close-limit}
The next stage in our computation is to recast the initial data just constructed
into data for a perturbed Schwarzschild black hole, which is the essence of the
CLA scheme. In this way we can extract initial data to be evolved by the
corresponding perturbation equations. Thanks to the expansions performed in the
previous section, the main task now becomes the extraction of the different
multipoles from the data.

The $3$-metric on the initial slice is conformally flat and hence determined by
the conformal factor $\Phi$.  If we look at the lowest-order contribution [see
Eq.~(\ref{cfactord3})] we realize that it coincides with the $3$-metric of
Schwarzschild spacetime associated with the $\{t=\mbox{const.}\}$-slicing in
isotropic coordinates, being $t$ the Schwarzschild time coordinate. However, in
order to make the connection with perturbation theory, it is very convenient to
reexpress the initial data in Schwarzschild coordinates:
\begin{equation}
ds^2 = f^{-1}dr^2 + r^2d\Omega^2\,,~~~~ f = 1 - \frac{2 M}{r}\,,
\end{equation}
where we recall that $M$ is the total ADM mass.  The transformation from
isotropic coordinates to Schwarzschild coordinates is given by the following
relations:
\begin{equation}
R = \frac{1}{4}(\sqrt{r} + \sqrt{r-2M})^2 \,,
\qquad 
r = R \left(1 + \frac{M}{2R} \right)^2\,,
\end{equation}
Applying this transformation to the 3-metric of our initial data we obtain:
\begin{equation}
ds^2 = {\cal F}^4\left(f^{-1}dr^2 + r^2 d\Omega^2\right)\,,
\end{equation}
where
\begin{equation}
{\cal F} = \frac{\Phi}{1+\frac{M}{2R}} \,.
\end{equation}

In order to construct initial data for the perturbations, evolve it, and compute
from the result all the relevant physical information, in the next subsections
we give a summary of (non-rotating) black-hole perturbation theory and the main
tools needed for the application of the CLA scheme. Afterwards, we apply this
machinery to the construction of the initial data and describe how the energy,
angular momentum, and linear momentum fluxes carried away by the gravitational
waves are evaluated.

\subsection{Black hole perturbation theory}

The CLA is based on the fact that, in the last stages of coalescence, the
gravitational field can be modeled, to a good degree of approximation, as the
gravitational field of a single perturbed black hole. Thus, perturbation theory
plays a key role in our calculations and it is worth reviewing its main concepts
and tools. The starting point is the assumption that the spacetime metric,
$\met^{}_{\mu\nu}\,,$ can be written as:
$\met^{}_{\mu\nu} = \met^{\sch}_{\mu\nu} + h^{}_{\mu\nu}$, where
$\met^{\sch}_{\mu\nu}$ denotes the background Schwarzschild metric and
$h^{}_{\mu\nu}$ the first-order perturbations. Then, we can take advantage of
the spherical symmetry of the Schwarzschild metric to simplify the structure of
the perturbations and of the equations that govern them. We can do this by
expanding the perturbations in tensor spherical harmonics. It turns out that the
linearized Einstein equations (in this case, around the Schwarzschild
background) decouple for each harmonic. Not only this, we can distinguish
between the perturbative modes with polar parity, which pick up a factor of
$(-1)^l$ under parity transformations, and the ones that have axial parity,
which pick up a factor of $(-1)^{l+1}$. This distinction is important because
polar and axial modes also decouple.

Following this discussion, we split the metric perturbations $h_{\mu\nu}$ into
polar and axial perturbations, $h_{\mu\nu} = h^{\mbox{\small a}}_{\mu\nu} +
h^{\mbox{\small p}}_{\mu\nu}$. And these perturbations can be expanded in tensor
spherical harmonics as
\begin{equation}
h^{\mbox{\small a}}_{\mu\nu} =
\sum_{\ell,m} h^{\mbox{\small a},\ell m}_{\mu\nu}\,,~~~
h^{\mbox{\small p}}_{\mu\nu} =
\sum_{\ell,m} h^{\mbox{\small p},\ell m}_{\mu\nu}\,,
\end{equation}
where
\begin{eqnarray}
h^{\mbox{\small a},\ell m}_{\mu\nu} = \left( \begin{array}{cc}
  0  &  h_A^{\ell m} \, S^{\ell m}_a \\
\\
 \ast  &  \  H^{\ell m} \, S_{ab}^{\ell m}
 \end{array}\right)\,, 
\label{maxial}
\end{eqnarray}
\begin{eqnarray}
h^{\mbox{\small p},\ell m}_{\mu\nu} = \left( \begin{array}{cc}
  h_{AB}^{\ell m}\, Y^{\ell m} &  p_A^{\ell m} \, Y^{\ell m}_a \\
\\
\ast  & \  r^2(K^{\ell m} \, Y^{\ell m}_{ab} +  G^{\ell m} \, Z_{ab}^{\ell m})
\end{array}\right)\,,
\label{mpolar}
\end{eqnarray}
where asterisks are used to denote components that are given by the symmetry of
these tensors.  $Y^{\ell m}$ are the scalar spherical harmonics [see 
Appendix~\ref{sphericalharmonics} for the conventions that we use and other
details].  $Y^{\ell m}_a$ and $S^{\ell m}_a$ are vector spherical harmonics and
are defined (for $l\ge 1$) in terms of the scalar spherical harmonics by
\begin{equation}
Y^{\ell m}_a\equiv Y^{\ell m}_{:a}\,, \qquad
S^{\ell m}_a\equiv \epsilon_a{}^bY^{\ell m}_b\,.
\label{vec-sph-harm}
\end{equation}
Finally, $Y^{\ell m}_{ab}\,,$ $Z_{ab}^{\ell m}\,,$ and $S_{ab}^{\ell m}$ are
(symmetric) tensor spherical harmonics, which can also be defined
($Z_{ab}^{\ell m}$ and $S_{ab}^{\ell m}$ only for $l\ge 2$) in terms of the
scalar spherical harmonics by
\begin{eqnarray}
Y_{ab}^{\ell m} \equiv Y^{\ell m}\Omega_{ab}\,,~~~
Z^{\ell m}_{ab} \equiv Y^{\ell m}_{:ab}+\frac{\ell(\ell+1)}{2}Y^{\ell m}
\Omega_{ab}\,,  \label{ten-sph-harm1} 
\end{eqnarray}
\begin{eqnarray}
S^{\ell m}_{ab} \equiv S^{\ell m}_{(a:b)}\,.  \label{ten-sph-harm2}
\end{eqnarray}
Here, the sign convention for the Levi-Civita tensor associated with the metric
of the 2-sphere is: $\epsilon^{}_{\theta\varphi} = \sin{\theta}$. In
Appendix~\ref{sphericalharmonics}, we give the orthogonality relations for the
different harmonic objects. All perturbative quantities, scalar
($h_{AB}^{\ell m}$), vectorial ($p_A^{\ell m}$ and $q_A^{\ell m}$), and
tensorial ($K^{\ell  m}\,$, $G^{\ell m}\,$, and $q_2^{\ell m}$), are functions
of $t$ and $r$ only.

The metric perturbations are in general not invariant under transformations of
the mapping between the background and perturbed spacetimes, or in other words,
they are in general not invariant under gauge transformations. However, for the
case of a spherically-symmetric background, like the Schwarzschild metric, there
is a complete set of perturbative quantities that are gauge invariant.  For
polar modes this set can be chosen as follows
\begin{eqnarray}
\tilde{h}^{\ell m}_{AB} & = & h_{AB}^{\ell m} - 2 v^{\ell m}_{A|B}\,,  \\
\tilde{K}^{\ell m} & = & K^{\ell m} + \frac{\ell(\ell+1)}{2}G^{\ell m}
- 2\frac{r^{|A}}{r}v^{\ell m}_A \,,
\end{eqnarray}
where $v^{\ell m}_A = p^{\ell m}_A - (r^2/2)G^{\ell m}_{|A}$.  And for axial
modes
\begin{equation}
\tilde{h}^{\ell m}_A = h^{\ell m}_A -\frac{1}{2}H^{\ell m}_{|A}
 + \frac{r^{}_{|A}}{r}H^{\ell m}\,,
\end{equation}

The equations for the metric perturbations decouple in terms of complex master
functions, so that once we solve the decoupled equations for these master
functions all the metric perturbations can be reconstructed from them. In the
case of axial modes, it was first done by Regge and Wheeler~\cite{Regge:1957rw},
and for polar modes by Zerilli~\cite{Zerilli:1970fj} and later by
Moncrief~\cite{Moncrief:1974vm}. These functions are made out of metric
perturbations and their first derivatives and they are gauge invariant. It is
also possible to express them in a covariant form. In the case of polar modes,
the Zerilli-Moncrief function can be written as follows~\cite{Martel:2005ir}
\begin{eqnarray}
\Psi^{\ell m}_{\ZM} = \frac{r}{1+\lambda^{}_\ell}\left\{\tilde{K}^{\ell m}
+\frac{1}{\Lambda^{}_\ell}\left[r^{|A}r^{|B}\tilde{h}^{\ell m}_{AB} -
r r^{|A}\tilde{K}^{\ell m}_{|A} \right] \right\}\,,
\end{eqnarray}
where $\lambda^{}_\ell = (\ell+2)(\ell-1)/2$ and $\Lambda^{}_\ell =
\lambda^{}_\ell + 3M/r$. For axial modes, instead of using the well-known
Regge-Wheeler master function
\begin{eqnarray}
\Psi^{\ell m}_{\RW} = -\frac{f}{r}r^{|A}\tilde{h}^{\ell m}_{|A}\,,
\end{eqnarray}
we are going to use the master function introduced by Cunningham, Price and
Moncrief~\cite{Cunningham:1978cp}, in the form used
in~\cite{Jhingan:2002kb,Martel:2005ir}. The main reason for this choice is that
it is simpler to evaluate the fluxes of energy, angular momentum, and linear
momentum. Moreover, the contributions of axial modes to these physical
quantities have the same form as the one of polar modes. Nevertheless, for the
sake of completeness, we provide formulae for both master functions. The
Cunningham-Price-Moncrief master function can be written in covariant form
as~\cite{Martel:2005ir}
\begin{equation}
\Psi^{\ell m}_{\CPM} = \frac{r}{\lambda^{}_\ell}\epsilon^{AB}
\left(\tilde{h}^{\ell m}_{B|A}- \frac{2}{r}r^{}_{|A}\tilde{h}^{\ell m}_B \right)\,.
\end{equation}
In Schwarzschild coordinates these functions take the following form (the 
connection with the Regge-Wheeler parameterization of the perturbations is
given Appendix~\ref{RWparametrization})
\begin{eqnarray}
\Psi^{\ell m}_{\ZM} &=& \frac{r}{1+\lambda^{}_\ell}\left\{ K^{\ell m}
  + (1+\lambda^{}_\ell)G^{\ell m}
  \right. 
\nonumber \\
& + & \left. \frac{f}{\Lambda^{}_\ell}\left[ f h^{\ell
  m}_{rr}-r\partial^{}_r K^{\ell m} -
  \frac{2}{r}(1+\lambda^{}_\ell)p^{\ell m}_r \right] \right\} \,, 
\label{PsiZM_sch}
\end{eqnarray}
\begin{eqnarray} 
\Psi^{\ell m}_{\RW} =  -\frac{f}{r} \left(h_r^{\ell m} - \frac{1}{2}
\partial^{}_r H^{\ell m} + \frac{1}{r} H^{\ell m} \right) \,, \label{PsiRW_sch}
\end{eqnarray}
\begin{equation}
\Psi^{\ell m}_{\CPM} = -\frac{r}{\lambda^{}_\ell}\left\{ \dot{h}^{\ell m}_r
- \partial^{}_r h^{\ell m}_t + \frac{2}{r}h^{\ell m}_t \right\}\,. \label{PsiCPM_sch}
\end{equation}

These master functions obey the following wave-type equation with a potential:
\begin{equation}
\left[-\partial^2_t + \partial^2_{r^{}_{\!\ast}} - V^{\RW/\ZM}_\ell(r)\right]
\Psi_{\CPM/\ZM}^{\ell m} = 0 \,, \label{masterequations}
\end{equation}
where $r^{}_{\!\ast}$ is the so-called {\em tortoise} coordinate ($r^{}_{\!\ast}
= r + 2M\ln(r/(2M)-1)$). The potential for the axial modes is the Regge-Wheeler
potential
\begin{equation}
V^{\RW}_\ell(r) = \frac{f}{r^2}\left( \ell(\ell+1)-\frac{6M}{r}\right)\,,
\end{equation}
and the one for polar modes is the Zerilli potential
\begin{equation}
V^{\ZM}_\ell(r) =\frac{f}{r^2\Lambda^2}\left[2\lambda^2_\ell\left(1+\lambda_\ell
+ \frac{3M}{r}\right)+18\frac{M^2}{r^2}\left(\lambda_\ell+\frac{M}{r}\right)
\right]\,,
\end{equation}

Once the different master functions have been computed we can estimate the
energy and angular momentum carried out by the radiation field to infinity. We
can do this by using the expressions of the energy and angular momentum fluxes
at infinity obtained from the Isaacson's averaged energy-momentum tensor for
gravitational waves~\cite{Isaacson:1968ra,Isaacson:1968gw} (see
also~\cite{Misner:1973cw,Thorne:1980rm}).  In terms of the axial and polar
master functions the expressions are
\begin{eqnarray}
\dot{E}^{}_{\GW} = \frac{1}{64\pi}\sum^{}_{\ell\geq 2, m}
\frac{(\ell+2)!}{(\ell-2)!}\left(|\dot{\Psi}_{\CPM}^{\ell m}|^2 +
|\dot{\Psi}_{\ZM}^{\ell m}|^2\right)\,,
\end{eqnarray}
\begin{eqnarray}
\dot{L}^{}_{\GW} = \frac{1}{64\pi}\sum^{}_{\ell\geq 2, m}i m
\frac{(\ell+2)!}{(\ell-2)!}\left(\bar{\Psi}_{\CPM}^{\ell m}
\dot{\Psi}_{\CPM}^{\ell m} + \bar{\Psi}_{\ZM}^{\ell m}
\dot{\Psi}_{\ZM}^{\ell m}\right)\,.
\end{eqnarray}
We can also construct the metric {\em waveforms} by using 
\begin{equation}
h^{}_{+} - i h^{}_{\times} = \frac{1}{2r}\sum^{}_{\ell\geq 2, m}
\sqrt{\frac{(\ell+2)!}{(\ell-2)!}}\left( \Psi_{\ZM}^{\ell m}+
i \Psi_{\CPM}^{\ell m} \right) {}^{}_{-2}Y^{\ell m} \,, \label{waveforms}
\end{equation}
where ${}^{}_{-2}Y^{\ell m}$ denotes the spherical harmonics of spin weight $-2$
(see, e.g.~\cite{Goldberg:1967sp} and Appendix~\ref{spinweighted} for details).
In this work we are interested in studying the gravitational recoil due to the
merger of unequal-mass black-hole binary systems and therefore, we want to
evaluate the flux of linear momentum emitted in gravitational waves. This
quantity can also be computed from the Isaacson's energy-momentum tensor and can
be written in terms of the metric waveforms as follows:
\begin{equation}
\dot{P}^{k}_{\GW} = \frac{r^2}{16\pi}\int d\Omega \; \hat{r}^k_{obs}\,
\left( \dot{h}^{2}_+ + \dot{h}^2_{\times}\right) \,, \label{linearmomentumflux}
\end{equation}
where $\hat{r}^k_{obs}$ is a unit vector that points from the source to the
observer. We can then express the components of $\hat{r}^k_{obs}$ in terms of
scalar spherical harmonics as
\begin{equation}
\hat{r}^k_{obs} = -2\sqrt{\frac{2\pi}{3}}\left(\Re(Y^{1,1})\,, \Im(Y^{1,1}) \,, 
-\frac{Y^{1,0}}{\sqrt{2}} \right)\,, \label{unitvectorobserver}
\end{equation}
where $\Re$ and $\Im$ denote the real and imaginary parts of a complex number.
By simple inspection of the linear momentum flux in
Eq.~(\ref{linearmomentumflux}), and taking into account the harmonic structure
of the metric waveforms in Eq.~(\ref{waveforms}) and of $\hat{r}^k_{obs}$ in
Eq.~(\ref{unitvectorobserver}), we realize that all terms in the flux contain
the product of three spherical harmonic objects. Therefore, in order to obtain a
practical expression for $\dot{P}^{k}_{\GW}$ we need to use the machinery for
studying coupled angular momenta common in quantum
physics~\cite{Condon:1970cs,Arfken}. The calculation goes along the lines
described in~\cite{Thorne:1980rm}, and some details are given in
Appendix~\ref{lin-mom}. The result can be written in the following form
\begin{widetext}
\begin{eqnarray}
\dot{P}^x_{\GW} & = & - \frac{1}{64\pi}\sum^{}_{\ell \geq 2, m}
\frac{(\ell+3)!}{(\ell-2)!} \frac{1}{(\ell+1)\sqrt{(2\ell+3)(2\ell+1)}}
\left\{ \sqrt{(\ell+m+2)(\ell+m+1)} \left(\dot{\Psi}^{\ell m}_{\ZM}
\dot{\bar{\Psi}}^{\ell+1,m+1}_{\ZM} + \dot{\Psi}^{\ell m}_{\CPM}
\dot{\bar{\Psi}}^{\ell+1,m+1}_{\CPM} \right) \right.
\nonumber \\
& - & \left. \sqrt{(\ell-m+2)(\ell-m+1)} \left(\dot{\Psi}^{\ell m}_{\ZM}
\dot{\bar{\Psi}}^{\ell+1,m-1}_{\ZM} + \dot{\Psi}^{\ell m}_{\CPM}
\dot{\bar{\Psi}}^{\ell+1,m-1}_{\CPM} \right) \right\}
\nonumber \\
& - & \frac{i}{64\pi} \sum^{}_{\ell \geq 2, m} (\ell+2)(\ell-1) \left\{
\sqrt{(\ell-m)(\ell+m+1)} \left(\dot{\Psi}^{\ell m}_{\ZM}
\dot{\bar{\Psi}}^{\ell,m+1}_{\CPM} - \dot{\Psi}^{\ell m}_{\CPM}
\dot{\bar{\Psi}}^{\ell,m+1}_{\ZM} \right) \right.
\nonumber \\
& + &  \left.  \sqrt{(\ell+m)(\ell-m+1)} \left(\dot{\Psi}^{\ell m}_{\ZM}
\dot{\bar{\Psi}}^{\ell,m-1}_{\CPM}  - \dot{\Psi}^{\ell m}_{\CPM}
\dot{\bar{\Psi}}^{\ell,m-1}_{\ZM} \right) \right\}
\,,  \label{recoilx}  
\end{eqnarray}
\begin{eqnarray}
\dot{P}^y_{\GW} & = &  \frac{i}{64\pi}\sum^{}_{\ell \geq 2, m}
\frac{(\ell+3)!}{(\ell-2)!} \frac{1}{(\ell+1)\sqrt{(2\ell+3)(2\ell+1)}}
\left\{ \sqrt{(\ell+m+2)(\ell+m+1)} \left(\dot{\Psi}^{\ell m}_{\ZM}
\dot{\bar{\Psi}}^{\ell+1,m+1}_{\ZM} + \dot{\Psi}^{\ell m}_{\CPM}
\dot{\bar{\Psi}}^{\ell+1,m+1}_{\CPM} \right) \right.
\nonumber \\
& + & \left. \sqrt{(\ell-m+2)(\ell-m+1)} \left(\dot{\Psi}^{\ell m}_{\ZM}
\dot{\bar{\Psi}}^{\ell+1,m-1}_{\ZM} + \dot{\Psi}^{\ell m}_{\CPM}
\dot{\bar{\Psi}}^{\ell+1,m-1}_{\CPM} \right) \right\}
\nonumber \\
& - &\frac{1}{64\pi} \sum^{}_{\ell \geq 2, m} (\ell+2)(\ell-1) \left\{
\sqrt{(\ell-m)(\ell+m+1)} \left(\dot{\Psi}^{\ell m}_{\ZM}
\dot{\bar{\Psi}}^{\ell,m+1}_{\CPM} - \dot{\Psi}^{\ell m}_{\CPM}
\dot{\bar{\Psi}}^{\ell,m+1}_{\ZM} \right) \right.
\nonumber \\ 
& - &  \left. \sqrt{(\ell+m)(\ell-m+1)} \left(\dot{\Psi}^{\ell m}_{\ZM}
\dot{\bar{\Psi}}^{\ell,m-1}_{\CPM} - \dot{\Psi}^{\ell m}_{\CPM}
\dot{\bar{\Psi}}^{\ell,m-1}_{\ZM} \right) \right\}
\,,  \label{recoily}
\end{eqnarray}
\begin{eqnarray}
\dot{P}^z_{\GW} & = &  \frac{1}{32\pi}\sum^{}_{\ell \geq 2, m}
\frac{(\ell+3)!}{(\ell-2)!} \sqrt{\frac{(\ell+m+1)(\ell-m+1)}
{(2\ell+3)(2\ell+1)(\ell+1)^2}} \left(\dot{\Psi}^{\ell m}_{\ZM}
\dot{\bar{\Psi}}^{\ell+1,m}_{\ZM} + \dot{\Psi}^{\ell m}_{\CPM}
\dot{\bar{\Psi}}^{\ell+1,m}_{\CPM} \right)
\nonumber \\
& - & \frac{i}{32\pi} \sum^{}_{\ell \geq 2, m} m\,(\ell+2)(\ell-1)
\left(\dot{\Psi}^{\ell m}_{\ZM}\dot{\bar{\Psi}}^{\ell,m}_{\CPM} -
\dot{\Psi}^{\ell m}_{\CPM}\dot{\bar{\Psi}}^{\ell,m}_{\ZM} \right)
\,. \label{recoilz}
\end{eqnarray}
\end{widetext}
In conclusion, all we need to extract relevant physical information is the
master functions. In the next subsections, we extract initial data for these
master functions.

\subsection{Relation between ADM variables and metric perturbation}
In section~\ref{bbhid}, we constructed initial data for a binary black hole
system in coalescence. The procedure used for this construction was based on the
$3+1$ ADM formalism~\cite{ArnDesMis:62} and, hence, the initial data is given in
terms of ADM variables [Eqs.~(\ref{cfactord3}), (\ref{K-first}) and
(\ref{pextrinsic})]. Then, in order to build initial data for the evolution of
the master functions, we need to first find the relation between the ADM
variables and the metric perturbations (see, e.g.~\cite{Abrahams:1995gn}). This
means that we need to use the relations between the components of the 3-metric
$\gamma^{}_{ij}$ and the metric perturbations, and also the relations between
the metric perturbations and their first derivatives and the components of the
extrinsic curvature $K^{}_{ij}\,.$ For the former, we use the fact that the
components of the $3$-metric are the spatial components of the orthogonal
projection operator on the hypersurfaces of the spacetime slicing, described by
a normal $n^{}_{\mu}$:
\begin{equation} 
\gamma^{}_{\mu\nu} = \met^{}_{\mu\nu} + n^{}_{\mu}n^{}_{\nu}\,. 
\label{3-metric_metric}
\end{equation}
Then, the different modes of the harmonically decomposed $3$-metric are related
to the metric perturbations via
\begin{eqnarray}
\gamma^{\ell m}_{tt} & = & -f\,\delta^{0,0}\,,  \label{gammatt} \\
\gamma^{\ell m}_{tr} & = & h^{\ell m}_{tr}Y^{\ell m}\,,  \label{gammatr} \\ 
\gamma^{\ell m}_{ta} & = & p^{\ell m}_t Y^{\ell m}_a + h^{\ell m}_t 
S^{\ell m}_a\,, \label{gammata} \\
\gamma^{\ell m}_{rr} & = & f^{-1}\,\delta^{0,0} + h^{\ell m}_{rr}Y^{\ell m} \,, 
\label{gammarr}  \\
\gamma^{\ell m}_{ra} & = & p^{\ell m}_r Y^{\ell m}_a + h^{\ell m}_r 
S^{\ell m}_a\,, \label{gammara} \\
\gamma^{\ell m}_{ab} & = & r^2 \Omega^{}_{ab}\,\delta^{0,0}
+ r^2\left(K^{\ell m}Y^{\ell m}_{ab} + G^{\ell m}Z^{\ell m}_{ab}\right)
\nonumber \\
& + & H^{\ell m} S^{\ell m}_{ab} \,.  \label{gammaab}
\end{eqnarray}
The first three equations are related to the choice of slicing, that is, to the
choice of shift vector $\beta^i$ and lapse $\alpha\,.$ Actually, the lapse and
shift at first order are given by
\begin{eqnarray}
\alpha^2 & = & f(1 - fh^{tt}) = f - h^{}_{tt} + {\cal{O}}(h^2)\,, 
\label{admgauge1} \\
\beta^i & = & h_t{}^i + {\cal{O}}(h^2)\,. \label{admgauge2}
\end{eqnarray}

The relation between the extrinsic curvature and the metric perturbations can be
found through the relation between the 3-metric and the extrinsic curvature
\begin{equation}
K^{}_{\mu\nu} = -\textstyle{\frac{1}{2}}\pounds^{}_{\mathrm{\scriptsize \bf n}}
\gamma^{}_{\mu\nu}\,, 
\end{equation}
where the symbol $\pounds$ denotes Lie differentiation, and
Eq.~(\ref{3-metric_metric}) between the 3-metric and the spacetime metric. In
this way, we find that the different modes of the harmonically decomposed
extrinsic curvature are related to the metric perturbations by the following
expressions:
\begin{widetext}
\begin{eqnarray}
K^{\ell m}_{rr} & = & \frac{1}{2\sqrt{f}}\left[ \dot{h}^{\ell m}_{rr}
-2 h_{tr}^{\ell m}{}' -\frac{f'}{f} h^{\ell m}_{tr} \right] Y^{\ell m}\,,
\label{Kphysrr}  \\
K^{\ell m}_{ra} & = & \frac{1}{2\sqrt{f}}\left\{ \left[\dot{p}^{\ell m}_r
- p_t^{\ell m}{}' +\frac{2}{r}p^{\ell m}_t - h^{\ell m}_{tr} \right]
Y^{\ell m}_a + \left[ \dot{h}^{\ell m}_r - h_t^{\ell m}{}'
+ \frac{2}{r}h^{\ell m}_t \right] S^{\ell m}_a \right\} \,,
\label{Kphysra}  \\
K^{\ell m}_{ab} & = & \frac{r^2}{2\sqrt{f}}\left\{ \left[\dot{K}^{\ell m}
+ \frac{\ell(\ell+1)}{r^2}p^{\ell m}_t -\frac{2f}{r}h^{\ell m}_{tr} \right]
Y^{\ell m}_{ab} + \left[ \dot{G}^{\ell m} - \frac{2}{r^2}p^{\ell m}_t \right]
Z^{\ell m}_{ab} + \frac{1}{r^2}\left[ \dot{H}^{\ell m} - 2h^{\ell m}_t\right]
S^{\ell m}_{ab}\right\}\,,  \label{Kphysab}
\end{eqnarray}
\end{widetext}
where the dots and primes denote partial differentiation with respect to time
$t$ and radial coordinate $r$ respectively.

\subsection{Initial data for the metric perturbations}

Before computing initial data for the master functions, we must find data for
the metric perturbations, that is, find $(h^{}_{\mu\nu},\dot{h}^{}_{\mu\nu})$,
in the parameterization given in Eqs.~(\ref{maxial}) and~(\ref{mpolar}) on the
initial slice $t=t^{}_o\,.$ To begin with, since our $3$-metric is conformally
flat, the following metric perturbations vanish on the initial slide:
\begin{equation}
p^{\ell m}_r = G^{\ell m} = 0\,, \qquad
h^{\ell m}_r = H^{\ell m} = 0\,.
\end{equation}
We have also seen that the conformal factor, the physical and the conformal
extrinsic curvatures can be formally expanded in powers of $d$ and $P$ as
follows
\begin{eqnarray}
\Phi & = & \Phi^{}_{(0)} + \Phi^{}_{(2)} d^2 + \Phi^{}_{(3)} d^3 +
{\cal O}(P^2 d^2,d^4) \,,   \\
\hat{K}^{}_{ij} & = & P d \; \hat{K}^{}_{(1)ij} + P d^2 \;
\hat{K}^{}_{(2)ij} + {\cal O}(P d^3) \,,
\end{eqnarray}
where $\hat{K}^{}_{(1)ij}$ and $\hat{K}^{}_{(2)ij}$ are the coefficients of the
terms of order $Pd$ and $Pd^2$ respectively in the expansion of
$\hat{K}^{}_{ij}\,$. Then, the physical extrinsic curvature, $K^{}_{ij}$, given
by Eq.~(\ref{pextrinsic}), can be formally expanded in the form
\begin{equation}
K^{}_{ij} = \Phi^{-2}_{(0)} \left(P d \; \hat{K}^{}_{(1)ij} + P d^2\;
\hat{K}^{}_{(2)ij} \right) + {\cal O}(P d^3)\,, \label{formalextrinsiccurvature}
\end{equation}
which means that in order to obtain the physical extrinsic curvature up to
${\cal{O}}(P d^2)$ we only need the zero-th order piece of the conformal factor.
The explicit expressions of the coefficients of these expansions are given by
equations (\ref{cfactord3}), (\ref{K-first}) and (\ref{pextrinsic}).

With this in mind, we are going to extract the remaining modes of the initial
data. To that end, we use the expression of the separation vector
$\hat{\mb{\epsilon}}$ in spherical coordinates, namely
\begin{equation}
{\hat{\epsilon}}^i = \left(\sin{\theta} \cos{\varphi}\,,
\frac{\cos{\theta}\cos{\varphi}}{R}\,,
-\frac{\sin{\varphi}}{R\cos{\theta}}\right)\,.
\label{sep-vec}
\end{equation}
Then, the non-zero components of the $3$-metric on the initial slice are given
by
\begin{eqnarray}
\gamma_{rr} & = & f^{-1} \left[ 1 + 2 \frac{q}{(1 + q)^2} \frac{M d^2}{r^3}
\frac{1}{\sigma^5} P_2(\xi) 
\right. \nonumber \\
& - & \left. 2 \frac{q(1 - q)}{(1 + q)^3} \frac{M d^3}{r^4} \frac{1}{\sigma^7}
  P_3(\xi) \right]+ {\cal{O}}(P^2 d^2,d^4)\,,  
\end{eqnarray}
\begin{eqnarray}
\gamma_{ab} & = & r^2 \Omega_{ab} \left[ 1 + 2 \frac{q}{(1 + q)^2}
  \frac{M d^2}{r^3} \frac{1}{\sigma^5} P_2(\xi) 
\right.
\nonumber \\
& - & \left. 2 \frac{q(1 - q)}{(1 + q)^3} \frac{Md^3}{r^4} \frac{1}{\sigma^7}
  P_3(\xi)\right] + {\cal{O}}(P^2 d^2,d^4)\,,
\end{eqnarray}
where we have introduced the following definitions
\begin{equation}
\xi = \sin{\theta}\cos{\varphi}\,,\qquad
\sigma = \frac{1 + \sqrt{f}}{2}\,. 
\end{equation}
We can now rewrite the 3-metric in terms of spherical harmonics as follows
\begin{widetext}
\begin{eqnarray}
\gamma_{rr} &=& f^{-1} \left\{ 1 - 2\sqrt{\frac{\pi}{5}}\, \frac{q}{(1 + q)^2}
\frac{M d^2}{r^3} \frac{1}{\sigma^5} \left[ Y^{2,0} - \sqrt{6}\, \Re(Y^{2,2})
\right] - 2\sqrt{\frac{\pi}{7}}\, \frac{q(1 - q)}{(1 + q)^3} \frac{Md^3}{r^4}
\frac{1}{\sigma^7}  \left[  \sqrt{3}\, \Re(Y^{3,1})  \right. \right.
\nonumber \\
&-& \left.  \left.  \sqrt{5}\,\Re(Y^{3,3})  \right] \right\}
+ {\cal{O}}(P^2 d^2,d^4) \,, \label{gammaharmonic1}
\end{eqnarray}
\begin{eqnarray}
\gamma^{}_{ab} &=& r^2 \Omega_{ab} - 2\sqrt{\frac{\pi}{5}}\, \frac{q}{(1 + q)^2}
\frac{Md^2}{r} \frac{1}{\sigma^5} \left[  Y_{ab}^{2,0} - \sqrt{6}\,
\Re(Y_{ab}^{2,2})\right] - 2 \sqrt{\frac{\pi}{7}}\, \frac{q(1 - q)}{(1 + q)^3}
\frac{Md^3}{r^2} \frac{1}{\sigma^7} \left[ \sqrt{3}\, \Re(Y_{ab}^{3,1})\right.
\nonumber \\
&-& \left. \sqrt{5}\, \Re(Y^{3,3}_{ab})  \right] +   {\cal{O}}(P^2 d^2,d^4)\,.
\label{gammaharmonic2}
\end{eqnarray}
\end{widetext}

In order to repeat this procedure with the extrinsic curvature, we first need to
compute the components of the conformal extrinsic curvature with the separation
vector $\mb{\hat{\epsilon}}$ of Eq.~(\ref{sep-vec}). The components of the
conformal extrinsic curvature are given by
\begin{widetext}
\begin{eqnarray}
\hat{K}^{}_{RR} &=& 3 \frac{P d}{R^3} \sin^2{\theta} \sin(2 \varphi) -
3 \frac{1 - q}{1 + q} \frac{P d^2}{R^4}  \sin{\varphi} \sin{\theta}
\left( 5 \sin^2{\theta} \cos^2{\varphi} - 2 \right) + {\cal{O}}(P d^3)\,, 
\\
\hat{K}^{}_{R\theta} &=& \frac{3}{4}  \frac{1 - q}{1 + q} \frac{P d^2}{R^3}
\cos{\theta} \sin{\varphi} \left(5 \sin^2{\theta} \cos^2{\varphi} + 3
\right) + {\cal{O}}(P d^3)\,, 
\\
\hat{K}^{}_{R\varphi} &=& 3 \frac{P d}{R^2} \sin^2{\theta} + \frac{3}{4}
\frac{1 - q}{1 + q} \frac{P d^2}{R^3} \sin{\theta} \cos{\varphi}  \left[
  \sin^2{\theta}  \left( 5 \cos^2{\varphi} - 14 \right) + 3 \right] +
{\cal{O}}(P d^3)\,,  
\\
\hat{K}^{}_{\theta\theta} &=& \frac{3}{8} \frac{P d }{R} \sin(2\varphi) \left[
  \cos(2 \theta) - 5 \right] - \frac{3}{4} \frac{1 - q}{1 +q} \frac{P
  d^2}{R^2}  \sin{\theta}\sin{\varphi} \left[ 5\cos^2{\varphi} 
\left( \cos^2{\theta} - 3 \right) + 3 \right] + {\cal{O}}(P d^3)\,, 
\\
\hat{K}^{}_{\theta\varphi} &=& - \frac{3}{4} \frac{Pd}{R} \sin(2\theta)
\cos(2\varphi) + \frac{3}{2} \frac{1 - q}{1 + q} \frac{P d^2 }{R^2}
\sin^2{\theta} \cos{\theta} \cos{\varphi} \left( 5 \cos^2{\varphi}
-2 \right) + {\cal{O}}(P d^3)\,,
\\
\hat{K}^{}_{\varphi\varphi} &=& \frac{3}{8} \frac{P d}{R} \sin^2\theta
\sin(2\varphi) \left[1 + 3 \cos(2\theta) \right] + \frac{15}{4}
\frac{1 - q}{1 + q} \frac{P d^2}{R^2}  \sin^3{\theta} \sin{\varphi}
\left[\cos^2{\varphi} \left(3 \sin^2{\theta} -2\right) - 1\right]
+ {\cal{O}}(P d^3)\,.
\end{eqnarray}
\end{widetext}
Here we have checked that the terms of ${\cal{O}}(Pd)$ agree with those
in~\cite{Khanna:2000dg} (up to a typo in their value of the
$\{\theta,\varphi\}$ component). The next step is the calculation of the
physical extrinsic curvature in terms of spherical harmonics and Schwarzschild
coordinates. This quantity is given by
\begin{widetext}
\begin{eqnarray}
K^{}_{rr} &=& 4 \sqrt{\frac{6\pi}{5}}\, \frac{P d}{r^3} \frac{1}{f}\Im(Y^{2,2}) 
- 2\sqrt{\frac{3\pi}{7}}\, \frac{1 - q}{1 + q} \frac{Pd^2}{r^4}
\frac{1}{f\sigma^2} \left[ \sqrt{14}\,\Im(Y^{1,1}) + \Im(Y^{3,1}) - \sqrt{15}\,
\Im(Y^{3,3})\right] + {\cal{O}}(P d^3)\,, \label{Kharmonic1}
\\
K^{}_{ra} &=& 2 \sqrt{3\pi}\, \frac{Pd}{r^2} \frac{1}{\sqrt{f}} S^{1,0}_a 
+ \frac{1}{2}\sqrt{\frac{\pi}{21}}\, \frac{1-q}{1+q}\frac{Pd^2}{r^3}
\frac{1}{\sqrt{f}\,\sigma^2}  \left[ 6\sqrt{14}\,\Im(Y^{1,1}_a)
+ 16\sqrt{\frac{14}{5}}\, \Re(S^{2,1}_a)  + \Im(Y^{3,1}_a) \right.
\nonumber \\
&-& \left. \sqrt{15}\,\Im(Y^{3,3}_{a})\right] + {\cal{O}}(Pd^3)\,,
\label{Kharmonic2} \\
K^{}_{ab} &=& -\sqrt{\frac{6\pi}{5}}\, \frac{P d}{r} \left[ 2\,\Im(Y_{ab}^{2,2})
+ \Im(Z_{ab}^{2,2}) \right] + \frac{1}{2}\sqrt{\frac{\pi}{21}}\,
\frac{1 - q}{1 + q} \frac{P d^2}{r^2} \frac{1}{\sigma^2} \left\{ 6\sqrt{14}\,
\Im(Y^{1,1}_{ab}) - 8\sqrt{\frac{14}{5}}\,\Re(S^{2,1}_{ab})\right.
\nonumber \\
&+& \left.   6\, \Im(Y^{3,1}_{ab}) + \Im(Z^{3,1}_{ab}) - \sqrt{15}\, \left[
6\,\Im(Y^{3,3}_{ab}) + \Re(Z^{3,3}_{ab}) \right] \right\} + {\cal{O}}(P d^3)\,. 
\label{Kharmonic3}
\end{eqnarray}
\end{widetext}
Eqs.~(\ref{gammaharmonic1},\ref{gammaharmonic2})
and~(\ref{Kharmonic1})-(\ref{Kharmonic3}) give the complete harmonic
decomposition of the initial data $(\gamma^{}_{ij},K^{}_{ij})$. We must now
extract the initial values of the metric perturbations and their time
derivative by comparing these expressions with
Eqs.~(\ref{gammarr})-(\ref{gammaab}) and~(\ref{Kphysrr})-(\ref{Kphysab}). To
simplify notation, we now drop the truncation error in all equations, since it
has already been given in the main expansions. Comparison of
Eqs.~(\ref{gammaharmonic1},\ref{gammaharmonic2}) with
Eqs.~(\ref{gammarr})-(\ref{gammaab}) yields the non-vanishing initial metric
perturbations, namely
\begin{eqnarray}
K^{2,0} &=& f h^{2,0}_{rr} = - 2 \sqrt{\frac{\pi}{5}} \frac{q}{(1+q)^2} \,, 
\label{id_hmunu1} \\
K^{2,\pm 2} &=&  f h^{2,\pm 2}_{rr} = \sqrt{\frac{6 \pi}{5}}
\frac{q}{(1+q)^2} \frac{Md^2}{r^3}\frac{1}{\sigma^5} \,,  \label{id_hmunu2} \\
K^{3,\pm 1} &=& f h_{rr}^{3,\pm 1} = \mp \sqrt{\frac{3\pi}{7}}
\frac{q(1 - q)}{(1 + q)^3} \frac{Md^3}{r^4} \frac{1}{\sigma^7}\,,
\label{id_hmunu3} \\
K^{3,\pm 3} &=& f h_{rr}^{3,\pm 3} = \pm \sqrt{\frac{5\pi}{7}} 
\frac{q(1 - q)}{(1 + q)^3} \frac{Md^3}{r^4} \frac{1}{\sigma^7}\,.
\label{id_hmunu4}
\end{eqnarray}
As we can see, all the axial metric perturbations initially vanish. Now, in
order to obtain the initial values of the time derivative of the metric
perturbations, we must compare Eqs.~(\ref{Kharmonic1})-(\ref{Kharmonic3}) with
Eqs.~(\ref{Kphysrr})-(\ref{Kphysab}). It is important to realize that in
Eqs.~(\ref{Kphysrr})-(\ref{Kphysab}) there are terms that are associated with
the gauge freedom of choosing the slicing, more specifically, terms associated
with components of the shift vector [see Eq.~(\ref{admgauge2})]. Moreover, there
is no unique way of assigning values to the different time derivatives of the
metric perturbations and the metric perturbations themselves. This reflects the
fact that the values of the metric perturbations are gauge dependent, since in
general the components of the metric perturbations (and their time derivatives)
are not gauge invariant. Keeping this in mind, we have assigned the following
initial values to the time derivatives of the metric perturbations: the
non-vanishing polar modes are
\begin{eqnarray}
\dot{h}_{rr}^{2,\pm 2} &=& \mp 24 i \sqrt{\frac{\pi}{30}}\,
\frac{Pd}{r^3} \frac{1}{\sqrt{f}} \,, \label{id_hmunudot1}  \\
\dot{h}_{rr}^{3,\pm 1} &=& 2 i \sqrt{\frac{3\pi}{7}} \,
\frac{1-q}{1 + q} \frac{Pd^2}{r^4} \frac{1}{\sqrt{f}\,\sigma^2}\,,  
\label{id_hmunudot2} \\
\dot{h}_{rr}^{3,\pm 3} &=& - 6 i \sqrt{\frac{5\pi}{7}} \,
\frac{1-q}{1 + q} \frac{P d^2}{r^4} \frac{1}{\sqrt{f}\,\sigma^2}\,,  
\label{id_hmunudot3} \\
\dot{p}_{r}^{3,\pm 1} &=& - \frac{i}{2} \sqrt{\frac{\pi}{21}}\,
\frac{1 - q}{1 + q} \frac{Pd^2}{r^3} \frac{1}{\sigma^2} \,,  
\label{id_hmunudot4} \\
\dot{p}_{r}^{3,\pm 3} &=& \frac{i}{2} \sqrt{\frac{5\pi}{7}} \,
\frac{1 - q}{1 + q} \frac{Pd^2}{r^3} \frac{1}{\sigma^2}\,,  
\label{id_hmunudot5} \\
\dot{G}^{2,\pm 2} &=& \pm i \sqrt{\frac{6\pi}{5}}\, 
\frac{Pd}{r^3} \sqrt{f} \,, \label{id_hmunudot6} \\
\dot{K}^{2,\pm 2} &=& \pm 2 i \sqrt{\frac{6\pi}{5}} \,
\frac{Pd}{r^3} \sqrt{f} \,, \label{id_hmunudot7} \\
\dot{G}^{3,\pm 1} &=& - \frac{i}{2}\sqrt{\frac{\pi}{21}}\,
\frac{1 - q}{1 + q} \frac{Pd^2}{r^4} \frac{\sqrt{f}}{\sigma^2} \,, 
\label{id_hmunudot8} \\
\dot{K}^{3,\pm 1} &=& - i \sqrt{\frac{3\pi}{7}} \,
\frac{1 - q}{1 + q} \frac{P d^2}{r^4} \frac{\sqrt{f}}{\sigma^2} \,,  
\label{id_hmunudot9} \\
\dot{G}^{3,\pm 3} &=& \frac{i}{2} \sqrt{\frac{5\pi}{7}} \,
\frac{1 - q}{1 + q} \frac{P d^2}{r^4} \frac{\sqrt{f}}{\sigma^2} \,, 
\label{id_hmunudot10} \\
\dot{K}^{3,\pm 3} &=&  3 i \sqrt{\frac{5\pi}{7}} \,
\frac{1 - q}{1 + q} \frac{P d^2}{r^4} \frac{\sqrt{f}}{\sigma^2} \,;
\label{id_hmunudot11}
\end{eqnarray}
and the axial ones are
\begin{eqnarray}
\dot{h}_{r}^{1,0} &=& 4 \sqrt{3\pi}\, \frac{P d}{r^2} \,,  
\label{id_hmunudot12} \\
\dot{h}_{r}^{2,\pm 1} &=& \pm 8 \sqrt{\frac{2\pi}{15}}\, 
\frac{1 - q}{1 + q} \frac{P d^2}{r^3} \frac{1}{\sigma^2} \,, 
\label{id_hmunudot13} \\
\dot{H}^{2,\pm 1} &=& \mp 4 \sqrt{\frac{2\pi}{15}}\,
\frac{1 - q}{1 + q} \frac{P d^2}{r^2} \frac{\sqrt{f}}{\sigma^2} \,. 
\label{id_hmunudot14}
\end{eqnarray}
By using the correspondence between our parameterization of the metric
perturbations and that of Regge Wheeler [Eqs.~(\ref{map1})-(\ref{map2}) in
Appendix~\ref{RWparametrization}] we have checked that up to ${\cal{O}}(d^2)$
and ${\cal{O}}(Pd)$ our expressions agree with those found
in~\cite{Khanna:2000dg} (up to a typo in their $\dot{h}^{1,0}_r$). We have
decided to assign values to the time derivatives of the metric perturbations and
the metric perturbations themselves by the following usual convention: all modes
with $\ell=1$ are assigned to metric perturbations associated with the shift
vector. These perturbations represent either translations or rotations of the
observers associated with the normal to the initial slice with respect to our
coordinate system.  In our case, these perturbations are given through the
following relationships
\begin{eqnarray}
f h^{1,\pm1'}_{tr} + \frac{f'}{2} h^{1,\pm1}_{tr}  =   - i \sqrt{6\pi}\,
\frac{1 - q}{1 + q} \frac{P d^2}{r^4} \frac{\sqrt{f}}{\sigma^2} \,,
\label{constraint1}
\end{eqnarray}
\begin{eqnarray}
p^{1,\pm1'}_{t} - \frac{2}{r} p^{1,\pm1}_{t} + h^{1,\pm1}_{tr}  =  i 
\sqrt{6\pi}\, \frac{P d^2}{r^3} \frac{1-q}{1+q} \frac{1}{\sigma^2} \,,  
\label{constraint2}
\end{eqnarray}
\begin{eqnarray}
p^{1,\pm1}_{t} - r f h^{1,\pm1}_{tr} &=& - i \sqrt{\frac{3\pi}{2}}\,
\frac{1-q}{1+q} \frac{P d^2}{r^2} \frac{\sqrt{f}}{\sigma^2}\,.
\label{constraint3}
\end{eqnarray}
Here, two comments are in order. First, one can see that these equations are
consistent in the sense that the derivative of Eq.~(\ref{constraint3}) with
respect to $r$ can be reduced to a trivial identity  by using
Eqs.~(\ref{constraint1}) and~(\ref{constraint2}). Second, from these equations
we can immediately see that the shift vector is different from zero
[see Eq.~(\ref{admgauge2})]. A non-zero shift could in principle be a problem if
we wanted to place observers at constant $r$ (in the wave zone), evaluate the
linear momentum flux, and then infer a recoil velocity of the final black hole
after the merger. If we were to do this, the measured velocity would have a
component due to the motion of the observers with respect to the position of the
final black hole, as described by the shift vector. This contribution would then
have to be subtracted, but it can be seen that the shift vector decays quite
fast as $r$ becomes large and, hence, this effect would be negligible.

\subsection{Initial data for the master functions}  \label{ZM-RW-sec}

Using the initial data for the metric perturbations 
[Eqs.~(\ref{id_hmunu1})-(\ref{id_hmunu4}),
(\ref{id_hmunudot1})-(\ref{id_hmunudot11}), and
Eqs.~(\ref{id_hmunudot12})-(\ref{id_hmunudot14})] in the master functions
[Eqs.~(\ref{PsiZM_sch})-(\ref{PsiCPM_sch})], we can compute initial data for
them: $(\Psi^{\ell m}_{\ZM}\,,\,\dot\Psi^{\ell m}_{\ZM})\,,$
$(\Psi^{\ell m}_{\RW}\,,\,\dot\Psi^{\ell m}_{\RW})\,,$ and
$(\Psi^{\ell m}_{\CPM}\,,\,\dot\Psi^{\ell m}_{\CPM})$ on the initial slice
$t=t^{}_o\,.$

The results for the Regge-Wheeler master function are:
\begin{eqnarray}
\Psi^{\ell m}_{\RW} =  0\,, 
\end{eqnarray} 
\begin{eqnarray}
\dot{\Psi}^{2\,, \pm 1}_{\RW} = \mp 2 \sqrt{\frac{2 \pi}{15}}
\frac{1 - q}{1 + q} \frac{P d^2}{r^4} \frac{\sqrt{f} (1 -
  \sqrt{f})}{\sigma^2} \left(7 \sigma - 3 \right) \,, 
\end{eqnarray}
and for the Cunningham-Price-Moncrief master function are
\begin{eqnarray}
\Psi^{2,\pm 1}_{\CPM} &=& \mp \frac{4}{3} \frac{\sqrt{30\pi}}{5}\,
\frac{1-q}{1+q} \frac{P d^2}{r^2} \frac{1}{\sigma^2} \,, \\
\dot{\Psi}^{\ell m}_{\CPM} &=& 0 \,.
\end{eqnarray}

In the same way, the non-vanishing initial data for the Zerilli-Moncrief master 
functions is given by
\begin{eqnarray}
\Psi^{2,0}_{\ZM} & = &  -\frac{2}{3} \sqrt{\frac{\pi}{5}}
  \frac{q}{(1 + q)^2} \frac{M d^2}{r^2} \frac{1 + 5 \sigma}{\Lambda_2
  \sigma^5} \,, \\
\Psi^{2,\pm 2}_{\ZM} & = & \sqrt{\frac{2 \pi}{15}} \frac{q}{(1+q)^2} 
\frac{M d^2}{r^2} \frac{1 + 5 \sigma}{\Lambda_2\sigma^5} \,,  \\
\Psi^{3,\pm 1}_{\ZM} & = & \mp \frac{1}{2} \sqrt{\frac{\pi}{21}}
  \frac{q(1 - q)}{(1 + q)^3} \frac{M d^3}{r^3} \frac{3 + 7
  \sigma}{\Lambda_3 \sigma^{7}} \,, \\
\Psi^{3,\pm 3}_{\ZM} & = & \pm \frac{1}{6} \sqrt{\frac{5 \pi}{7}}
  \frac{q (1 - q)}{(1 + q)^3} \frac{M d^3}{r^3} \frac{3 + 7
  \sigma}{\Lambda_3 \sigma^{7}} \,, 
\end{eqnarray}
\begin{eqnarray}
\dot{\Psi}^{2,\pm 2}_{\ZM} =  \pm i \sqrt{\frac{6 \pi}{5}}
\frac{P d}{r^2} \frac{\sqrt{f}}{\Lambda_2} \left(4 + \frac{3 M}{r} \right)\,, 
\end{eqnarray}
\begin{eqnarray}
\dot{\Psi}^{3,\pm 1}_{\ZM} = - \frac{i}{2}
  \sqrt{\frac{\pi}{21}} \frac{1-q}{1 + q} \frac{P d^2}{r^3}
  \frac{\sqrt{f}}{\Lambda_3 \sigma^3} \left[ 10 \sigma
+ \left(1 + 3 \sigma  \right) \frac{M}{r} \right] \,, 
\end{eqnarray}
\begin{eqnarray}
\dot{\Psi}^{3,\pm 3}_{\ZM} = \frac{i}{2}
  \sqrt{\frac{5 \pi}{7}} \frac{1-q}{1 + q} \frac{P d^2}{r^3}
  \frac{\sqrt{f}}{\Lambda_3 \sigma^3} \left[ 10 \sigma
+ \left(1 + 3 \sigma  \right) \frac{M}{r} \right] \,. 
\end{eqnarray}
Note that the master equations do not have the same uncontrolled remainders as
its derivatives, since they come from different components of the initial data.
In the case of unboosted head-on collisions~\cite{Andrade:1996pc}, the initial
data scales in powers of $d^{N}$. Therefore, one only needs to perform one
single numerical evolution of the master functions for some reference value of
$d=d^{\star}$, and the results for any other value of $d$ can be found using the
scaling relation. For non-time-symmetric data, such as for quasicircular or
boosted sets, such scaling does not exist. In our case, for example, although
$\Psi^{\ell m}_{\ZM}$ still scales as $d^{N}$, its time derivative
$\dot\Psi^{\ell m}_{\ZM}$ scales as $P d^{N}$. Therefore, the master functions
themselves are not straightforwardly scalable and several runs with different
values of the initial parameters must be performed.

\section{Results from the CLA} \label{resultscla}

In this section, we evolve the master functions with the initial data obtained
in the previous sections in the CLA scheme and report the results for the main
physical quantities, in particular for the gravitational recoil velocities. We
first need to choose appropriately the parameters that completely determine the
initial data, such that it describes a binary black hole system merging from a
quasicircular orbit (subsection~\ref{physicalparameters}). Then, in
subsection~\ref{numericalevolution}, we use a numerical code to evolve the
different master equations [Eqs.~(\ref{masterequations})] and compute the
relevant physical quantities.  We discuss the results and compare with previous
ones in the literature when possible.

\subsection{Determining the parameters of the initial data}
\label{physicalparameters}

Our initial data depends on the following parameters: 
\begin{itemize}
\item The total (ADM) mass of the system, $M$;
\item The mass ratio, where one can use either the {\em bare} mass ratio $q$ or 
the {\em physical} one $Q$, related by Eqs.~(\ref{mass-ratio1}) 
and~(\ref{mass-ratio2}).
\item The initial separation, where again one can use the {\em bare} separation 
$d$ or the {\em physical} one $D$, related by Eq.~(\ref{D});
\item The linear momentum parameter $P$ of each individual hole.
\end{itemize}

Within the family of initial data spanned by these four parameters, we need to 
single out the subset that corresponds to configurations in {\em quasicircular} 
orbital motion. In numerical relativity this is done by looking at the minimum 
in the {\em binding} energy of the system with respect to the distance, while 
keeping the total ADM angular momentum constant~(see, e.g.~\cite{Cook:2000vr}).
We here follow the same procedure without using the {\em slow motion} 
approximation. The binding energy that we minimize is 
\begin{equation}
E_b = {\cal M}^{}_{\ADM} - {\cal M}^{}_1 - {\cal M}^{}_2\,. \label{bindingenergy}
\end{equation}
where ${\cal M}^{}_{\ADM}$ is the total ADM mass and it is computed in the 
asymptotically-flat region containing the two holes ($\Sigma^{}_0$).  This mass 
is given by (see, e.g.~\cite{Beig:2000ei})
\begin{equation}
{\cal M}^{}_{\ADM} = M + \frac{5P^2}{8\mu}\,,
\end{equation}
where $M$ is given in Eq.~(\ref{themass}) and $\mu$ is the reduced {\em bare} 
mass, i.e. $\mu = m^{}_1 m^{}_2/M\,.$ Moreover, in Eq.~(\ref{bindingenergy}), 
${\cal M}^{}_1$ and ${\cal M}^{}_2$ denote the masses computed in the 
asymptotically-flat regions $\Sigma^{}_1$ and $\Sigma^{}_2$. These masses are 
given by~(see, e.g.~\cite{Beig:2000ei})
\begin{equation}
{\cal M}^{}_\Lambda = M^{}_\Lambda + \frac{P^2}{8m^{}_\Lambda}\qquad 
(\Lambda=1,2)\,,
\end{equation}
where $M^{}_\Lambda$ is given in Eq.~(\ref{physicalmasses}).  Then, the binding
energy can be written in the following form
\begin{equation}
E_b = - \frac{m_1 m_2}{d} + \frac{J^2}{2 \mu d^2}\,,
\end{equation}
where $J$ is the ADM angular momentum, given by $J = P d\,$. This binding energy 
is formally the same as the one corresponding to a binary system in Newtonian 
gravity. One can then minimize this binding energy with respects to $d$, while 
keeping $J$ fixed, to obtain the condition for {\em quasicircular} motion (note 
that in our context there is no such a thing as an ISCO)
\begin{equation}
d = \frac{J^2}{\mu^2 M}\,. \label{minimum}
\end{equation}
In the same way, one can calculate the associated orbital frequency of such 
orbital motion by differentiating the binding energy with respect to $J$, while 
keeping $d$ fixed.  The result is 
\begin{equation}
\Omega = \frac{J}{\mu d^2}\,.
\end{equation}
From Eq.~(\ref{minimum}) we can write the linear momentum $P$ is terms of the
other parameters of our initial data as 
\begin{equation}
P = \mu \sqrt{\frac{M}{d}}\,.
\label{P}
\end{equation}
The binding energy and other quantities derived from it have a Newtonian form 
because of the particular type of initial data that we are using: a 
conformally-flat 3-metric with a Bowen-York extrinsic curvature and a 
Brill-Lindquist conformal factor. The PN metric produced by a binary system
differs from conformal flatness at 
${\cal O}(v^4)$~(see~\cite{Blanchet:1998vx,Blanchet:2003kz} for the argument in 
the case of time-symmetric initial data), and, hence, the binding energy used 
above differs from the PN binding energy at that order. Note, however, that 
although the binding energy, linear and angular momentum used here have a 
Newtonian form, they are not strictly Newtonian. This is mainly because the 
distance parameter $d$ is not the physical distance $D$, which is related to the 
parameter $d$ via Eq.~(\ref{D}).

Adopting Eq.~(\ref{P}) for the linear momentum parameter in our initial data and 
leaving the total mass $M$ fixed (which defines a system of units), we reduce 
our initial parameter space to a 2-dimensional one.  The final parameter space 
can be parameterized either by the bare quantities $(q,d)$ or by the physical 
quantities $(Q,D)\,.$  The range of $q$, or $Q$, is the obvious one, i.e. 
$[0,1]\,$, while the range for the bare distance parameter $d$ is 
$[d^{}_{\MIN},d^{}_{\CLA}]$, where $d^{}_{\CLA}$ is an estimate of the maximum 
distance for which the CLA is expected to be valid. For the case of equal-mass 
head-on collisions, it has been shown~\cite{Gleiser:1996yc}, by comparing with 
second-order calculations and with fully numerical relativistic simulations, 
that $d^{}_{\CLA}\sim 1.7\,M$, which roughly corresponds to 
$D^{}_{\CLA}\sim 4\,M\,.$ On the other hand, in principle $d^{}_{\MIN}$ could be 
just zero however, if we adopt the prescription (\ref{P}) for the linear 
momentum parameter, then we are limited by the {\em slow motion} approximation 
that we are using, which means that $d^{}_{\MIN}$ should be bigger than 
$qM/(1+q)^2\,.$ Finally, we should remark that the CLA also is expected to fail 
in the point-particle limit~\cite{Lousto:1996sx}, but, as we will see, the 
recoil is very small when $Q \ll 1$.

In order to study the gravitational recoil in the CLA scheme, we are going to 
evaluate the recoil velocity for a representative number of (physical) mass 
ratios for a given fixed physical distance $D\,.$ In particular, we study the 
cases $D = 3\,,\,3.5\,,\,4\,M\,$ and instead of using $Q$ we use the physical 
symmetric mass ratio
\begin{equation}
\eta = \frac{M_1 M_2}{(M_1 + M_2)^2} = \frac{Q}{(1 + Q)^2}\,.
\end{equation}
The inverse relation is given by
\begin{equation}
Q = \frac{1}{2 \eta} \left(1 - \sqrt{1 - 4\eta} \right) - 1\,. \label{eta_to_Q}
\end{equation}
However, the parameters that appear in our expressions for the initial data are 
the bare ones. Then, in order to obtain a plot of the recoil velocity in terms 
of the physical mass ratio, we need to translate from the set $(Q,D)$ to 
$(q,d)$. This, however, is not a trivial calculation because the definition of 
$D$ [Eq.~(\ref{D})] is quite intricate, involving $x^{}_1$ and $x^{}_2$. These 
numbers are the values of the coordinate $x$ in the conformal flat space of the 
intersections of the extremal surfaces (marginally trapped surfaces or apparent 
horizons depending on the parameters of each particular configuration) 
surrounding each individual hole with the $X$-axis. The translation has to be 
done numerically through the following iteration scheme in which the physical 
distance $D$ is kept fixed:
\begin{enumerate}
\item Given a value of $\eta$ we pick an initial guess for the bare distance, 
say $d^{}_\ast\,$.  
\item By solving the equations that determine the extremal surfaces surrounding 
each individual holes (they are given in Appendix~\ref{extremalsurfaces}) we 
find some intersection points $x^{\ast}_1$ and $x^{\ast}_2\,.$ This requires
another iteration, since we do not know a priori where these surfaces are 
located. What we do is to start, for each individual hole, with an initial guess 
for the intersection of the extremal surface at the other end of the $X$ axis 
(the intersection more distance to the other hole) and integrate the 
corresponding ODEs by using an extrapolation 
Bulirsch-Stoer scheme~\cite{Bulirsch:1966bs,Stoer:1993sb,Press:1992nr}. Then we 
study whether the integration ends far away from the $X$ axis or converges 
towards it. We repeat the iteration until we find the intersection points 
$x^{\ast}_1$ and $x^{\ast}_2\,$ with enough accuracy. 
\item Using Eq.~(\ref{D}) we compute the physical distance associated with these 
values of the intersection points and $(q^{}_\ast,d^{}_\ast)$ [where $q^{}_\ast$ 
is computed in terms of $\eta$ and $d^{}_\ast$ using expressions 
(\ref{eta_to_Q}) and (\ref{mass-ratio2})], $D^{}_\ast\,$. We compare $D^{}_\ast$ 
and $D$ and stop the iteration if the absolute difference between them is 
smaller than $10^{-4}\,M\,.$ Otherwise, we go back to point (i) changing the 
ansatz depending on whether $D^{}_\ast$ is bigger or smaller than $D\,$.  
\end{enumerate}

\begin{figure}[htb]
\begin{center}
\includegraphics[width=85mm]{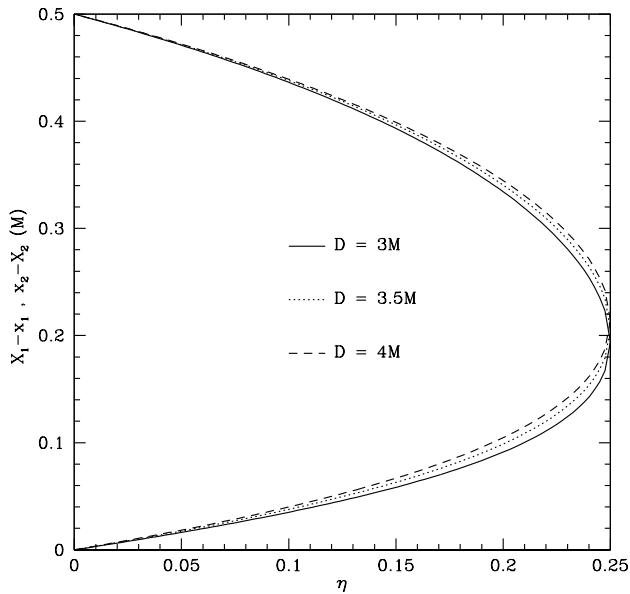}
\caption{Plot of the coordinate distance between the holes and the intersection 
points of the extremal surfaces with the $X$-axis in terms of the symmetric mass 
ratio $\eta$ for three values of the physical distance: 
$D=3\,,\,3.5\,,\,4\,M\,$. \label{intersectionpoints}}
\end{center}
\end{figure}

\begin{figure}[htb]
\begin{center}
\includegraphics[width=85mm]{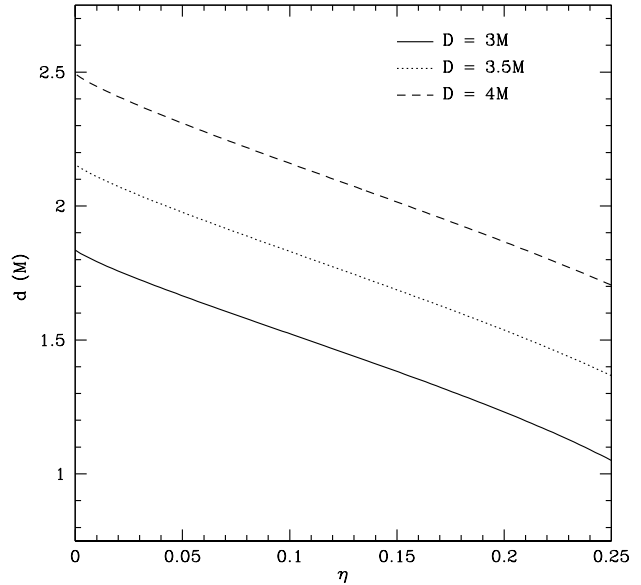}
\caption{Plot of the bare distance $d$ in terms of the symmetric mass ratio 
$\eta$ for three values of the physical distance: $D=3\,,\,3.5\,,\,4\,M\,$. 
\label{baredistancefigure}}
\end{center}
\end{figure}

We have carried out this iteration for 101 values of $\eta\,$. The coordinate 
distance (in the conformally related flat space) from the holes to the 
intersection points of the extremal surfaces is shown in 
Figure~\ref{intersectionpoints}. Here, we observe how these distances move from 
equal values (top right) to the values corresponding to the {\em point particle} 
limit, $M^{}_2 \rightarrow m^{}_2 \rightarrow 0\,$ (top left). The bare distance 
as a function of $\eta$ is shown in Figure~\ref{baredistancefigure} for the 
three values of fixed proper separation.

\subsection{Results from the numerical evolution of the master equations}
\label{numericalevolution}

We now have initial data for the master equations and also a method to prescribe 
the initial data parameters in a meaningful way. Then, the next step is to 
evolve the master equations [Eq.~(\ref{masterequations})]. In this paper we use 
a numerical code, based on Finite Element methods, that was developed 
in~\cite{Sopuerta:2005gz} for calculations of the gravitational radiation 
emitted by a point particle orbiting a non-rotating black hole. This method is 
based on linear elements and hence it has a second order convergence rate with 
respect to the spatial resolution. The time-evolution algorithms that it uses 
are second-order and unconditionally stable, since they are based on implicit 
methods. Apart from the tests of the numerical code carried out 
in~\cite{Sopuerta:2005gz}, we have also done some checks to validate the 
additional infrastructure added for the gravitational recoil calculations in the 
CLA scheme. First, we have checked that the energy and angular momentum emitted 
in an equal-mass grazing collision coincide with the ones found by Khanna 
{\em et al} in~\cite{Khanna:2000dg}. Second, we have checked that the recoil 
velocities that we obtain are consistent with the plots shown by Andrade and 
Price~\cite{Andrade:1996pc} for the case of head-on collisions from rest of 
unequal mass black holes using BL initial data.

We have then performed evolutions for 101 equally-spaced values of the symmetric 
mass ratio $\eta$ covering the whole range $[0,0.25]$ for the three values of 
the physical distance mentioned above, i.e. $D=3\,,\,3.5\,,\,4\,M\,$. The 
procedure to calculate the bare distance $d$ has been described in the previous 
subsection. Finally, the linear momentum parameter $P$ is obtained through 
Eq.~(\ref{P}). For each evolution we have computed the fluxes of energy, angular 
momentum and linear momentum carried by the gravitational waves to infinity.

\begin{figure}[htb]
\begin{center}
\includegraphics[width=85mm]{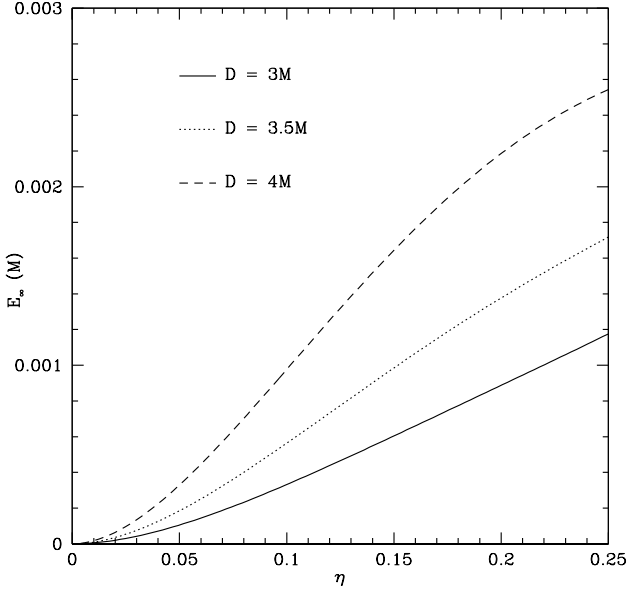}
\caption{Energy radiated to infinity in terms of the symmetric mass ratio $\eta$ 
for three values of the physical distance: $D=3\,,\,3.5\,,\,4\,M\,$. 
\label{energy_infinity}}
\end{center}
\end{figure}

Our initial data only has a few non-zero multipoles contributing to the 
gravitational radiation emitted. Thus, the expressions for the different fluxes 
simplify dramatically. The energy flux is given by 
\begin{eqnarray}
\dot{E}^{}_{\GW} &=& \frac{3}{8 \pi} \left[ (\dot{\Psi}^{2,0}_{\ZM})^2 + 
2 |\dot{\Psi}^{2,1}_{\CPM}|^2 + 2 |\dot{\Psi}^{2,2}_{\ZM}|^2  \right.
\nonumber \\
&+& \left. 10 \left( |\dot{\Psi}^{3,1}_{\ZM}|^2 + |\dot{\Psi}^{3,3}_{\ZM}|^2 
\right) \right]\,,  \label{Eflux}
\end{eqnarray}
Figure~\ref{energy_infinity} shows the total energy to infinity, given by the 
integral of Eq.~(\ref{Eflux}) over time, as a function of the symmetric mass 
ratio.

\begin{figure}[htb]
\begin{center}
\includegraphics[width=85mm]{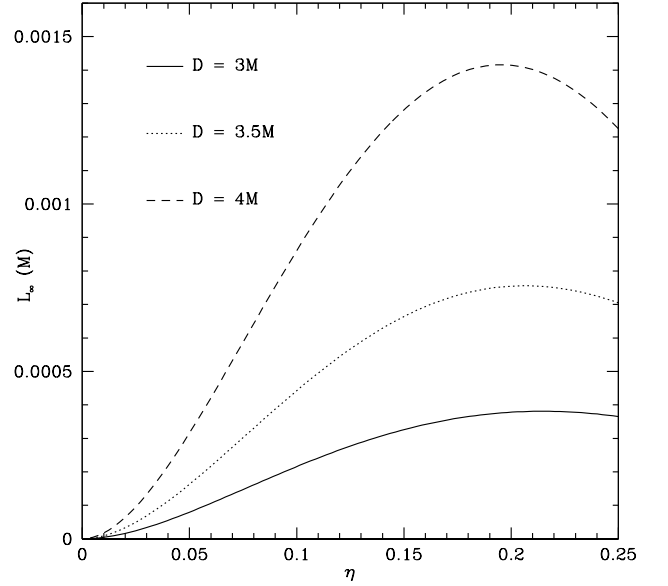}
\caption{Angular momentum radiated to infinity in terms of the symmetric mass 
ratio $\eta$ for three values of the physical distance: 
$D=3\,,\,3.5\,,\,4\,M\,$. \label{angular_momentum_infinity}}
\end{center}
\end{figure}

The angular momentum flux also simplifies greatly and becomes 
\begin{eqnarray}
\dot{L}^{}_{\GW} &=& \frac{3}{2 \pi} \left\{ \Re(\dot{\Psi}^{2,2}_{\ZM}) 
\Im(\Psi^{2,2}_{\ZM}) - \Im(\dot{\Psi}^{2,2}_{\ZM}) \Re(\Psi^{2,2}_{\ZM}) 
\right. \nonumber \\
&+& \left. \frac{5}{2} \left[ \Re(\dot{\Psi}^{3,1}_{\ZM})\Im(\Psi^{3,1}_{\ZM})-
\Im(\dot{\Psi}^{3,1}_{\ZM}) \Re(\Psi^{3,1}_{\ZM}) \right] \right.
\nonumber \\
&+& \left. \frac{15}{2} \left[ \Re(\dot{\Psi}^{3,3}_{\ZM}) 
\Im(\Psi^{3,3}_{\ZM}) - \Im(\dot{\Psi}^{3,3}_{\ZM}) \Re(\Psi^{3,3}_{\ZM}) 
\right] \right\}\,.  \label{Lflux}
\end{eqnarray}
Note that Eq.~(\ref{Lflux}) does not contain any contributions from the axial 
modes, since the only non-zero axial mode, $\Psi^{2,1}_{\CPM}$, is purely real. 
Figure~\ref{angular_momentum_infinity} shows the total angular momentum 
radiated to infinity, given by the integral of Eq.~(\ref{Lflux}), as a function 
of the symmetric mass ratio.

\begin{figure}[htb]
\begin{center}
\includegraphics[width=85mm]{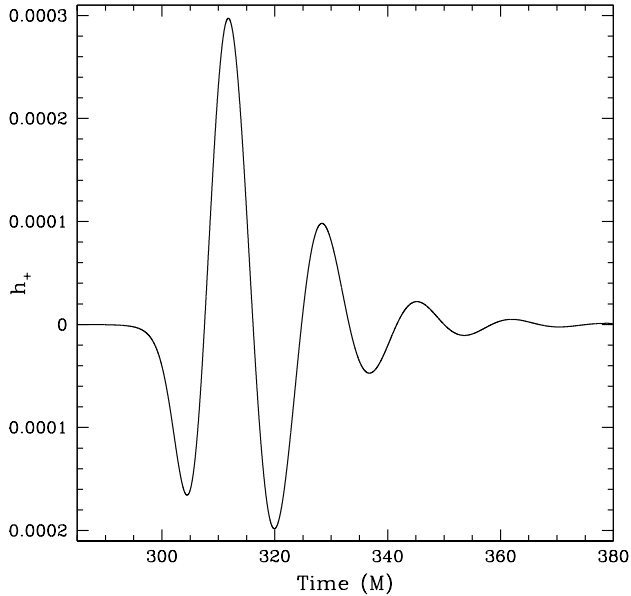}
\caption{Metric waveform $h^{}_+$ for the case $D=3.5M$ and $\eta = 0.185\,$ as 
a function of time. The observer is located at $\sim 300 M$ on the $Z$-axis
\label{metric_waveform_plus}}
\end{center}
\end{figure}

Finally, the gravitational waveform also simplifies, and we obtain  
\begin{eqnarray}
h^{}_+ &=&  \frac{\sqrt{6}}{r} \left\{ \left[ \Psi^{2,0}_{\ZM}\, 
{}_{-2}Y^{2,0}_{} \right. \right.  \nonumber \\
&+& \left. \left. 2 \Re(\Psi^{22}_{\ZM}) \, \Re({}_{-2}Y^{2,2}_{}) 
- 2 \Im(\Psi^{2,2}_{\ZM}) \, \Im({}_{-2}Y^{2,2}_{})\right] \right. 
\nonumber \\
&+& \left.  2 \sqrt{5} \left[ \Re(\Psi^{3,1}_{\ZM}) \, \Re({}_{-2}Y^{3,1}_{}) 
 - \Im(\Psi^{3,1}_{\ZM}) \, \Im({}_{-2}Y^{3,1}_{}) \right. \right.
\nonumber \\
&+& \left. \left. \Re(\Psi^{3,3}_{\ZM}) \, \Re({}_{-2}Y^{3,3}_{})
 - \Im(\Psi^{3,3}_{\ZM}) \, \Im({}_{-2}Y^{3,3}_{}) \right] \right\}, \\
h^{}_{\times} &=& -\frac{2\sqrt{6}}{r} \Re(\Psi^{2,1}_{\CPM}) \, 
\Re({}_{-2}Y^{2,1}_{})\,,
\end{eqnarray}
where the definition and some properties of the spin-weighted spherical 
harmonics ${}^{}_{s}Y^{\ell m}$ are given in Appendix~\ref{spinweighted}. Note 
that the $\times$-polarization consists purely of the axial modes, while the 
$+$-polarization contains only polar contributions. 
Figure~\ref{metric_waveform_plus} shows a typical metric waveform, namely $h_+$
as a function of time, for an observer located at $\sim 300 M$ on the $Z$-axis 
(the cross polarization vanishes on this axis).

\begin{figure}[htb]
\begin{center}
\includegraphics[width=85mm]{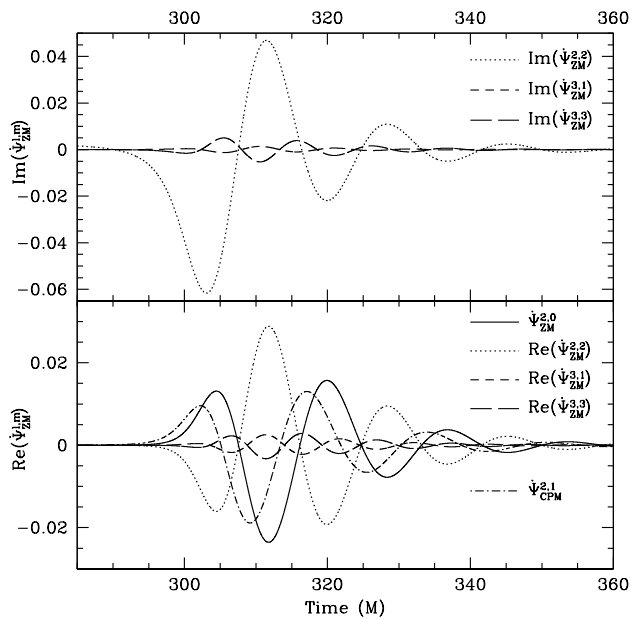}
\caption{Time derivate of the Zerilli-Moncrief and Cunningham-Price-Moncrief master functions as a function of time,
for the case $D=3.5M$ and $\eta = 0.185$.
The plots in the top panel represent the imaginary parts whereas the ones in
the bottom panel represent the real parts.  \label{dotmasterfunctions}}
\end{center}
\end{figure}

Let us now concentrate on the main physical quantity of interest, namely the 
linear momentum flux, given in  Eqs.~(\ref{recoilx}-\ref{recoilz}). These 
expressions reduce to  
\begin{eqnarray}
\dot{P}^x_{\GW} & = & -\frac{1}{2\pi}\sqrt{\frac{5}{14}} \left\{ \sqrt{6}\,
  \dot\Psi_{\ZM}^{2,0} \, \Re(\dot\Psi_{\ZM}^{3,1})  \right. 
\nonumber \\
& + & \left. \sqrt{15} \left[\Re(\dot\Psi_{\ZM}^{2,2}) \Re(\dot\Psi_{\ZM}^{3,3}) 
+ \Im(\dot\Psi_{\ZM}^{2,2}) \Im(\dot\Psi_{\ZM}^{3,3}) \right] \right. 
\nonumber \\
& - & \left. \left[ \Re(\dot\Psi_{\ZM}^{2,2})  \Re(\dot\Psi_{\ZM}^{3,1}) +
  \Im(\dot\Psi_{\ZM}^{2,2}) \Im(\dot\Psi_{\ZM}^{3,1}) \right] \right\}
\nonumber \\  
& + & \frac{1}{2\pi}\dot\Psi^{2,1}_{\CPM}\Im(\dot\Psi^{2,2}_{\ZM}) \,, 
\end{eqnarray}
\begin{eqnarray}
\dot{P}^y_{\GW} & = & \frac{1}{2\pi}\sqrt{\frac{5}{14}} \left\{ \sqrt{6}\,
  \dot\Psi_{\ZM}^{2,0} \, \Im(\dot\Psi_{\ZM}^{3,1})  \right. 
\nonumber \\
& + & \left. \sqrt{15}\left[\Re(\dot\Psi_{\ZM}^{2,2})\Im(\dot\Psi_{\ZM}^{3,3}) 
-   \Im(\dot\Psi_{\ZM}^{2,2}) \Re(\dot\Psi_{\ZM}^{3,3}) \right] \right. 
\nonumber \\
& + & \left. \left[ \Re(\dot\Psi_{\ZM}^{2,2}) \Im(\dot\Psi_{\ZM}^{3,1}) 
- \Im(\dot\Psi_{\ZM}^{2,2}) \Re(\dot\Psi_{\ZM}^{3,1}) \right] \right\}  
\nonumber \\
& + & \frac{1}{2\pi}\dot\Psi^{2,1}_{\CPM}\left[
\Re(\dot\Psi^{2,2}_{\ZM}) - \sqrt{\frac{3}{2}}\,\dot\Psi^{2,0}_{\ZM}\right]\,,
\end{eqnarray}
\begin{eqnarray}
\dot{P}^z_{\GW} & = & 0\,.
\end{eqnarray}
As we can see, there is only contributions from the overlap of polar modes with 
different $\ell$ and $m$. From this flux, the recoil velocity can be obtained by 
performing the following integration
\begin{equation}
v^i_{\recoil} = -\frac{1}{M}\int^{t^{}_f}_{t^{}_i} dt\; \dot{P}^i \,,
\end{equation}
where the integration times, $t^{}_i$ and $t^{}_f$, are such that the time 
interval includes essentially all the contribution from the waves to the flux.  
We can then calculate the magnitude of the recoil velocity simply by 
\begin{equation}
v^{}_{\recoil} = \sqrt{(v^x)^2 + (v^y)^2 + (v^z)^2}\,,
\end{equation}
where $v^z = 0$ in our case, due to the choices made in the initial setup. 
Figure~\ref{dotmasterfunctions} shows the time derivatives of the master 
functions that contribute to the recoil velocity for a typical evolution. In 
this figure, we have separated the real (bottom panel) and the imaginary (top 
panel) parts of these time derivatives. Observe that the magnitude of the 
$\ell=2$ modes is much bigger than the one of the $\ell=3$ modes, as expected.  
This also gives an indication that the {\em superposition} of the $\ell=2$ modes
with $\ell=2$ and $\ell=3$ modes is going to be the dominant contribution to the 
gravitational recoil. The contribution from superpositions involving higher 
$\ell$'s is going to be much smaller.

\begin{figure}[htb]
\begin{center}
\includegraphics[width=85mm]{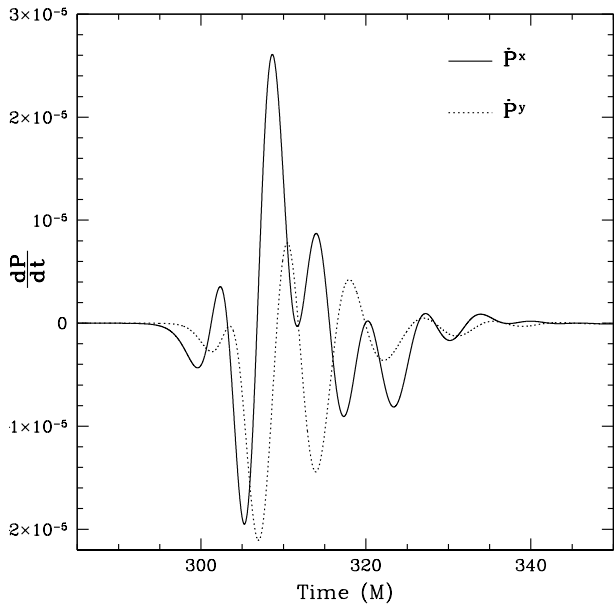}
\caption{Linear momentum fluxes, $\dot{P}^x_{\GW}$ and $\dot{P}^y_{\GW}$, 
as a function of time, for the case $D=3.5M$ and $\eta = 0.185\,$. \label{pdot}}
\end{center}
\end{figure}

Figure~\ref{pdot} shows the the linear momentum flux as a function of time.
Observe that the magnitude of the $x$-component is bigger than the 
$y$-component, which reflects the fact that our configuration corresponds to the 
transition from merger to plunge.

\begin{figure}[htb]
\begin{center}
\includegraphics[width=85mm]{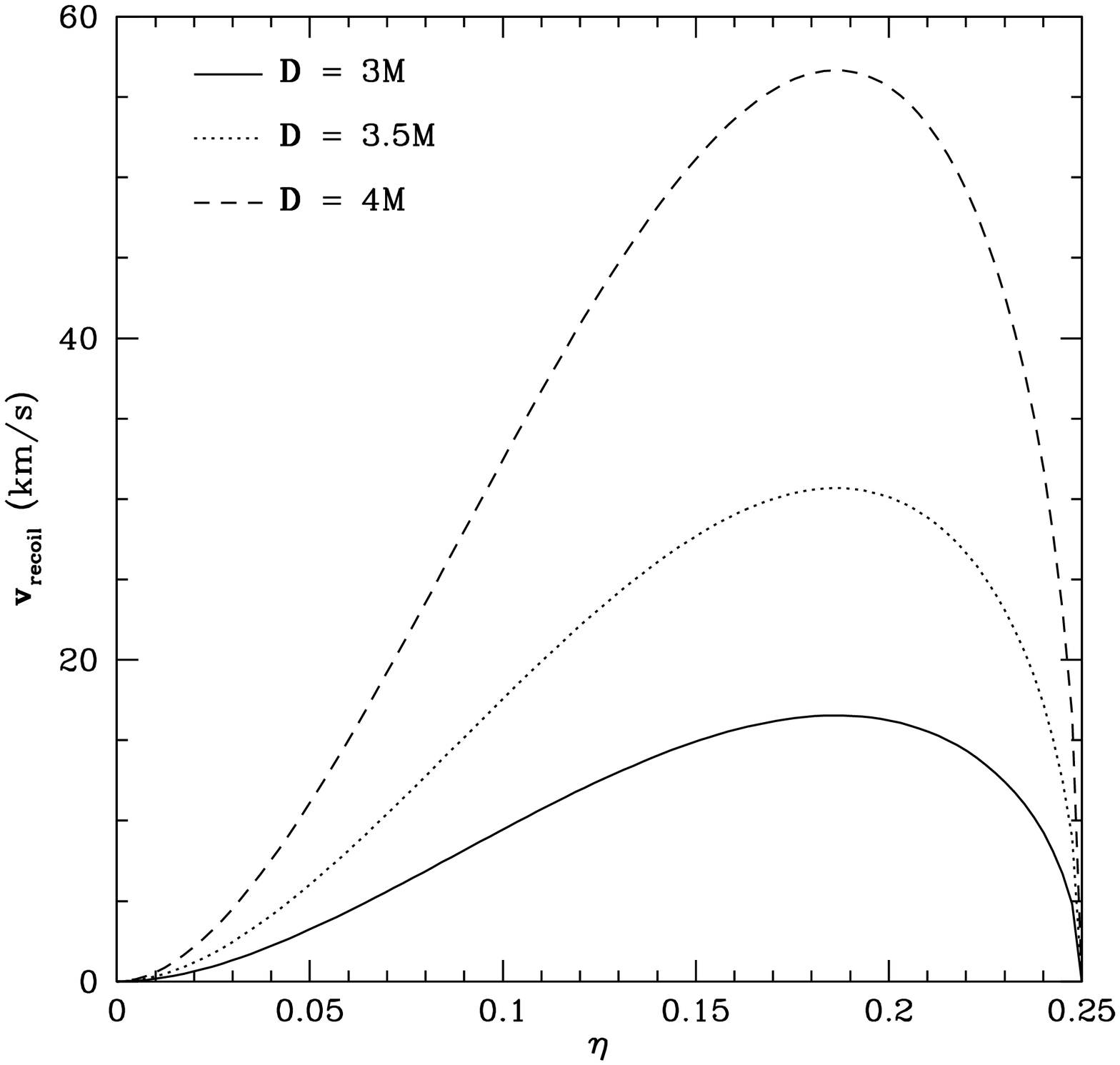}
\caption{Magnitude of the recoil velocity in terms of the symmetric mass ratio 
$\eta$ for three values of the physical distance: $D=3\,,\,3.5\,,\,4\,M\,$. 
\label{recoilvelocities}}
\end{center}
\end{figure}

Finally, Figure~\ref{recoilvelocities} presents the magnitude of the recoil 
velocity as a function of the symmetric mass ratio, for the following initial 
physical separations: $D=3\,,\,3.5\,,\,4\,M\,$. For all cases studied, the 
maximum velocity is reached for a symmetric mass ratio of $\eta \sim 0.19$, 
which agrees with the value reported in 
Refs.~\cite{Blanchet:2005rj,Damour:2006tr} up to uncontrolled remainders. 
Observe that this maximum is not a strong peak, but instead resembles a 
{\em plateau}, where this maximum is spread out for a number of $eta$s, as also 
seen in other calculations~\cite{Blanchet:2005rj,Damour:2006tr}.

\section{Estimating the total recoil} \label{totalrecoil}

In this section, we discuss the recoil velocities obtained from the evolution of 
the master functions and produce lower and upper limits for the total recoil 
velocity. In particular, we will provide analytic approximations to the data and 
we will also compare these results to other ones already present in the 
literature.

One of the limitations, and at the same time an advantage, of the CLA scheme is 
that the initial separation of the black holes must be {\em sufficiently} small 
in some well-defined sense. Apart from numerical relativity, this method is the 
only known one to be capable of producing accurate estimates of physical 
quantities near plunge. This advantage, however, is a double-edged sword since 
the method cannot account for the inspiral phase. Actually, the initial 
separation must even be smaller than that for which the last ISCO exists. Thus, 
not only is the inspiral phase neglected but also the beginning part of the 
merger phase.

Due to these limitations, an approximate value for the {\em total} recoil 
velocity cannot be provided by the CLA alone, without supplementing it with some 
other scheme valid when the system is well separated. The PN scheme is well 
suited for this task and extensive studies have been recently carried
out~\cite{Blanchet:2005rj,Damour:2006tr} to estimate the recoil velocity. The 
approximate recoil velocity accumulated from infinity up to some final 
separation in the PN scheme is given by~\cite{Blanchet:2005rj}
\begin{widetext}
\begin{eqnarray}
v_{PN} = \frac{464}{105} \eta^2 \frac{\delta m^{}_{\PN}}{m^{}_{\PN}} x_f^4 
\left[ 1 - \left(\frac{452}{87} + \frac{1139}{522} \eta \right) x_f 
+ \frac{309}{58} \pi x_f^{3/2} + \left( \frac{71345}{229968} +
    \frac{36761}{2088} \eta + \frac{147101}{68904} \eta^2 \right) x_f^2
\right]\,, \label{v-PN}
\end{eqnarray}
\end{widetext}
with remainders of ${\cal{O}}(v^5)$. In Eq.~(\ref{v-PN}), 
$x_f=(m \omega_f)^{2/3}$ is a PN parameter, $m^{}_{\PN}=m^{}_{1,\PN} 
+ m^{}_{2,\PN}$ is the total mass, $m^{}_{1,2,\PN}$ are the masses of the PN 
point particles, $\eta=m_1 m_2/m^2$ is the symmetric mass ratio and 
$\delta m_{\PN} = m^{}_{1,\PN} - m^{}_{2,\PN}$ is the mass difference. The
PN masses $m_{1,2,\PN}$ have been shown to agree, within the PN approximation,
with the horizon masses $M_{1,2}$~\cite{Blanchet:1998vx,Blanchet:2003kz} and we 
make this identification here. The angular velocity $\omega$ is given to 
${\cal{O}}(v^4)$ by
\begin{equation}
\omega^2 = \frac{m^{}_{\PN}}{b^3} \left[1 + \frac{m^{}_{\PN}}{b} \left(
\eta - 3\right) + \frac{m^{2}_{\PN}}{b^2} \left(6 + \frac{41}{4} \eta +
\eta^2 \right) \right]\,.     \label{PN-w}
\end{equation}
and $\omega_f$ is the angular velocity evaluated at some final coordinate 
separation $b_f$. Post-Newtonian theory is usually carried out in harmonic 
coordinates, which are different from the Schwarzschild coordinate system we use 
in the CLA scheme. However, sufficiently far from the holes, $D \sim b$, to 
${\cal{O}}(v^2)$.

Supplementing the CLA estimate with the PN estimate, we can obtain upper and 
lower limits on the possible values of the magnitude of the recoil velocity. 
A lower limit can be obtained via
\begin{equation}
v^{}_{\LOW} = v^{}_{\CLA}[0\,,4M] + v^{}_{\PN}[6M\,,\infty]\,, \label{v-ll}
\end{equation}
where $v^{}_{\PN}[D^{}_2,\infty]$ is the PN estimate for the recoil velocity of 
Eq.~(\ref{v-PN}) evaluated at $b_f = D^{}_2$. For this lower limit, we evaluate 
the PN estimate at the edge of the region of validity of the PN approximation, 
{\textit{i.e.}} $b_f = 6 M$, or equivalently $x_f = 6^{-3/2}$, as done in 
Ref.~\cite{Blanchet:2005rj}. This location corresponds to the ISCO of a test 
particle around a Schwarzschild hole of mass $M$. One obtains this value of 
$x_f(b_f = 6 M)$ by neglecting terms of ${\cal{O}}(v^2)$ and higher in 
Eq.~(\ref{PN-w}). If we had included these higher order terms in $\omega_f$ and 
$x_f$, the upper bounds would have decreased by approximately $50 \,$ km/s. 
These higher order terms, however, become large as $b$ becomes smaller, and 
thus we choose to neglect them to have a conservative upper bound. In 
Eq.~(\ref{v-ll}), $v^{}_{\CLA}[0\,,D^{}_1]$ is the estimate of the recoil 
velocity in the CLA approximation with an initial proper separation of 
$D=D^{}_1$.

The estimate of $v^{}_{\LOW}$ is a lower limit because it does not take into 
account the contribution to the gravitational recoil in the region $b\in(4,6)M$.  
In this region neither the CLA, nor the PN scheme, is guaranteed to provide an 
accurate estimate for the recoil. However, it is possible to construct upper 
limits by modelling either the entire region or part of it with PN and CLA
estimates. Such upper limits are given by 
\begin{eqnarray}
v^{}_{\UPUNO} &=& v^{}_{\CLA}[0,4M] + v^{}_{\PN}[4M,\infty]\,, 
\label{v-up1}
\\
v^{}_{\UPDOS} &=& v^{}_{\CLA}[0,5M] + v^{}_{\PN}[5M,\infty]\,.
\label{v-up2}
\end{eqnarray}
These expressions are upper limits because the contribution to the recoil 
estimated either with PN theory or the CLA approximation in the region $b_f \in 
(4,6)M$ is monotonically increasing with $b_f$.

Equations (\ref{v-ll}), (\ref{v-up1}), and (\ref{v-up2}) require some extra 
justification and clarification. In general, it is not true that the magnitude 
of the total recoil can be estimated by adding the magnitude of the integrated 
momentum flux in the region $[4M,\infty]$ to that in the region $[0,4 M]$. The 
important point is that the cut is made at a separation $D = 4 M$ in the 
regime where the main contribution to the recoil comes from. Then, the main 
contribution to each recoil velocity vector comes from the region near the cut 
and hence, the error we made by adding the norms will be relatively small.
Independent of this argument we have the inequality $v[0,\infty] \leq
v[0,D_{\CUT}] + v[D_{\CUT},\infty]$ (with equality when $v[0,D_{\CUT}]$ and 
$v[D_{\CUT},\infty]$ are aligned). In this sense, the proposed upper limit is 
indeed always an upper limit, irrespective of the orientation of the vectors. As 
for the lower limit, neglecting the accumulated recoil in the region $[4M,6M]$ 
is a very conservative estimate, because there the recoil accumulates greatly. 
Thus, the issue of the orientation of the vectors will not affect the fact that 
this is a lower limit, as other recent estimates in the literature confirm.

\begin{figure}[t]
\includegraphics[scale=0.33,clip=true]{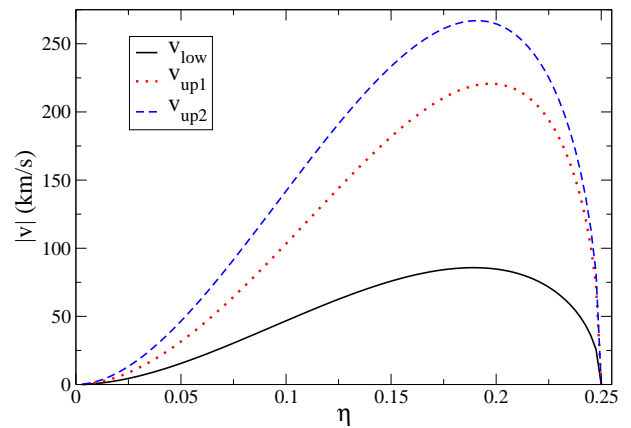}
\caption{\label{v-up-ll} Estimated lower (black solid curve) and upper limits 
  $1$ (dotted red line) and $2$ (dashed blue line) for the recoil velocity after 
  a binary black hole merger as a function of the physical symmetric mass ratio. 
  Note that the maximum occurs roughly in the same place, namely $\eta^{\star}
  \sim \{0.19,0.2\}$.}
\end{figure}

Figure~\ref{v-up-ll} shows the behavior of these upper limits (red dotted and 
blue dashed lines respectively) and the lower limit (black solid curve as a 
function of $\eta$.
The maximum in these curves occurs roughly at the same symmetric mass ratio,
namely $\eta \sim \{0.19,0.2\}$. The slight disagreement in this maximum is 
within error bars and rooted in that PN theory predicts it at approximately 
$\eta \sim 0.2$, while the CLA predicts it at $\eta \sim 0.19$. We should note
that the maximum recoil from $v^{}_{\CLA}[0,4M]$ and $v^{}_{\CLA}[0,5M]$ is 
approximately $64\,$ km/s and $215\,$ km/s, while the maximum recoil from 
$v^{}_{\PN}[4M,\infty]$, $v^{}_{\PN}[5M,\infty]$ and $v^{}_{\PN}[6M,\infty]$ is 
approximately $160\,$ km/s, $50\,$ km/s, and $20\,$ km/s respectively.

A non-linear fit can be performed to these curves via Eq.~(\ref{v-fit})
\begin{eqnarray}
v_{\FIT}  = a \eta^2 \sqrt{ 1 - 4 \eta} \left(1 + b \eta + c \eta^2
\right)\,, \nonumber
\end{eqnarray}
where the fitting parameters $a$, $b$ and $c$ are listed in Table~\ref{fitting}.

\begin{table}
\caption{\label{fitting} Values of the parameter of the non-linear fitting for 
the following models: the CLA with initial separations of $D=\{3,3.5,4\}M$; the 
lower and upper limits of Eqs.~(\ref{v-ll})-(\ref{v-up2}); Taylor PN (BQW) and 
EOB PN (DG) calculations~\cite{Blanchet:2005rj,Damour:2006tr}.} 
\begin{ruledtabular}
\begin{tabular}{c|r|r|r|r}
Model  & $a$ (km/s)~~ & $b$~~~ & $c$~~~ & Mean square error \\
\hline
$v^{}_{\CLA}[0,3M]$    &  $1841$  &  $-3.31$  &  $3.45$  &  $0.001$ \\
$v^{}_{\CLA}[0,3.5M]$  &  $3548$  &  $-3.15$  &  $3.33$  &  $0.003$ \\
$v^{}_{\CLA}[0,4M]$    &  $6576$  &  $-2.98$  &  $3.21$  &  $0.008$ \\
$v^{}_{\LOW}$          &  $7782$  &  $-2.51$  &  $2.73$  &  $0.008$ \\
$v^{}_{\UPUNO}$        &  $14802$ &  $-1.13$  &  $1.48$  &  $0.008$ \\
$v^{}_{\UPDOS}$        &  $23124$ &  $-2.33$  &  $2.61$  &  $0.06$ \\
$v^{}_{BQW}$           &  $12891$ &  $0.25$   &  $0$     &  $10^{-8}$ \\
$v^{}_{DG}$            &  $4483$  &  $-0.95$  &  $2.68$  &  $10^{-10}$ \\
\end{tabular}
\end{ruledtabular}
\end{table}

Observe that the mean square error for all cases is small, which is an 
indication that Eq.~(\ref{v-fit}) is a good analytic model for the functional 
form of the recoil velocity. In this table, we also present the values 
corresponding to the estimates of Refs.~\cite{Blanchet:2005rj} (BQW) 
and~\cite{Damour:2006tr} (DG.) Since the predictions of these references are 
based on analytic formulae, the mean square error can be made arbitrarily small 
by increasing the number of points in the discretization of the analytic curve.

With the analytic fits to the upper and lower limits, we can construct a curve 
that is in between these limits with an error given by the distance from the 
curve to the upper or lower bound. Such a curve is given by Eq.~(\ref{v-fit}) 
with the following fitting parameters
\begin{eqnarray}
a &=& \frac{a_{\LOW} + a_{\UP}}{2}\,, \\
b &=& \frac{a_{\LOW} b_{\LOW} + a_{\UP} b_{\UP}}{a_{\LOW} + a_{\UP}}\,, \\
c &=& \frac{a_{\LOW} c_{\LOW} + a_{\UP} c_{\UP}}{a_{\LOW} + a_{\UP}}\,,
\label{ansatz}
\end{eqnarray}
while the error on this curve is also given by Eq.~(\ref{v-fit}) with the 
following fitting parameters
\begin{eqnarray}
a &=& \frac{a_{\LOW} - a_{\UP}}{2}\,, \\
b &=& \frac{a_{\LOW} b_{\LOW} - a_{\UP} b_{\UP}}{a_{\LOW} - a_{\UP}}\,, \\
c &=& \frac{a_{\LOW} c_{\LOW} - a_{\UP} c_{\UP}}{a_{\LOW} - a_{\UP}}\,,
\label{ansatz2}
\end{eqnarray}
where the subscript $low$ and $up$ stand for the fitting parameters of the lower 
or upper limit respectively. This curve is only an alternative way to visualize 
the upper and lower limits of Fig.~\ref{v-up-ll}. The curve is not to be 
interpreted as the best guess in this work, since in principle, the recoil 
velocities present in nature could be closer to either upper or lower limit.

\begin{figure}[t]
  \includegraphics[scale=0.33,clip=true]{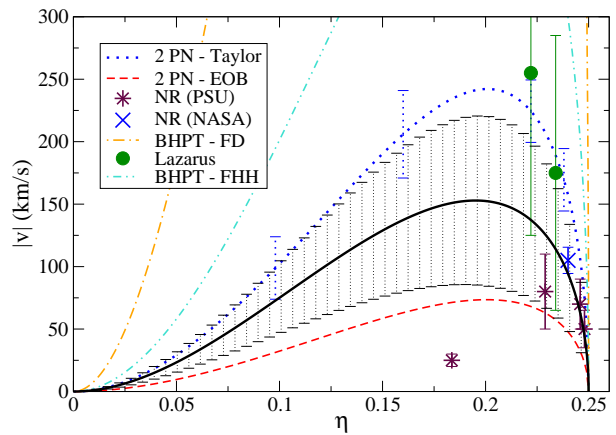}
\caption{\label{other-calc} Estimates for the recoil velocity (km/s) of the 
  inspiral and merger of a binary system of compact object as a function of the 
  physical symmetric mass ratio parameter. The symbols used are the following: 
  squares and triangles stand for black hole perturbation theory 
  results~\cite{Fitchett:1984fd,Favata:2004wz}; circles stand for Lazarus 
  results~\cite{Campanelli:2004zw}; stars and crosses correspond to full 
  numerical relativistic simulation~\cite{Herrmann:2006ks,Baker:2006vn}; the 
  dotted line and the dashed line correspond to $2$ PN Taylor
  expansions~\cite{Blanchet:2005rj} and $2$ PN effective-one-body
  expansions~\cite{Damour:2006tr} respectively. The solid black line corresponds 
  to the estimate of this paper, which, together with the error bars, condense 
  both upper and lower limits. Other error bars, when present, correspond to an 
  estimate of some of the error contained in the calculation. }
\end{figure}

We can now compare these estimates of the recoil velocity with those present in 
the literature. Fig.~\ref{other-calc} shows the recoil velocity in units of km/s 
as a function of the physical symmetric mass ratio as estimated in the 
literature and by this paper. As is clear from the figure, there are many 
approaches to calculate this velocity and not all of them agree. The symbols 
used are the following: squares and triangles stand for the results obtained 
using black hole perturbation theory in the extreme-mass ratio approximation in 
Refs.~\cite{Fitchett:1984fd,Favata:2004wz} respectively; circles stand for the 
calculations carried out via the Lazarus approach~\cite{Campanelli:2004zw}; 
stars and crosses correspond to the results coming from a full numerical 
relativistic simulation (PSU stands for Ref.~\cite{Herrmann:2006ks} and NASA 
stands for Ref.~\cite{Baker:2006vn}); the dotted line and the dashed line
correspond to the $2$ PN Taylor expansion approach~\cite{Blanchet:2005rj} and 
the $2$ PN effective-one-body (EOB) approach~\cite{Damour:2006tr} respectively. 
The solid line with error bars is the estimate of 
Eq.~(\ref{ansatz},\ref{ansatz2}) that properly condenses the lower and upper 
limits into one curve. We briefly describe each approach below.

In the PN calculations of Ref.~\cite{Blanchet:2005rj}, the recoil velocity 
(dotted line in the figure) and momentum flux are estimated by studying the $2$ 
PN Taylor-expanded radiative moments of a binary system of compact objects, 
while in Ref.~\cite{Damour:2006tr} an effective-one-body approach is used 
(dashed line in the figure.) Post-Newtonian calculations are usually valid only 
when the binary is weakly gravitating, or equivalently when the orbital 
separation is greater than the ISCO. In this regime, the recoil velocity 
(Eq.~(\ref{v-PN}) has been found to be small for any mass ratio (usually less 
than $20 \,$ km/s), since, as expected, most of the contribution to the recoil 
comes from the merger part of the inspiral. In Ref.~\cite{Blanchet:2005rj} the 
calculation is extended through the merger by integrating the $2$ PN 
Taylor-expanded momentum flux along a geodesic of the Schwarzschild metric. On 
the other hand, Ref.~\cite{Damour:2006tr} uses the effective-one-body 
Hamiltonian to extend the inspiral through the merger. Both of these approaches 
have inherent errors that are difficult to estimate without calculating the $3$ 
PN contributions to the recoil velocity.

Black hole perturbation theory has also been used to estimate the recoil 
velocity in Refs.~\cite{Fitchett:1984fd,Favata:2004wz}. In these studies, the 
extreme-mass ratio approximation is adopted ({\textit{i.e.}}, $Q \ll 1$) and 
then the system is approximated as a point particle orbiting a black hole. The 
first study of the recoil velocity using this formalism was performed in 
Ref.~\cite{Fitchett:1984fd} (squares in the figure), but there the gravitational 
force was treated as Newtonian and only the lowest multipoles were considered. 
In Ref.~\cite{Favata:2004wz}, these relativistic effects were taken into 
account, as well as spin, and the velocity estimates were improved (triangles in 
the figure.) The extreme-mass ratio approximation, however, requires $Q \ll 1$, 
which allows the exploration of a limited section of the $\eta$-space.

A combination of black hole perturbation theory and full numerical relativity 
(the so-called Lazarus approach) has also been implemented to estimate the 
recoil velocity~\cite{Campanelli:2004zw}. In this case, a full numerical 
relativistic simulation is carried out until the black holes merge and a single 
perturbed spinning black hole has formed. Then, this spacetime is used as 
initial data in a Teukolsky evolution to determine the recoil velocity (circles 
in the figure.) The error in this calculation is rooted in the interpretation of 
the initial data as that of a perturbed Kerr spacetime. Finally, there have also
been recently full numerical relativistic simulations of binary black hole 
coalescence~\cite{Herrmann:2006ks,Baker:2006vn} (shown as stars and crosses 
respectively in the figure.)  In this case, the error shown in the figure is 
assumed to be given only by finite differencing, while the error due to initial 
data is neglected.

Even though there has been much work in the calculation of the recoil velocity 
there is still some disagreement. In Fig.~\ref{other-calc} we observe that there 
seem to be three groups of results: one that clusters around the $2$ PN Taylor 
expanded result; another that is close to the $2$ PN effective-one-body result; 
and a third one that is in between the first two. This disagreement, however, is 
misleading in several ways. First, some estimates of the recoil velocity quote 
no error bars, as is the case of the first perturbation theory 
approach~\cite{Fitchett:1984fd} and the $2$ PN effective-one-body 
approach~\cite{Damour:2006tr}. Second, the error bars that do exist in other 
calculation are only estimates and could very well have been underestimated.  A 
surprising disagreement is between the PN approaches, since when used to 
calculate other quantities, such as the angular frequency at the ISCO, they do 
agree. This disagreement seems to be rooted in the fact that the greatest 
contribution to the recoil velocity comes from the merger part of the inspiral, 
where neither extension of the PN approach is guaranteed to be accurate.

Our estimates seem to agree with most results if one accounts for error bars. 
The results of Refs.~\cite{Fitchett:1984fd,Campanelli:2004zw,Favata:2004wz}
(squares, triangles and circles in the figure) seem to overestimate the recoil, 
which is expected in the case of the extreme-mass ratio approximation. The PN 
results of Refs.~\cite{Blanchet:2005rj,Damour:2006tr} seem to overestimate and 
underestimate the recoil respectively, but they are consistent with our bounds 
if one takes their error bars (not shown in the figure) into account. The full 
numerical relativity results seem to overlap with our bounds, although there are 
only a few of them. We should note that our bound seems to disagree with the 
full numerical relativistic result for $\eta \sim 0.18$, but that result seems 
to be an underestimate because of the small initial separation~\cite{frank}.

\section{Conclusions and Discussion}\label{discussion}

We have calculated the recoil velocity after the merger of an unequal mass 
binary black hole system using the CLA scheme. This approximation assumes that 
the black holes are close enough that the system can be approximated by a single 
perturbed black hole spacetime. In contrast to other approaches, except for full 
numerical relativity, this approximation allows us to make valid statement about 
physical process when the system is close to plunge. Therefore, it is of great 
interest to use this method for the study of gravitational recoil. However, the 
CLA has the disadvantage that it cannot be used during the beginning of the 
merger or the inspiral phases.

Initial data for the CLA can be constructed analytically by mapping data 
suitable for a binary black hole inspiral to that of a single perturbed hole. 
With such initial data, the Cunningham-Price-Moncrief and the Zerilli-Moncrief 
master functions can be numerically evolved from some initial proper separation 
through ringdown. These gauge-invariant master functions contain all the 
information necessary to evaluate the gravitational metric waveforms and, thus, 
the energy, angular momentum and linear momentum fluxes carried away from the 
system.

The results obtained can be summarized as follows. First, the maximum recoil 
velocity obtained in the CLA scheme is of $v \sim 64 \,$ km/s for the maximum 
initial separation allowed by this method ($D=4 M$). This maximum occurs at a 
symmetric mass ratio of $\eta \sim 0.19$. By supplementing this estimate with PN 
ones valid in the inspiral regime, we obtain lower and upper bounds with maxima 
of $v \sim 84 \,$ km/s and $v \sim 220 \,$ km/s respectively. We have further 
provided non-linear analytic fit functions that conveniently parameterize these 
bounds, together with the results from the CLA, and other results in the 
literature. These results also suggest that there is a region around 
$D \sim \{4,6\} \,M$ that greatly contributes to the recoil, but can only be 
poorly modelled by current approximation schemes.

Ultimately, the estimates presented here suffer of the same predicaments as 
other calculations. Due to its analytical nature, the CLA relies on certain 
assumptions that do not hold over the entire history of the binary. Such 
assumptions introduce an error in the estimated recoil that is difficult to 
quantify. In particular, the assumptions made here are the following: close 
separations ; slow-motion; simple initial data. The close-limit assumption is 
essential to allow a mapping of a binary inspiral to a single perturbed 
spacetime. The slow-motion approximation supplements the close-limit assumption 
and can, in principle, be improved on in future extensions of this work. The 
choice of initial data is assumed to represent the gravitational content of some 
initial slice, although we know that this fails even at large separation because 
it does not agree with the deviations from conformal flatness predicted by PN
theory. Moreover, it does not contain any radiation, which is not what it should 
be expected for initial data corresponding to a snapshot of the orbital 
evolution. Due to these assumptions, the estimate of the recoil velocity will be 
contaminated by some error. However, experience in CLA calculations indicates 
that the error made only overestimates the physical quantities calculated, 
relative to full numerical 
simulations~\cite{Nicasio:1998aj,Khanna:1999mh,Khanna:2000dg}.

Future work will concentrate on extending this approach to second order in $P$ 
and to other, more realistic, initial data 
sets~\cite{Yunes:2005nn,Yunes:2006iw}. Ultimately, it would be interesting to 
compare the CLA approach directly to full numerical relativistic simulations in 
an attempt to determine the region of validity of the CLA more accurately. 
Another possibility is to use a multi-parameter perturbation scheme 
(see~\cite{Bruni:2002sm,Sopuerta:2003rg,Passamonti:2004je,Passamonti:2005cz})
where perturbations in the linear momentum and separation parameters can be 
cleanly separated at the different perturbative orders.

\acknowledgments
The authors acknowledge the support of the Center for Gravitational Wave Physics 
funded by the National Science Foundation under Cooperative Agreement 
PHY-0114375. This work was partially supported by NSF grants PHY0218750, 
PHY0244788, PHY0245649, PHY0555436, and PHY0555628. The authors also wish to 
thank the Information Technology Services at Penn State University for the use 
of the LION-XO and LION-XM computer clusters in some of the calculations
presented in this paper. Other calculations used the computer algebra systems 
MATHEMATICA and MAPLE (in combination with the GRTensorII 
package~\cite{grtensor}). We would also like to thank Frank Herrmann, Carlos 
Lousto, Ben Owen and Ulrich Sperhake for enlightening discussions and comments.

~

\noindent{\bf Note added at revision:} After this paper was submitted, full
numerical relativity estimates have appeared~\cite{Gonzalez:2006md}, and a
best-bet estimate based on our scheme, which includes some refinements in the PN 
part with respect to the present paper, has been compared to them 
in~\cite{Sopuerta:2006et}. The agreement between them is remarkable and supports 
the approach presented in this paper.

\appendix

\section{Conventions for Special Functions} \label{spec-func}
In this appendix, we describe the conventions we use for the special functions 
presented in this paper, and we also present some important properties of such 
functions.

\subsection{Special polynomials}
The expression for the associated Legendre polynomials that we are using in this
paper is the following 
\begin{equation}
P^m_\ell(x) = \frac{(-1)^{\ell+m}}{2^\ell\,\ell!}(1-x^2)^{m/2}
\frac{d^{\ell+m}}{dx^{\ell+m}}(1-x^2)^\ell \,,
\end{equation}
where $\ell$ is a non-negative integer and $m$ is an integer restricted to the 
following range: $m \in (-\ell\,,-(\ell-1)\,,\ldots\,,\ell-1\,,\ell)$.

Gegenbauer polynomials, also known as ultraspherical harmonics~\cite{Arfken}, 
are generalization of Legendre polynomial for higher dimensional spaces. They 
can be written in terms of other special functions, as in
\begin{equation}
C^{(\lambda)}_n = \frac{\Gamma(\lambda + 1/2)}{\Gamma(2 \lambda)}
\frac{\Gamma(n + 2 \lambda)}{\Gamma(n + \lambda + 1/2)} P^{(\lambda -
  1/2,\lambda - 1/2)}_n\,,
\end{equation}
where $\lambda$ is a real number, $n$ is a positive integer, $\Gamma$ is the 
Gamma function, and $P^{(\lambda^{}_1,\lambda^{}_2)}_n$ are the Jacobi 
polynomials~\cite{Arfken}. There are also recursion relations for these
polynomials, but we will not present them here. Instead, we will provide the 
first few Gegenbauer polynomials
\begin{eqnarray}
C^{(\lambda)}_0(x) &=& 1\,,  \\ 
C^{(\lambda)}_1(x) &=& 2 \lambda x\,,   \\
C^{(\lambda)}_2(x) &=& -\lambda\left[1 - 2(1+\lambda)x^2\right]\,,  \\
C^{(\lambda)}_3(x) &=& -2\lambda(1+\lambda)x\left[1 -
\frac{2}{3}(2+\lambda)x^2 \right] \,.
\end{eqnarray}
These polynomials allow for the far field expansion of potentials that scale as 
$|\vec{x} - \vec{x}_0|^{-\alpha}$ to arbitrary order.

\subsection{Spherical Harmonics}\label{sphericalharmonics}
The scalar spherical harmonics are solutions of the eigenvalue problem described 
by the following equation
\begin{equation}
\Omega^{ab}Y^{\ell m}_{:ab} + \ell(\ell+1)Y^{\ell m} = 0\,, \label{Yequation}
\end{equation}
where $(\ell,m)$ have the same range of values as in the associated Legendre
polynomials above.  In this paper we use the conventions 
of~\cite{Abramowitz:1970as,Press:1992nr} to define specify the solutions of 
Eq.~(\ref{Yequation}).  The precise expression is given by
\begin{equation}
\label{scalar}
Y^{\ell m}(\theta,\varphi) = \sqrt{\frac{2 \ell + 1}{4 \pi} 
      \frac{\left(\ell - m\right)!}{\left(\ell + m\right)!}}\,
      P^m_\ell(\cos{\theta}) e^{i m \varphi}\,.
\end{equation}
The scalar spherical harmonics form an orthonormal basis on the two-sphere, that 
is
\begin{eqnarray}
\int^{}_{S^2} \hspace{-3mm} d\Omega\, Y^{\ell m}\bar{Y}^{\ell' m'} & =
& \delta^{\ell \ell'} \delta^{m m'}\,,
\end{eqnarray}
where $\delta^{a b}$ denotes the Kronecker delta.
The vector spherical harmonics are defined in terms of the scalar ones as in
Eq.~(\ref{vec-sph-harm}), and from this definition we can derive the following 
orthogonality relations:
\begin{eqnarray}
\int^{}_{S^2} \hspace{-3mm} d\Omega\, \Omega^{ab}Y^{\ell  m}_a
\bar{Y}^{\ell' m'}_b & = & \ell(\ell+1)\delta^{\ell \ell'}
\delta^{m m'}\,,   \\
\int^{}_{S^2} \hspace{-3mm} d\Omega\, \Omega^{ab}S^{\ell  m}_a
\bar{S}^{\ell' m'}_b & = & \ell(\ell+1)\delta^{\ell \ell'}
\delta^{m m'}\,,  \\
\int^{}_{S^2} \hspace{-3mm} d\Omega\, \Omega^{ab}Y^{\ell m}_a
\bar{S}^{\ell' m'}_b & = & 0\,.
\end{eqnarray}
The (symmetric) tensor spherical harmonics used in this paper are also 
constructed from the scalar ones by means of Eqs.~(\ref{ten-sph-harm1}) 
and~(\ref{ten-sph-harm2}), from where the following orthogonality relations can 
be deduced:
\begin{eqnarray}
\int^{}_{S^2} \hspace{-3mm} d\Omega\, \Omega^{ac}\Omega^{bd}Y^{\ell
  m}_{ab}\bar{Y}^{\ell' m'}_{cd} & = &  2\delta^{\ell \ell'} \delta^{m
  m'}\,,  \\
\int^{}_{S^2} \hspace{-3mm} d\Omega\, \Omega^{ac}\Omega^{bd}Z^{\ell
  m}_{ab}\bar{Z}^{\ell' m'}_{cd} & = &
\frac{(\ell+2)!}{2(\ell-2)!}\delta^{\ell \ell'} \delta^{m m'}\,,  \\
\int^{}_{S^2} \hspace{-3mm} d\Omega\, \Omega^{ac}\Omega^{bd}S^{\ell
  m}_{ab}\bar{S}^{\ell' m'}_{cd} & = &
\frac{(\ell+2)!}{2(\ell-2)!}\delta^{\ell \ell'} \delta^{m m'}\,,  \\
\int^{}_{S^2} \hspace{-3mm} d\Omega\, \Omega^{ac}\Omega^{bd}Z^{\ell
  m}_{ab}\bar{S}^{\ell' m'}_{cd} & = &  0\,,  
\end{eqnarray}
and
\begin{equation}
\Omega^{ac}\Omega^{bd}Z^{\ell m}_{ab}Y^{\ell' m'}_{cd}  =
\Omega^{ac}\Omega^{bd}S^{\ell m}_{ab}Y^{\ell' m'}_{cd} =0\,. 
\end{equation}
%

\subsection{Spin-weighted scalar spherical harmonics}\label{spinweighted}
Spin-weighted scalar spherical harmonics are another basis to expand 
functions on the 2-sphere.  They can be defined by the following general
formula~\cite{Stewart:1990}
\begin{equation}
{}^{}_sY^{lm}(\theta,\varphi) = 
\left\{ \begin{array}{lr}
\left[ \frac{\left(l - s\right)!}{\left(l+s\right)!}
\right]^{1/2} \hat\partial^{s} Y^{lm}, & 0 < s < l, \\
(-1)^s \left[ \frac{\left(l + s\right)!}{\left(l-s\right)!}
\right]^{1/2} \check{\partial}^{-s} Y^{lm}, & -l < s < 0, \\
0, & l < |s|, \\
\end{array} \right.
\end{equation}
where $\hat\partial$ ($\check{\partial}$) is a ladder operator, usually called 
the edth operator, that raises (lowers) in a unity the spin weight of any 
quantity.  Its action on a scalar $Q$ can be expressed in the following 
way~\cite{Stewart:1990,Dray:1984gy}
\begin{eqnarray}
\hat\partial Q &=& m^a \partial^{}_a Q - \frac{s}{2} 
(\bar{m}^a m^b\nabla^{}_b m^{}_a) Q\,,  \\
\check\partial Q &=& \bar{m}^a \partial^{}_a Q + \frac{s}{2} 
(m^a\bar{m}^b\nabla_b \bar{m}_a) Q \,,
\end{eqnarray}
where $s$ is the spin weight of $Q$ and $(m^a,\bar{m}^a)$ is a null complex 
basis on the 2-sphere ($\Omega^{}_{ab} m^a m^b = 0\,,$ 
$\Omega^{}_{ab} m^a \bar{m}^b = 1$). It is worth noting that the action of the 
edth depends explicitly on the spin weight of the quantity on which it acts.  
Taking $m^a = \frac{1}{\sqrt{2}}\left[1, \frac{i}{\sin{\theta}} \right]$,
we can write the edth-operator as~\cite{Dray:1984gy}
\begin{eqnarray}
\; \hat\partial &=& \frac{1}{\sqrt{2}}
\left[\partial^{}_{\theta} + \frac{i}{\sin{\theta}} \partial^{}_{\varphi} -
  \frac{s}{2} \frac{\cos{\theta}}{\sin{\theta}} \right] \\
\; \check{\partial} &=& \frac{1}{\sqrt{2}}
\left[\partial^{}_{\theta} - \frac{i}{\sin{\theta}} \partial^{}_{\varphi} +
  \frac{s}{2} \frac{\cos{\theta}}{\sin{\theta}} \right]
\end{eqnarray}
It is important to mention that these definitions are applicable only to integer 
powers of the spin-weight. A generalized definition for half-integer powers of 
$s$ exists but will not be discussed here (see~\cite{Stewart:1990}).

In this paper we are interested in the $s=-2$ case, for which the definition of 
the spin-weighted spherical harmonics reduces to~\cite{Martel:2003th}
\begin{equation}
{}^{}_{-2}Y^{\ell m} = 2 \sqrt{\frac{(\ell - 2)!}{(\ell+2)!}} Z^{\ell m}_{ab}
\bar{m}^a \bar{m}^b \,.  
\end{equation}
%

\section{Relations with the Regge-Wheeler Parameterization of the Perturbations}
\label{RWparametrization}
For the sake of completeness, we give here the relations between our 
parameterization of the metric perturbations and the one used by Regge and 
Wheeler~\cite{Regge:1957rw}. For polar modes (our notation is on the left column 
and the one of Regge and Wheeler is on the right one) we have
\begin{eqnarray}
\label{map1}
(h_{AB}^{\ell m}) & \leftrightarrow & \left( \begin{array}{cc}
f H^{\ell m}_0 & H^{\ell m}_1  \\
\ast  & f^{-1} H^{\ell m}_2
 \end{array}\right)\,,  \\
p_{A}^{\ell m} & \leftrightarrow & \left(h_{t}^{\ell m}, h_{r}^{\ell
    m} \right)\,,  \\
K^{\ell m} & \leftrightarrow & K^{\ell m} - \frac{\ell(\ell+1)}{2}G^{\ell m}\,, 
\\
G^{\ell m} & \leftrightarrow & G^{\ell m} \,.
\end{eqnarray}
and for axial modes
\begin{eqnarray}
\label{map2}
(h^{\ell m}_A) & \leftrightarrow &  - ( h^{\ell m}_0, h^{\ell m}_1)\,,  \\
H^{\ell m} & \leftrightarrow & - h^{\ell m}_2 \,.
\end{eqnarray}

The expressions for the master functions, in Schwarzschild coordinates, in terms 
of the parameterization of Regge and Wheeler are:
\begin{eqnarray}
\Psi^{\ell m}_{\ZM} &=& \frac{r}{1+\lambda^{}_\ell}\left[ K^{\ell m}
  + \frac{f}{\Lambda^{}_\ell}\left( H^{\ell m}_2 -r\partial^{}_r
  K^{\ell m}\right) \right]
\nonumber \\ 
& - & \frac{2f}{\Lambda^{}_\ell}\left( h^{\ell m}_1
  -\frac{r^2}{2}\partial^{}_rG^{\ell m}\right)\,, 
\end{eqnarray}
\begin{eqnarray}
\Psi^{\ell m}_{\RW} &=&  \frac{f}{r} \left(h_1^{\ell m} - \frac{1}{2}
\partial^{}_r h_2^{\ell m} + \frac{1}{r} h_2^{\ell m} \right) \,,
\end{eqnarray}
\begin{equation}
\Psi^{\ell m}_{\CPM} = \frac{r}{\lambda^{}_\ell}\left( \dot{h}^{\ell m}_1
- \partial^{}_r h^{\ell m}_0 + \frac{2}{r}h^{\ell m}_0 \right)\,.
\end{equation}
%

\section{On the Derivation of the Linear Momentum Flux Formula}\label{lin-mom}
In order to obtain Eqs.~(\ref{recoilx})-(\ref{recoilz}) from 
Eq.~(\ref{linearmomentumflux}) we need to use the decompositions of products of 
spherical harmonics in single harmonics typical of problems that deal with 
angular momentum coupling (for accounts dealing with this problem 
see~\cite{Thorne:1980rm}, where multipole expansions of gravitational radiation 
in different sets of harmonics are described; see~\cite{Brizuela:2006ne} for a 
recent systematic treatment of higher-order perturbation theory where these 
issues are also treated).

Introducing Eq.~(\ref{waveforms}) into Eq.~(\ref{linearmomentumflux}), using 
that any spherical harmonic ${\cal S}^{\ell m}$ has the property 
${\cal S}^{\ell, -m} = (-1)^m \bar{\cal S}^{\ell m}\,,$  and the fact that
$Z^{\ell m}_{ab}$ is symmetric and traceless we get
\begin{widetext}
\begin{eqnarray}
\dot{P}_{GW}^k & = & \frac{1}{32 \pi} 
\mathop{\sum^{}_{\ell \geq 2, m}}_{\ell' \geq 2, m'} 
\left(\dot{\Psi}_{\ZM}^{\ell m}\dot{\bar{\Psi}}^{\ell' m'}_{\ZM}  +  
\dot{\Psi}_{\CPM}^{\ell m}\dot{\bar{\Psi}}^{\ell' m'}_{\CPM} \right) 
\int^{}_{S^2} d\Omega \,\hat{r}^k_{obs} Z^{\ell m}_{ab} \bar{Z}^{\ell' m'}_{cd}
\Omega^{ac}\Omega^{bd}
\nonumber \\
& - & \frac{1}{32 \pi} \mathop{\sum^{}_{\ell \geq 2, m}}_{\ell' \geq 2, m'}
\left(\dot{\Psi}_{\CPM}^{\ell m}\dot{\bar{\Psi}}^{\ell' m'}_{\ZM}  -
\dot{\Psi}_{\ZM}^{\ell m}\dot{\bar{\Psi}}^{\ell' m'}_{\CPM} \right)
\int^{}_{S^2} d\Omega \,\hat{r}^k_{obs} Z^{\ell m}_{ab} \bar{Z}^{\ell' m'}_{cd}
\Omega^{ac}\epsilon^{bd}
\,.  \label{intermediatepdot}
\end{eqnarray}
\end{widetext}
At this point everything reduces to evaluating the integrals on the 2-sphere.
To that end, the starting point is the well-known formula~\cite{Messiah:1962am}
\begin{widetext}
\begin{eqnarray}
Y^{\ell m} Y^{\ell' m'} &=& \sum_{L, M} \sqrt{\frac{(2 \ell + 1)(2
    \ell' + 1)(2 L + 1)}{4 \pi}} 
\left( \begin{array}{ccc}
\ell & \ell' & L \\
m & m' & M \\
\end{array} \right) 
\left( \begin{array}{ccc}
\ell & \ell' & L \\
0 & 0 & 0 \\
\end{array} \right) \bar{Y}^{LM}\,, \label{keyharmonic}
\end{eqnarray}
\end{widetext}
where the objects with the round brackets are the $3j$-Wigner symbols, which are
related to the Clebsch-Gordon coefficients.  They are subject to certain 
{\em selection} rules, namely, $(\ell, m)\,,$ $(\ell',m')$, and $(L,M)$ are 
integers with the usual ranges of values; $m + m' + M = 0\,;$ and the triangular 
inequality $|\ell - \ell'| \leq L \leq \ell + \ell'\,.$ By using 
(\ref{keyharmonic}) and the definition of $Z^{\ell m}_{ab}$ 
[Eq.~(\ref{ten-sph-harm1})] we find the following relationship
\begin{widetext}
\begin{eqnarray}
Z_{ab}^{\ell m} \bar{Z}_{cd}^{\ell' m'} \Omega^{ac} \Omega^{bd} = \sum_{L,M} 
C(\ell,\ell',L) \sqrt{\frac{(2 \ell + 1)(2 \ell' + 1) (2 L + 1)}{4 \pi}} 
\left( \begin{array}{ccc}
\ell & \ell' & L \\
m & m' & M \\
\end{array} \right)
\left( \begin{array}{ccc}
\ell & \ell' & L \\
0 & 0 & 0 \\
\end{array} \right)
\bar{Y}^{LM}. 
\end{eqnarray}
\end{widetext}
where $C(\ell,\ell',L)$ is a constant given by 
\begin{eqnarray}
C(\ell,\ell',L) = \frac{1}{4} \left\{ L^2 (L+1)^2 + \ell^2 (\ell+1)^2 + \ell'^2 
(\ell' + 1)^2 \right. \nonumber \\
+ \left. 2 L (L + 1)  - 2 \left[ \ell (\ell+1) + \ell' (\ell' + 1) \right]
\left[L (L+1) + 1\right] \right\}\,. 
\end{eqnarray}
A similar formula can be found for the second term in~(\ref{intermediatepdot}).
Then, since the components of $\hat{r}^k_{obs}$ are linear in $Y^{1,m}$, the
integral that appears in Eq.~(\ref{intermediatepdot}) is now straightforward.  
In order to get Eqs.~(\ref{recoilx})-(\ref{recoilz}) we just need to use the 
selection rules of the $3-j$ Wigner symbols and the following  additional 
properties:
\begin{widetext}
\begin{eqnarray}
\left( \begin{array}{ccc}
j_1 & j_2 & j_1 + j_2 \\
m_1 & m_2 & -M \\
\end{array} \right) 
&=& (-1)^{j_1 - j_2 + M} \left[ \frac{(2 j_1)! (2 j_2)!}{(2 j_1 + 2 j_2
    + 1)!} \frac{(j_1 + j_2 + M)! (j_1 + j_2 - M)!}{(j_1 + m_1)!(j_1 -
    m_1)!(j_2 + m_2)!(j_2 - m_2)!} \right]^{1/2}\,, \\
\left( \begin{array}{ccc}
j_1 & j_2 & j \\
0 & 0 & 0 \\
\end{array} \right) 
&=& 
\left\{ \begin{array}{l}
(-1)^g \left[ \frac{(2 g - 2 j_1)! (2 g - 2 j_2)! (2 g - 2 j)!}{(2 g +
    1)!} \right]^{1/2} \frac{g!}{(g - j_1)! (g - j_2)! (g - j)!},
\quad \textrm{if} \quad J = 2g\,,
\\
0, \qquad \qquad \qquad \qquad \qquad \qquad \qquad \qquad \qquad
\qquad \qquad \quad \; \; \textrm{if} \quad J = 2 g + 1\,,
\end{array} \right.
\end{eqnarray}
\end{widetext}
where $J = j_1 + j_2 + j\,.$

Alternatively, one can use the following relation for the integral of three
spin-weighted spherical harmonics:
\begin{widetext}
\begin{equation}
\int^{}_{S^2} d\Omega \; {}^{}_SY^{LM} {}^{}_{s}Y^{\ell m} {}^{}_{s'}Y^{\ell' m'}
= \sqrt{\frac{(2L+1)(2\ell+1)(2\ell'+1)}{4\pi}}\,
\left( \begin{array}{ccc}
L  & \ell & \ell' \\
-S &  -s  & -s'  \\
\end{array} \right)\,
\left( \begin{array}{ccc}
L  & \ell & \ell' \\
M  &  m   & m'  \\
\end{array} \right)\,.
\end{equation}
\end{widetext}
%

\section{Determining Extremal Surfaces in the Brill-Lindquist Binary Black Hole 
Data}\label{extremalsurfaces}
We are interested in finding the location of the extremal surfaces that surround 
the individual holes in the BL binary black hole initial data that we are using
in this paper. When the holes are separated enough these surfaces form the 
apparent horizon of the initial data (since we are neglecting the extrinsic 
curvature the data is time symmetric). If we put the two holes close enough, 
another maximal surface enclosing the two holes appears and becomes the apparent 
horizon, and then, the two individual maximal surfaces are called marginally 
trapped surfaces.

Following Bishop~\cite{Bishop:1982nt,Bishop:1984nt}, in order to look for 
maximal surfaces in the BL data it is very convenient to exploit the cylindrical
symmetry of the configuration by expressing the metric in cylindrical 
coordinates (for coherence with the conventions of the paper we choose the axis 
of symmetry to be the $X$ axis): $x = x\,,$ $y = \rho \cos\vartheta\,,$ $z = \rho 
\sin\vartheta\,.$  Then, the line element of Eq.~(\ref{bl_metric}) becomes:
\begin{equation}
ds^2 = \Phi^4(\rho,x)\left(d\rho^2 + \rho^2 d\vartheta^2 + dx^2  \right)\,,
\end{equation}
where now
\begin{equation}
\Phi = 1+ \frac{m^{}_1}{2\sqrt{\rho^2+(x-X^{}_1)^2}}+ 
\frac{m^{}_2}{2\sqrt{\rho^2+(x-X^{}_2)^2}}\,.
\end{equation}
Moreover, the problem of finding the extremal surfaces reduces to that of 
finding a path $(\rho(\lambda)\,,x(\lambda))$ in the subspace $(\rho\,,x)$. The 
area of a surface with cylindrical symmetry can be written as
\begin{equation}
{\cal A} = 2\pi\int^{\lambda^{}_2}_{\lambda^{}_1} \rho \Phi^4 
\sqrt{\dot{\rho}^2+\dot{x}^2}\;d\lambda \,,
\end{equation}
where the dots denote differentiation along the path, that is $d/d\lambda\,$, 
and $(\lambda^{}_1,\lambda^{}_2)$ are the intersections of the surface with the 
symmetry axis $X\,.$ The equations for the path $(\rho(\lambda)\,,x(\lambda))$ 
is found by extremizing the area, $\delta{\cal A} =0\,,$ 
Bishop~\cite{Bishop:1982nt,Bishop:1984nt} took $\lambda$ to be an affine 
parameter, that is, such that ${\cal A} = 2\pi\,(\lambda^{}_2-\lambda^{}_1)\,.$
We have fixed $\lambda$ in such a way the ordinary differential equations for 
$(\rho(\lambda)\,,x(\lambda))$ are simple and amenable for numerical 
computations. In particular, we have chosen $\lambda$ so that 
\begin{equation}
\sqrt{\dot{\rho}^2+\dot{x}^2} = \rho\, \Phi^4\,,
\end{equation}
which is a constraint preserved by the Euler-Lagrange equations that we obtain 
from $\delta{\cal A}=0$:
\begin{eqnarray}
\ddot{\rho} & = & - \partial^{}_\rho {\cal V}\,, \\
\ddot{x}    & = & - \partial^{}_x {\cal V}\,,
\end{eqnarray}
where the {\em potential} ${\cal V}$ is given by
\begin{equation}
{\cal V}(\rho,x) = \frac{1}{2}\rho^2\,\Phi^8\,.
\end{equation}
%



\begin{thebibliography}{97}
\expandafter\ifx\csname natexlab\endcsname\relax\def\natexlab#1{#1}\fi
\expandafter\ifx\csname bibnamefont\endcsname\relax
  \def\bibnamefont#1{#1}\fi
\expandafter\ifx\csname bibfnamefont\endcsname\relax
  \def\bibfnamefont#1{#1}\fi
\expandafter\ifx\csname url\endcsname\relax
  \def\url#1{\texttt{#1}}\fi
\expandafter\ifx\csname urlprefix\endcsname\relax\def\urlprefix{URL }\fi
\providecommand*{\bibinfo}[2]{#2}
\providecommand*{\eprint}[1]{#1}
\providecommand*{\url}[1]{#1}
\begingroup\makeatletter
 \@temptokena{%
  \expandafter\ifx\csname citenamefont\endcsname\relax
   \DeclareRobustCommand\citenamefont{\@firstofone}%
   \global\let\citenamefont\citenamefont
   \global\expandafter\let\csname citenamefont \expandafter\endcsname\csname
  citenamefont \endcsname
  \fi
 }\if@filesw\immediate\write\@auxout{\the\@temptokena}\fi
\expandafter\endgroup\the\@temptokena

\bibitem[{lig()}]{ligo}
\emph{\bibinfo{title}{LIGO}}, \bibinfo{note}{{\tt
  www.ligo.caltech.edu}}.

\bibitem[{vir()}]{virgo}
\emph{\bibinfo{title}{VIRGO}}, \bibinfo{note}{{\tt www.virgo.infn.it}}.

\bibitem[{geo()}]{geo}
\emph{\bibinfo{title}{GEO600}}, \bibinfo{note}{{\tt
  www.geo600.uni-hannover.de}}.

\bibitem[{tam()}]{tama}
\emph{\bibinfo{title}{TAMA}}, \bibinfo{note}{{\tt
  tamago.mtk.nao.ac.jp}}.

\bibitem[{lis()}]{lisa}
\emph{\bibinfo{title}{LISA}}, \bibinfo{note}{{\tt lisa.jpl.nasa.gov}}.

\bibitem[{\citenamefont{{Binggeli} and {Jerjen}}(1998)}]{Binggeli:1998bj}
\bibinfo{author}{\bibfnamefont{B.}~\bibnamefont{{Binggeli}}} \bibnamefont{and}
  \bibinfo{author}{\bibfnamefont{H.}~\bibnamefont{{Jerjen}}},
  \bibinfo{journal}{\aap} \textbf{\bibinfo{volume}{333}}, \bibinfo{pages}{17}
  (\bibinfo{year}{1998}), \eprint{astro-ph/9704027}.

\bibitem[{\citenamefont{{Binggeli}}
  \emph{et~al.}(2000)\citenamefont{{Binggeli}, {Barazza}, and
  {Jerjen}}}]{Binggeli:2000bb}
\bibinfo{author}{\bibfnamefont{B.}~\bibnamefont{{Binggeli}}},
  \bibinfo{author}{\bibfnamefont{F.}~\bibnamefont{{Barazza}}},
  \bibnamefont{and} \bibinfo{author}{\bibfnamefont{H.}~\bibnamefont{{Jerjen}}},
  \bibinfo{journal}{\aap} \textbf{\bibinfo{volume}{359}}, \bibinfo{pages}{447}
  (\bibinfo{year}{2000}).

\bibitem[{\citenamefont{Merritt} \emph{et~al.}(2004)\citenamefont{Merritt,
  Milosavljevic, Favata, Hughes, and Holz}}]{Merritt:2004xa}
\bibinfo{author}{\bibfnamefont{D.}~\bibnamefont{Merritt}},
  \bibinfo{author}{\bibfnamefont{M.}~\bibnamefont{Milosavljevic}},
  \bibinfo{author}{\bibfnamefont{M.}~\bibnamefont{Favata}},
  \bibinfo{author}{\bibfnamefont{S.~A.} \bibnamefont{Hughes}},
  \bibnamefont{and} \bibinfo{author}{\bibfnamefont{D.~E.} \bibnamefont{Holz}},
  \bibinfo{journal}{\apj} \textbf{\bibinfo{volume}{607}}, \bibinfo{pages}{L9}
  (\bibinfo{year}{2004}), \eprint{astro-ph/0402057}.

\bibitem[{\citenamefont{{Haehnelt}}
  \emph{et~al.}(2006)\citenamefont{{Haehnelt}, {Davies}, and
  {Rees}}}]{Haehnelt:2006hd}
\bibinfo{author}{\bibfnamefont{M.~G.} \bibnamefont{{Haehnelt}}},
  \bibinfo{author}{\bibfnamefont{M.~B.} \bibnamefont{{Davies}}},
  \bibnamefont{and} \bibinfo{author}{\bibfnamefont{M.~J.}
  \bibnamefont{{Rees}}}, \bibinfo{journal}{\mnras}
  \textbf{\bibinfo{volume}{366}}, \bibinfo{pages}{L22} (\bibinfo{year}{2006}),
  \eprint{astro-ph/0511245}.

\bibitem[{\citenamefont{Haiman}(2004)}]{Haiman:2004ve}
\bibinfo{author}{\bibfnamefont{Z.}~\bibnamefont{Haiman}},
  \bibinfo{journal}{Astrophys. J.} \textbf{\bibinfo{volume}{613}},
  \bibinfo{pages}{36} (\bibinfo{year}{2004}), \eprint{astro-ph/0404196}.

\bibitem[{\citenamefont{Bonnor and Rotenberg}(1961)}]{Bonnor:1961br}
\bibinfo{author}{\bibfnamefont{W.}~\bibnamefont{Bonnor}} \bibnamefont{and}
  \bibinfo{author}{\bibfnamefont{M.}~\bibnamefont{Rotenberg}},
  \bibinfo{journal}{Proc. R. Soc. London} \textbf{\bibinfo{volume}{A265}},
  \bibinfo{pages}{109} (\bibinfo{year}{1961}).

\bibitem[{\citenamefont{Papapetrou}(1962)}]{Papapetrou:1962ap}
\bibinfo{author}{\bibfnamefont{A.}~\bibnamefont{Papapetrou}},
  \bibinfo{journal}{Ann. Inst. Henri Poincare} \textbf{\bibinfo{volume}{14}},
  \bibinfo{pages}{79} (\bibinfo{year}{1962}).

\bibitem[{\citenamefont{{Peres}}(1962)}]{Peres:1962ap}
\bibinfo{author}{\bibfnamefont{A.}~\bibnamefont{{Peres}}},
  \bibinfo{journal}{Physical Review} \textbf{\bibinfo{volume}{128}},
  \bibinfo{pages}{2471} (\bibinfo{year}{1962}).

\bibitem[{\citenamefont{{Bekenstein}}(1973)}]{Bekenstein:1973jd}
\bibinfo{author}{\bibfnamefont{J.~D.} \bibnamefont{{Bekenstein}}},
  \bibinfo{journal}{\apj} \textbf{\bibinfo{volume}{183}}, \bibinfo{pages}{657}
  (\bibinfo{year}{1973}).

\bibitem[{\citenamefont{{Redmount} and {Rees}}(1989)}]{Redmount:1989rr}
\bibinfo{author}{\bibfnamefont{I.~H.} \bibnamefont{{Redmount}}}
  \bibnamefont{and} \bibinfo{author}{\bibfnamefont{M.~J.}
  \bibnamefont{{Rees}}}, \bibinfo{journal}{Comments on Astrophysics}
  \textbf{\bibinfo{volume}{14}}, \bibinfo{pages}{165} (\bibinfo{year}{1989}).

\bibitem[{\citenamefont{Wiseman}(1992)}]{Wiseman:1992dv}
\bibinfo{author}{\bibfnamefont{A.~G.} \bibnamefont{Wiseman}},
  \bibinfo{journal}{Phys. Rev.} \textbf{\bibinfo{volume}{D46}},
  \bibinfo{pages}{1517} (\bibinfo{year}{1992}).

\bibitem[{\citenamefont{Favata} \emph{et~al.}(2004)\citenamefont{Favata,
  Hughes, and Holz}}]{Favata:2004wz}
\bibinfo{author}{\bibfnamefont{M.}~\bibnamefont{Favata}},
  \bibinfo{author}{\bibfnamefont{S.~A.} \bibnamefont{Hughes}},
  \bibnamefont{and} \bibinfo{author}{\bibfnamefont{D.~E.} \bibnamefont{Holz}},
  \bibinfo{journal}{\apj} \textbf{\bibinfo{volume}{607}}, \bibinfo{pages}{L5}
  (\bibinfo{year}{2004}), \eprint{astro-ph/0402056}.

\bibitem[{\citenamefont{{Fitchett}}(1983)}]{Fitchett:1983fc}
\bibinfo{author}{\bibfnamefont{M.~J.} \bibnamefont{{Fitchett}}},
  \bibinfo{journal}{\mnras} \textbf{\bibinfo{volume}{203}},
  \bibinfo{pages}{1049} (\bibinfo{year}{1983}).

\bibitem[{\citenamefont{{Fitchett} and {Detweiler}}(1984)}]{Fitchett:1984fd}
\bibinfo{author}{\bibfnamefont{M.~J.} \bibnamefont{{Fitchett}}}
  \bibnamefont{and}
  \bibinfo{author}{\bibfnamefont{S.}~\bibnamefont{{Detweiler}}},
  \bibinfo{journal}{\mnras} \textbf{\bibinfo{volume}{211}},
  \bibinfo{pages}{933} (\bibinfo{year}{1984}).

\bibitem[{\citenamefont{{Oohara} and {Nakamura}}(1983)}]{Oohara:1983on}
\bibinfo{author}{\bibfnamefont{K.-I.} \bibnamefont{{Oohara}}} \bibnamefont{and}
  \bibinfo{author}{\bibfnamefont{T.}~\bibnamefont{{Nakamura}}},
  \bibinfo{journal}{Physics Letters A} \textbf{\bibinfo{volume}{94}},
  \bibinfo{pages}{349} (\bibinfo{year}{1983}).

\bibitem[{\citenamefont{{Nakamura} and {Haugan}}(1983)}]{Nakamura:1983nh}
\bibinfo{author}{\bibfnamefont{T.}~\bibnamefont{{Nakamura}}} \bibnamefont{and}
  \bibinfo{author}{\bibfnamefont{M.~P.} \bibnamefont{{Haugan}}},
  \bibinfo{journal}{\apj} \textbf{\bibinfo{volume}{269}}, \bibinfo{pages}{292}
  (\bibinfo{year}{1983}).

\bibitem[{\citenamefont{Blanchet} \emph{et~al.}(2005)\citenamefont{Blanchet,
  Qusailah, and Will}}]{Blanchet:2005rj}
\bibinfo{author}{\bibfnamefont{L.}~\bibnamefont{Blanchet}},
  \bibinfo{author}{\bibfnamefont{M.~S.~S.} \bibnamefont{Qusailah}},
  \bibnamefont{and} \bibinfo{author}{\bibfnamefont{C.~M.} \bibnamefont{Will}},
  \bibinfo{journal}{\apj} \textbf{\bibinfo{volume}{635}}, \bibinfo{pages}{508}
  (\bibinfo{year}{2005}), \eprint{astro-ph/0507692}.

\bibitem[{\citenamefont{Damour and Gopakumar}(2006)}]{Damour:2006tr}
\bibinfo{author}{\bibfnamefont{T.}~\bibnamefont{Damour}} \bibnamefont{and}
  \bibinfo{author}{\bibfnamefont{A.}~\bibnamefont{Gopakumar}},
  \bibinfo{journal}{Phys. Rev.} \textbf{\bibinfo{volume}{D73}},
  \bibinfo{pages}{124006} (\bibinfo{year}{2006}), \eprint{gr-qc/0602117}.

\bibitem[{\citenamefont{Hughes} \emph{et~al.}(2005)\citenamefont{Hughes,
  Favata, and Holz}}]{Hughes:2004ck}
\bibinfo{author}{\bibfnamefont{S.~A.} \bibnamefont{Hughes}},
  \bibinfo{author}{\bibfnamefont{M.}~\bibnamefont{Favata}}, \bibnamefont{and}
  \bibinfo{author}{\bibfnamefont{D.~E.} \bibnamefont{Holz}}, in
  \emph{\bibinfo{booktitle}{Growing Black Holes: Accretion in a Cosmological
  Context}}, edited by
  \bibinfo{editor}{\bibfnamefont{A.}~\bibnamefont{{Merloni}}},
  \bibinfo{editor}{\bibfnamefont{S.}~\bibnamefont{{Nayakshin}}},
  \bibnamefont{and} \bibinfo{editor}{\bibfnamefont{R.~A.}
  \bibnamefont{{Sunyaev}}} (\bibinfo{year}{2005}), pp.
  \bibinfo{pages}{333--339}, \eprint{astro-ph/0408492}.

\bibitem[{\citenamefont{Campanelli}(2005)}]{Campanelli:2004zw}
\bibinfo{author}{\bibfnamefont{M.}~\bibnamefont{Campanelli}},
  \bibinfo{journal}{Class. Quant. Grav.} \textbf{\bibinfo{volume}{22}},
  \bibinfo{pages}{S387} (\bibinfo{year}{2005}), \eprint{astro-ph/0411744}.

\bibitem[{\citenamefont{Anninos and Brandt}(1998)}]{Anninos:1998wt}
\bibinfo{author}{\bibfnamefont{P.}~\bibnamefont{Anninos}} \bibnamefont{and}
  \bibinfo{author}{\bibfnamefont{S.}~\bibnamefont{Brandt}},
  \bibinfo{journal}{Phys. Rev. Lett.} \textbf{\bibinfo{volume}{81}},
  \bibinfo{pages}{508} (\bibinfo{year}{1998}), \eprint{gr-qc/9806031}.

\bibitem[{\citenamefont{Brandt and Anninos}(1999)}]{Brandt:1999zh}
\bibinfo{author}{\bibfnamefont{S.}~\bibnamefont{Brandt}} \bibnamefont{and}
  \bibinfo{author}{\bibfnamefont{P.}~\bibnamefont{Anninos}},
  \bibinfo{journal}{Phys. Rev.} \textbf{\bibinfo{volume}{D60}},
  \bibinfo{pages}{084005} (\bibinfo{year}{1999}), \eprint{astro-ph/9907075}.

\bibitem[{\citenamefont{Pretorius}(2005)}]{Pretorius:2005gq}
\bibinfo{author}{\bibfnamefont{F.}~\bibnamefont{Pretorius}},
  \bibinfo{journal}{Phys. Rev. Lett.} \textbf{\bibinfo{volume}{95}},
  \bibinfo{pages}{121101} (\bibinfo{year}{2005}), \eprint{gr-qc/0507014}.

\bibitem[{\citenamefont{Campanelli}
  \emph{et~al.}(2006)\citenamefont{Campanelli, Lousto, Marronetti, and
  Zlochower}}]{Campanelli:2005dd}
\bibinfo{author}{\bibfnamefont{M.}~\bibnamefont{Campanelli}},
  \bibinfo{author}{\bibfnamefont{C.~O.} \bibnamefont{Lousto}},
  \bibinfo{author}{\bibfnamefont{P.}~\bibnamefont{Marronetti}},
  \bibnamefont{and}
  \bibinfo{author}{\bibfnamefont{Y.}~\bibnamefont{Zlochower}},
  \bibinfo{journal}{Phys. Rev. Lett.} \textbf{\bibinfo{volume}{96}},
  \bibinfo{pages}{111101} (\bibinfo{year}{2006}), \eprint{gr-qc/0511048}.

\bibitem[{\citenamefont{Baker} \emph{et~al.}(2006)\citenamefont{Baker,
  Centrella, Choi, Koppitz, and van Meter}}]{Baker:2005vv}
\bibinfo{author}{\bibfnamefont{J.~G.} \bibnamefont{Baker}},
  \bibinfo{author}{\bibfnamefont{J.}~\bibnamefont{Centrella}},
  \bibinfo{author}{\bibfnamefont{D.-I.} \bibnamefont{Choi}},
  \bibinfo{author}{\bibfnamefont{M.}~\bibnamefont{Koppitz}}, \bibnamefont{and}
  \bibinfo{author}{\bibfnamefont{J.}~\bibnamefont{van Meter}},
  \bibinfo{journal}{Phys. Rev. Lett.} \textbf{\bibinfo{volume}{96}},
  \bibinfo{pages}{111102} (\bibinfo{year}{2006}), \eprint{gr-qc/0511103}.

\bibitem[{\citenamefont{Herrmann} \emph{et~al.}(2006)\citenamefont{Herrmann,
  Shoemaker, and Laguna}}]{Herrmann:2006ks}
\bibinfo{author}{\bibfnamefont{F.}~\bibnamefont{Herrmann}},
  \bibinfo{author}{\bibfnamefont{D.}~\bibnamefont{Shoemaker}},
  \bibnamefont{and} \bibinfo{author}{\bibfnamefont{P.}~\bibnamefont{Laguna}}
  (\bibinfo{year}{2006}), \eprint{gr-qc/0601026}.

\bibitem[{Baker \emph{et~al.}(2006)\citenamefont{Baker}
  \emph{et~al.}}]{Baker:2006vn}
\bibinfo{author}{\bibfnamefont{J.~G.} \bibnamefont{Baker}} \emph{et~al.}
  (\bibinfo{year}{2006}), \eprint{astro-ph/0603204}.

\bibitem[{\citenamefont{Price and Pullin}(1994)}]{Price:1994pm}
\bibinfo{author}{\bibfnamefont{R.~H.} \bibnamefont{Price}} \bibnamefont{and}
  \bibinfo{author}{\bibfnamefont{J.}~\bibnamefont{Pullin}},
  \bibinfo{journal}{Phys. Rev. Lett.} \textbf{\bibinfo{volume}{72}},
  \bibinfo{pages}{3297} (\bibinfo{year}{1994}), \eprint{gr-qc/9402039}.

\bibitem[{\citenamefont{Anninos} \emph{et~al.}(1993)\citenamefont{Anninos,
  Hobill, Seidel, Smarr, and Suen}}]{Anninos:1993zj}
\bibinfo{author}{\bibfnamefont{P.}~\bibnamefont{Anninos}},
  \bibinfo{author}{\bibfnamefont{D.}~\bibnamefont{Hobill}},
  \bibinfo{author}{\bibfnamefont{E.}~\bibnamefont{Seidel}},
  \bibinfo{author}{\bibfnamefont{L.}~\bibnamefont{Smarr}}, \bibnamefont{and}
  \bibinfo{author}{\bibfnamefont{W.-M.} \bibnamefont{Suen}},
  \bibinfo{journal}{Phys. Rev. Lett.} \textbf{\bibinfo{volume}{71}},
  \bibinfo{pages}{2851} (\bibinfo{year}{1993}), \eprint{gr-qc/9309016}.

\bibitem[{\citenamefont{Anninos} \emph{et~al.}(1995)\citenamefont{Anninos,
  Price, Pullin, Seidel, and Suen}}]{Anninos:1995vf}
\bibinfo{author}{\bibfnamefont{P.}~\bibnamefont{Anninos}},
  \bibinfo{author}{\bibfnamefont{R.~H.} \bibnamefont{Price}},
  \bibinfo{author}{\bibfnamefont{J.}~\bibnamefont{Pullin}},
  \bibinfo{author}{\bibfnamefont{E.}~\bibnamefont{Seidel}}, \bibnamefont{and}
  \bibinfo{author}{\bibfnamefont{W.-M.} \bibnamefont{Suen}},
  \bibinfo{journal}{Phys. Rev.} \textbf{\bibinfo{volume}{D52}},
  \bibinfo{pages}{4462} (\bibinfo{year}{1995}), \eprint{gr-qc/9505042}.

\bibitem[{\citenamefont{Gleiser} \emph{et~al.}(1996)\citenamefont{Gleiser,
  Nicasio, Price, and Pullin}}]{Gleiser:1996yc}
\bibinfo{author}{\bibfnamefont{R.~J.} \bibnamefont{Gleiser}},
  \bibinfo{author}{\bibfnamefont{C.~O.} \bibnamefont{Nicasio}},
  \bibinfo{author}{\bibfnamefont{R.~H.} \bibnamefont{Price}}, \bibnamefont{and}
  \bibinfo{author}{\bibfnamefont{J.}~\bibnamefont{Pullin}},
  \bibinfo{journal}{Phys. Rev. Lett.} \textbf{\bibinfo{volume}{77}},
  \bibinfo{pages}{4483} (\bibinfo{year}{1996}), \eprint{gr-qc/9609022}.

\bibitem[{\citenamefont{Gleiser} \emph{et~al.}(2000)\citenamefont{Gleiser,
  Nicasio, Price, and Pullin}}]{Gleiser:1998rw}
\bibinfo{author}{\bibfnamefont{R.~J.} \bibnamefont{Gleiser}},
  \bibinfo{author}{\bibfnamefont{C.~O.} \bibnamefont{Nicasio}},
  \bibinfo{author}{\bibfnamefont{R.~H.} \bibnamefont{Price}}, \bibnamefont{and}
  \bibinfo{author}{\bibfnamefont{J.}~\bibnamefont{Pullin}},
  \bibinfo{journal}{Phys. Rept.} \textbf{\bibinfo{volume}{325}},
  \bibinfo{pages}{41} (\bibinfo{year}{2000}), \eprint{gr-qc/9807077}.

\bibitem[{\citenamefont{Krivan and Price}(1999)}]{Krivan:1998er}
\bibinfo{author}{\bibfnamefont{W.}~\bibnamefont{Krivan}} \bibnamefont{and}
  \bibinfo{author}{\bibfnamefont{R.~H.} \bibnamefont{Price}},
  \bibinfo{journal}{Phys. Rev. Lett.} \textbf{\bibinfo{volume}{82}},
  \bibinfo{pages}{1358} (\bibinfo{year}{1999}), \eprint{gr-qc/9810080}.

\bibitem[{\citenamefont{Nicasio} \emph{et~al.}(1999)\citenamefont{Nicasio,
  Gleiser, Price, and Pullin}}]{Nicasio:1998aj}
\bibinfo{author}{\bibfnamefont{C.~O.} \bibnamefont{Nicasio}},
  \bibinfo{author}{\bibfnamefont{R.~J.} \bibnamefont{Gleiser}},
  \bibinfo{author}{\bibfnamefont{R.~H.} \bibnamefont{Price}}, \bibnamefont{and}
  \bibinfo{author}{\bibfnamefont{J.}~\bibnamefont{Pullin}},
  \bibinfo{journal}{Phys. Rev.} \textbf{\bibinfo{volume}{D59}},
  \bibinfo{pages}{044024} (\bibinfo{year}{1999}), \eprint{gr-qc/9802063}.

\bibitem[{\citenamefont{Gleiser and Dominguez}(2002)}]{Gleiser:2001in}
\bibinfo{author}{\bibfnamefont{R.~J.} \bibnamefont{Gleiser}} \bibnamefont{and}
  \bibinfo{author}{\bibfnamefont{A.~E.} \bibnamefont{Dominguez}},
  \bibinfo{journal}{Phys. Rev.} \textbf{\bibinfo{volume}{D65}},
  \bibinfo{pages}{064018} (\bibinfo{year}{2002}), \eprint{gr-qc/0109018}.

\bibitem[{\citenamefont{Khanna}(2001)}]{Khanna:2001ch}
\bibinfo{author}{\bibfnamefont{G.}~\bibnamefont{Khanna}},
  \bibinfo{journal}{Phys. Rev.} \textbf{\bibinfo{volume}{D63}},
  \bibinfo{pages}{124007} (\bibinfo{year}{2001}), \eprint{gr-qc/0101015}.

\bibitem[{\citenamefont{Sarbach} \emph{et~al.}(2002)\citenamefont{Sarbach,
  Tiglio, and Pullin}}]{Sarbach:2001tj}
\bibinfo{author}{\bibfnamefont{O.}~\bibnamefont{Sarbach}},
  \bibinfo{author}{\bibfnamefont{M.}~\bibnamefont{Tiglio}}, \bibnamefont{and}
  \bibinfo{author}{\bibfnamefont{J.}~\bibnamefont{Pullin}},
  \bibinfo{journal}{Phys. Rev.} \textbf{\bibinfo{volume}{D65}},
  \bibinfo{pages}{064026} (\bibinfo{year}{2002}), \eprint{gr-qc/0110085}.

\bibitem[{\citenamefont{Khanna}(2002)}]{Khanna:2002qp}
\bibinfo{author}{\bibfnamefont{G.}~\bibnamefont{Khanna}},
  \bibinfo{journal}{Phys. Rev.} \textbf{\bibinfo{volume}{D66}},
  \bibinfo{pages}{064004} (\bibinfo{year}{2002}), \eprint{gr-qc/0206010}.

\bibitem[{\citenamefont{Andrade and Price}(1997)}]{Andrade:1996pc}
\bibinfo{author}{\bibfnamefont{Z.}~\bibnamefont{Andrade}} \bibnamefont{and}
  \bibinfo{author}{\bibfnamefont{R.~H.} \bibnamefont{Price}},
  \bibinfo{journal}{Phys. Rev.} \textbf{\bibinfo{volume}{D56}},
  \bibinfo{pages}{6336} (\bibinfo{year}{1997}), \eprint{gr-qc/9611022}.

\bibitem[{\citenamefont{Gerlach and Sengupta}(1979)}]{Gerlach:1979rw}
\bibinfo{author}{\bibfnamefont{U.~H.} \bibnamefont{Gerlach}} \bibnamefont{and}
  \bibinfo{author}{\bibfnamefont{U.~K.} \bibnamefont{Sengupta}},
  \bibinfo{journal}{Phys. Rev.} \textbf{\bibinfo{volume}{D19}},
  \bibinfo{pages}{2268} (\bibinfo{year}{1979}).

\bibitem[{\citenamefont{Gerlach and Sengupta}(1980)}]{Gerlach:1980tx}
\bibinfo{author}{\bibfnamefont{U.~H.} \bibnamefont{Gerlach}} \bibnamefont{and}
  \bibinfo{author}{\bibfnamefont{U.~K.} \bibnamefont{Sengupta}},
  \bibinfo{journal}{Phys. Rev.} \textbf{\bibinfo{volume}{D22}},
  \bibinfo{pages}{1300} (\bibinfo{year}{1980}).

\bibitem[{\citenamefont{Khanna} \emph{et~al.}(2000)\citenamefont{Khanna,
  Gleiser, Price, and Pullin}}]{Khanna:2000dg}
\bibinfo{author}{\bibfnamefont{G.}~\bibnamefont{Khanna}},
  \bibinfo{author}{\bibfnamefont{R.}~\bibnamefont{Gleiser}},
  \bibinfo{author}{\bibfnamefont{R.}~\bibnamefont{Price}}, \bibnamefont{and}
  \bibinfo{author}{\bibfnamefont{J.}~\bibnamefont{Pullin}},
  \bibinfo{journal}{New J. Phys.} \textbf{\bibinfo{volume}{2}},
  \bibinfo{pages}{3} (\bibinfo{year}{2000}), \eprint{gr-qc/0003003}.

\bibitem[{\citenamefont{Lichnerowicz}(1944)}]{Lichnerowicz:1944al}
\bibinfo{author}{\bibfnamefont{A.}~\bibnamefont{Lichnerowicz}},
  \bibinfo{journal}{J. Math. Pures et Appl.} \textbf{\bibinfo{volume}{23}},
  \bibinfo{pages}{37} (\bibinfo{year}{1944}).

\bibitem[{\citenamefont{York}(1971)}]{York:1971hw}
\bibinfo{author}{\bibfnamefont{J.}~\bibnamefont{York, Jr.}},
  \bibinfo{journal}{Phys. Rev. Lett.} \textbf{\bibinfo{volume}{26}},
  \bibinfo{pages}{1656} (\bibinfo{year}{1971}).

\bibitem[{\citenamefont{York}(1972)}]{York:1972sj}
\bibinfo{author}{\bibfnamefont{J.}~\bibnamefont{York, Jr.}},
  \bibinfo{journal}{Phys. Rev. Lett.} \textbf{\bibinfo{volume}{28}},
  \bibinfo{pages}{1082} (\bibinfo{year}{1972}).

\bibitem[{\citenamefont{York}(1973)}]{York:1973jw}
\bibinfo{author}{\bibfnamefont{J.}~\bibnamefont{York, Jr.}},
  \bibinfo{journal}{J. Math. Phys.} \textbf{\bibinfo{volume}{14}},
  \bibinfo{pages}{456} (\bibinfo{year}{1973}).

\bibitem[{\citenamefont{Cook}(2000)}]{Cook:2000vr}
\bibinfo{author}{\bibfnamefont{G.~B.} \bibnamefont{Cook}},
  \bibinfo{journal}{Living Rev. Rel.} \textbf{\bibinfo{volume}{3}},
  \bibinfo{pages}{5} (\bibinfo{year}{2000}), \eprint{gr-qc/0007085}.

\bibitem[{\citenamefont{Bowen and York}(1980)}]{Bowen:1980yu}
\bibinfo{author}{\bibfnamefont{J.~M.} \bibnamefont{Bowen}} \bibnamefont{and}
  \bibinfo{author}{\bibfnamefont{J.~W.} \bibnamefont{York}},
  \bibinfo{journal}{Phys. Rev.} \textbf{\bibinfo{volume}{D21}},
  \bibinfo{pages}{2047} (\bibinfo{year}{1980}).

\bibitem[{\citenamefont{{Misner}}(1960)}]{Misner:1960cw}
\bibinfo{author}{\bibfnamefont{C.~W.} \bibnamefont{{Misner}}},
  \bibinfo{journal}{Phys. Rev.} \textbf{\bibinfo{volume}{118}},
  \bibinfo{pages}{1110} (\bibinfo{year}{1960}).

\bibitem[{\citenamefont{{Brill} and {Lindquist}}(1963)}]{Brill:1963bl}
\bibinfo{author}{\bibfnamefont{D.~R.} \bibnamefont{{Brill}}} \bibnamefont{and}
  \bibinfo{author}{\bibfnamefont{R.~W.} \bibnamefont{{Lindquist}}},
  \bibinfo{journal}{Phys. Rev.} \textbf{\bibinfo{volume}{131}},
  \bibinfo{pages}{471} (\bibinfo{year}{1963}).

\bibitem[{\citenamefont{Condon and Shortley}(1970)}]{Condon:1970cs}
\bibinfo{author}{\bibfnamefont{E.~U.} \bibnamefont{Condon}} \bibnamefont{and}
  \bibinfo{author}{\bibfnamefont{G.~H.} \bibnamefont{Shortley}},
  \emph{\bibinfo{title}{The Theory of Atomic Spectra}}
  (\bibinfo{publisher}{Cambridge University Press},
  \bibinfo{address}{Cambridge}, \bibinfo{year}{1970}).

\bibitem[{\citenamefont{Arfken and Weber}(2001)}]{Arfken}
\bibinfo{author}{~\bibnamefont{Arfken}} \bibnamefont{and}
  \bibinfo{author}{~\bibnamefont{Weber}}, \emph{\bibinfo{title}{Mathematical
  Methods for Physicists}} (\bibinfo{publisher}{Academic Press},
  \bibinfo{address}{California}, \bibinfo{year}{2001}).

\bibitem[{\citenamefont{Regge and {Wheeler}}(1957)}]{Regge:1957rw}
\bibinfo{author}{\bibfnamefont{T.}~\bibnamefont{Regge}} \bibnamefont{and}
  \bibinfo{author}{\bibfnamefont{J.~A.} \bibnamefont{{Wheeler}}},
  \bibinfo{journal}{Phys. Rev.} \textbf{\bibinfo{volume}{108}},
  \bibinfo{pages}{1063} (\bibinfo{year}{1957}).

\bibitem[{\citenamefont{Zerilli}(1970)}]{Zerilli:1970fj}
\bibinfo{author}{\bibfnamefont{F.~J.} \bibnamefont{Zerilli}},
  \bibinfo{journal}{\prl} \textbf{\bibinfo{volume}{24}}, \bibinfo{pages}{737}
  (\bibinfo{year}{1970}).

\bibitem[{\citenamefont{Moncrief}(1974)}]{Moncrief:1974vm}
\bibinfo{author}{\bibfnamefont{V.}~\bibnamefont{Moncrief}},
  \bibinfo{journal}{Ann. Phys. (N.Y.)} \textbf{\bibinfo{volume}{88}},
  \bibinfo{pages}{323} (\bibinfo{year}{1974}).

\bibitem[{\citenamefont{Martel and Poisson}(2005)}]{Martel:2005ir}
\bibinfo{author}{\bibfnamefont{K.}~\bibnamefont{Martel}} \bibnamefont{and}
  \bibinfo{author}{\bibfnamefont{E.}~\bibnamefont{Poisson}},
  \bibinfo{journal}{Phys. Rev.} \textbf{\bibinfo{volume}{D71}},
  \bibinfo{pages}{104003} (\bibinfo{year}{2005}), \eprint{gr-qc/0502028}.

\bibitem[{\citenamefont{{Cunningham}}
  \emph{et~al.}(1978)\citenamefont{{Cunningham}, {Price}, and
  {Moncrief}}}]{Cunningham:1978cp}
\bibinfo{author}{\bibfnamefont{C.~T.} \bibnamefont{{Cunningham}}},
  \bibinfo{author}{\bibfnamefont{R.~H.} \bibnamefont{{Price}}},
  \bibnamefont{and}
  \bibinfo{author}{\bibfnamefont{V.}~\bibnamefont{{Moncrief}}},
  \bibinfo{journal}{\apj} \textbf{\bibinfo{volume}{224}}, \bibinfo{pages}{643}
  (\bibinfo{year}{1978}).

\bibitem[{\citenamefont{Jhingan and Tanaka}(2003)}]{Jhingan:2002kb}
\bibinfo{author}{\bibfnamefont{S.}~\bibnamefont{Jhingan}} \bibnamefont{and}
  \bibinfo{author}{\bibfnamefont{T.}~\bibnamefont{Tanaka}},
  \bibinfo{journal}{Phys. Rev.} \textbf{\bibinfo{volume}{D67}},
  \bibinfo{pages}{104018} (\bibinfo{year}{2003}), \eprint{gr-qc/0211060}.

\bibitem[{\citenamefont{{Isaacson}}(1968{\natexlab{a}})}]{Isaacson:1968ra}
\bibinfo{author}{\bibfnamefont{R.~A.} \bibnamefont{{Isaacson}}},
  \bibinfo{journal}{Phys. Rev.} \textbf{\bibinfo{volume}{166}},
  \bibinfo{pages}{1263} (\bibinfo{year}{1968}{\natexlab{a}}).

\bibitem[{\citenamefont{{Isaacson}}(1968{\natexlab{b}})}]{Isaacson:1968gw}
\bibinfo{author}{\bibfnamefont{R.~A.} \bibnamefont{{Isaacson}}},
  \bibinfo{journal}{Phys. Rev.} \textbf{\bibinfo{volume}{166}},
  \bibinfo{pages}{1272} (\bibinfo{year}{1968}{\natexlab{b}}).

\bibitem[{\citenamefont{Misner} \emph{et~al.}(1973)\citenamefont{Misner,
  Thorne, and Wheeler}}]{Misner:1973cw}
\bibinfo{author}{\bibfnamefont{C.~W.} \bibnamefont{Misner}},
  \bibinfo{author}{\bibfnamefont{K.}~\bibnamefont{Thorne}}, \bibnamefont{and}
  \bibinfo{author}{\bibfnamefont{J.~A.} \bibnamefont{Wheeler}},
  \emph{\bibinfo{title}{Gravitation}} (\bibinfo{publisher}{W. H. Freeman \&
  Co.}, \bibinfo{address}{San Francisco}, \bibinfo{year}{1973}).

\bibitem[{\citenamefont{Thorne}(1980)}]{Thorne:1980rm}
\bibinfo{author}{\bibfnamefont{K.~S.} \bibnamefont{Thorne}},
  \bibinfo{journal}{Rev. Mod. Phys.} \textbf{\bibinfo{volume}{52}},
  \bibinfo{pages}{299} (\bibinfo{year}{1980}).

\bibitem[{\citenamefont{Goldberg} \emph{et~al.}(1967)\citenamefont{Goldberg,
  Macfarlane, Newman, Rohrlich, and Sudarshan}}]{Goldberg:1967sp}
\bibinfo{author}{\bibfnamefont{J.~N.} \bibnamefont{Goldberg}},
  \bibinfo{author}{\bibfnamefont{A.~J.} \bibnamefont{Macfarlane}},
  \bibinfo{author}{\bibfnamefont{E.~T.} \bibnamefont{Newman}},
  \bibinfo{author}{\bibfnamefont{F.}~\bibnamefont{Rohrlich}}, \bibnamefont{and}
  \bibinfo{author}{\bibfnamefont{E.~C.~G.} \bibnamefont{Sudarshan}},
  \bibinfo{journal}{J. Math. Phys.} \textbf{\bibinfo{volume}{8}},
  \bibinfo{pages}{2155} (\bibinfo{year}{1967}).

\bibitem[{\citenamefont{Arnowitt} \emph{et~al.}(1962)\citenamefont{Arnowitt,
  Deser, and Misner}}]{ArnDesMis:62}
\bibinfo{author}{\bibfnamefont{R.}~\bibnamefont{Arnowitt}},
  \bibinfo{author}{\bibfnamefont{S.}~\bibnamefont{Deser}}, \bibnamefont{and}
  \bibinfo{author}{\bibfnamefont{C.~W.} \bibnamefont{Misner}}, in
  \emph{\bibinfo{booktitle}{Gravitation: An introduction to current research}},
  edited by \bibinfo{editor}{\bibfnamefont{L.}~\bibnamefont{Witten}}
  (\bibinfo{publisher}{Wiley}, \bibinfo{address}{New York},
  \bibinfo{year}{1962}), pp. \bibinfo{pages}{227--265}.

\bibitem[{\citenamefont{Abrahams and Price}(1996)}]{Abrahams:1995gn}
\bibinfo{author}{\bibfnamefont{A.~M.} \bibnamefont{Abrahams}} \bibnamefont{and}
  \bibinfo{author}{\bibfnamefont{R.~H.} \bibnamefont{Price}},
  \bibinfo{journal}{Phys. Rev.} \textbf{\bibinfo{volume}{D53}},
  \bibinfo{pages}{1963} (\bibinfo{year}{1996}), \eprint{gr-qc/9508059}.

\bibitem[{\citenamefont{Beig}(2000)}]{Beig:2000ei}
\bibinfo{author}{\bibfnamefont{R.}~\bibnamefont{Beig}}, \bibinfo{journal}{LNP
  Vol.~537: Mathematical and Quantum Aspects of Relativity and Cosmology}
  \textbf{\bibinfo{volume}{537}}, \bibinfo{pages}{55} (\bibinfo{year}{2000}),
  \eprint{gr-qc/0005043}.

\bibitem[{\citenamefont{Blanchet} \emph{et~al.}(1998)\citenamefont{Blanchet,
  Faye, and Ponsot}}]{Blanchet:1998vx}
\bibinfo{author}{\bibfnamefont{L.}~\bibnamefont{Blanchet}},
  \bibinfo{author}{\bibfnamefont{G.}~\bibnamefont{Faye}}, \bibnamefont{and}
  \bibinfo{author}{\bibfnamefont{B.}~\bibnamefont{Ponsot}},
  \bibinfo{journal}{Phys. Rev.} \textbf{\bibinfo{volume}{D58}},
  \bibinfo{pages}{124002} (\bibinfo{year}{1998}), \eprint{gr-qc/9804079}.

\bibitem[{\citenamefont{Blanchet}(2003)}]{Blanchet:2003kz}
\bibinfo{author}{\bibfnamefont{L.}~\bibnamefont{Blanchet}},
  \bibinfo{journal}{Phys. Rev.} \textbf{\bibinfo{volume}{D68}},
  \bibinfo{pages}{084002} (\bibinfo{year}{2003}), \eprint{gr-qc/0304080}.

\bibitem[{\citenamefont{Lousto and Price}(1997)}]{Lousto:1996sx}
\bibinfo{author}{\bibfnamefont{C.~O.} \bibnamefont{Lousto}} \bibnamefont{and}
  \bibinfo{author}{\bibfnamefont{R.~H.} \bibnamefont{Price}},
  \bibinfo{journal}{Phys. Rev.} \textbf{\bibinfo{volume}{D55}},
  \bibinfo{pages}{2124} (\bibinfo{year}{1997}), \eprint{gr-qc/9609012}.

\bibitem[{\citenamefont{Bulirsch and Stoer}(1966)}]{Bulirsch:1966bs}
\bibinfo{author}{\bibfnamefont{R.}~\bibnamefont{Bulirsch}} \bibnamefont{and}
  \bibinfo{author}{\bibfnamefont{J.}~\bibnamefont{Stoer}},
  \bibinfo{journal}{Num. Math.} \textbf{\bibinfo{volume}{8}},
  \bibinfo{pages}{1} (\bibinfo{year}{1966}).

\bibitem[{\citenamefont{Stoer and Bulirsch}(1993)}]{Stoer:1993sb}
\bibinfo{author}{\bibfnamefont{J.}~\bibnamefont{Stoer}} \bibnamefont{and}
  \bibinfo{author}{\bibfnamefont{R.}~\bibnamefont{Bulirsch}},
  \emph{\bibinfo{title}{Introduction to Numerical Analysis}}
  (\bibinfo{publisher}{Springer-Verlag}, \bibinfo{address}{New York},
  \bibinfo{year}{1993}).

\bibitem[{\citenamefont{Press} \emph{et~al.}(1992)\citenamefont{Press,
  Flannery, Teukolsky, and Vetterling}}]{Press:1992nr}
\bibinfo{author}{\bibfnamefont{W.~H.} \bibnamefont{Press}},
  \bibinfo{author}{\bibfnamefont{B.~P.} \bibnamefont{Flannery}},
  \bibinfo{author}{\bibfnamefont{S.~A.} \bibnamefont{Teukolsky}},
  \bibnamefont{and} \bibinfo{author}{\bibfnamefont{W.~T.}
  \bibnamefont{Vetterling}}, \emph{\bibinfo{title}{Numerical Recipes: The Art
  of Scientific Computing}} (\bibinfo{publisher}{Cambridge University Press},
  \bibinfo{address}{Cambridge (UK) and New York}, \bibinfo{year}{1992}).

\bibitem[{\citenamefont{Sopuerta and Laguna}(2006)}]{Sopuerta:2005gz}
\bibinfo{author}{\bibfnamefont{C.~F.} \bibnamefont{Sopuerta}} \bibnamefont{and}
  \bibinfo{author}{\bibfnamefont{P.}~\bibnamefont{Laguna}},
  \bibinfo{journal}{Phys. Rev.} \textbf{\bibinfo{volume}{D73}},
  \bibinfo{pages}{044028} (\bibinfo{year}{2006}), \eprint{gr-qc/0512028}.

\bibitem[{\citenamefont{Herrmann}()}]{frank}
\bibinfo{author}{\bibfnamefont{F.}~\bibnamefont{Herrmann}},
  \emph{\bibinfo{title}{private communication}}.

\bibitem[{Khanna \emph{et~al.}(1999)\citenamefont{Khanna}
  \emph{et~al.}}]{Khanna:1999mh}
\bibinfo{author}{\bibfnamefont{G.}~\bibnamefont{Khanna}} \emph{et~al.},
  \bibinfo{journal}{Phys. Rev. Lett.} \textbf{\bibinfo{volume}{83}},
  \bibinfo{pages}{3581} (\bibinfo{year}{1999}), \eprint{gr-qc/9905081}.

\bibitem[{\citenamefont{Yunes} \emph{et~al.}(2005)\citenamefont{Yunes, Tichy,
  Owen, and Bruegmann}}]{Yunes:2005nn}
\bibinfo{author}{\bibfnamefont{N.}~\bibnamefont{Yunes}},
  \bibinfo{author}{\bibfnamefont{W.}~\bibnamefont{Tichy}},
  \bibinfo{author}{\bibfnamefont{B.~J.} \bibnamefont{Owen}}, \bibnamefont{and}
  \bibinfo{author}{\bibfnamefont{B.}~\bibnamefont{Bruegmann}}
  (\bibinfo{year}{2005}), \eprint{gr-qc/0503011}.

\bibitem[{\citenamefont{Yunes and Tichy}(2006)}]{Yunes:2006iw}
\bibinfo{author}{\bibfnamefont{N.}~\bibnamefont{Yunes}} \bibnamefont{and}
  \bibinfo{author}{\bibfnamefont{W.}~\bibnamefont{Tichy}}
  (\bibinfo{year}{2006}), \eprint{gr-qc/0601046}.

\bibitem[{\citenamefont{Bruni} \emph{et~al.}(2003)\citenamefont{Bruni,
  Gualtieri, and Sopuerta}}]{Bruni:2002sm}
\bibinfo{author}{\bibfnamefont{M.}~\bibnamefont{Bruni}},
  \bibinfo{author}{\bibfnamefont{L.}~\bibnamefont{Gualtieri}},
  \bibnamefont{and} \bibinfo{author}{\bibfnamefont{C.~F.}
  \bibnamefont{Sopuerta}}, \bibinfo{journal}{Class. Quant. Grav.}
  \textbf{\bibinfo{volume}{20}}, \bibinfo{pages}{535} (\bibinfo{year}{2003}),
  \eprint{gr-qc/0207105}.

\bibitem[{\citenamefont{Sopuerta} \emph{et~al.}(2004)\citenamefont{Sopuerta,
  Bruni, and Gualtieri}}]{Sopuerta:2003rg}
\bibinfo{author}{\bibfnamefont{C.~F.} \bibnamefont{Sopuerta}},
  \bibinfo{author}{\bibfnamefont{M.}~\bibnamefont{Bruni}}, \bibnamefont{and}
  \bibinfo{author}{\bibfnamefont{L.}~\bibnamefont{Gualtieri}},
  \bibinfo{journal}{Phys. Rev.} \textbf{\bibinfo{volume}{D70}},
  \bibinfo{pages}{064002} (\bibinfo{year}{2004}), \eprint{gr-qc/0306027}.

\bibitem[{\citenamefont{Passamonti}
  \emph{et~al.}(2005)\citenamefont{Passamonti, Bruni, Gualtieri, and
  Sopuerta}}]{Passamonti:2004je}
\bibinfo{author}{\bibfnamefont{A.}~\bibnamefont{Passamonti}},
  \bibinfo{author}{\bibfnamefont{M.}~\bibnamefont{Bruni}},
  \bibinfo{author}{\bibfnamefont{L.}~\bibnamefont{Gualtieri}},
  \bibnamefont{and} \bibinfo{author}{\bibfnamefont{C.~F.}
  \bibnamefont{Sopuerta}}, \bibinfo{journal}{Phys. Rev.}
  \textbf{\bibinfo{volume}{D71}}, \bibinfo{pages}{024022}
  (\bibinfo{year}{2005}), \eprint{gr-qc/0407108}.

\bibitem[{\citenamefont{Passamonti}
  \emph{et~al.}(2006)\citenamefont{Passamonti, Bruni, Gualtieri, Nagar, and
  Sopuerta}}]{Passamonti:2005cz}
\bibinfo{author}{\bibfnamefont{A.}~\bibnamefont{Passamonti}},
  \bibinfo{author}{\bibfnamefont{M.}~\bibnamefont{Bruni}},
  \bibinfo{author}{\bibfnamefont{L.}~\bibnamefont{Gualtieri}},
  \bibinfo{author}{\bibfnamefont{A.}~\bibnamefont{Nagar}}, \bibnamefont{and}
  \bibinfo{author}{\bibfnamefont{C.~F.} \bibnamefont{Sopuerta}},
  \bibinfo{journal}{Phys. Rev.} \textbf{\bibinfo{volume}{D73}},
  \bibinfo{pages}{084010} (\bibinfo{year}{2006}), \eprint{gr-qc/0601001}.

\bibitem[{grt()}]{grtensor}
\emph{\bibinfo{title}{GRTensorII}}, \bibinfo{note}{this is a package
  which runs within Maple but distinct from packages distributed with Maple. It
  is distributed freely on the World-Wide-Web from the address: {\tt
  http://grtensor.org}}.

\bibitem[{\citenamefont{Gonzalez} \emph{et~al.}(2006)\citenamefont{Gonzalez,
  Sperhake, Bruegmann, Hannam, and Husa}}]{Gonzalez:2006md}
\bibinfo{author}{\bibfnamefont{J.~A.} \bibnamefont{Gonzalez}},
  \bibinfo{author}{\bibfnamefont{U.}~\bibnamefont{Sperhake}},
  \bibinfo{author}{\bibfnamefont{B.}~\bibnamefont{Bruegmann}},
  \bibinfo{author}{\bibfnamefont{M.}~\bibnamefont{Hannam}}, \bibnamefont{and}
  \bibinfo{author}{\bibfnamefont{S.}~\bibnamefont{Husa}}
  (\bibinfo{year}{2006}), \eprint{gr-qc/0610154}.

\bibitem[{\citenamefont{Sopuerta} \emph{et~al.}(2006)\citenamefont{Sopuerta,
  Yunes, and Laguna}}]{Sopuerta:2006et}
\bibinfo{author}{\bibfnamefont{C.~F.} \bibnamefont{Sopuerta}},
  \bibinfo{author}{\bibfnamefont{N.}~\bibnamefont{Yunes}}, \bibnamefont{and}
  \bibinfo{author}{\bibfnamefont{P.}~\bibnamefont{Laguna}}
  (\bibinfo{year}{2006}), \eprint{astro-ph/0611110}.

\bibitem[{\citenamefont{Abramowitz and Stegun}(1972)}]{Abramowitz:1970as}
\bibinfo{author}{\bibfnamefont{M.}~\bibnamefont{Abramowitz}} \bibnamefont{and}
  \bibinfo{author}{\bibfnamefont{I.~A.} \bibnamefont{Stegun}},
  \emph{\bibinfo{title}{Handbook of Mathematical Functions with Formulas,
  Graphs, and Mathematical Tables}} (\bibinfo{publisher}{Dover},
  \bibinfo{address}{New York}, \bibinfo{year}{1972}).

\bibitem[{\citenamefont{Stewart}(1990)}]{Stewart:1990}
\bibinfo{author}{\bibfnamefont{J.}~\bibnamefont{Stewart}},
  \emph{\bibinfo{title}{Advanced General Relativity}}
  (\bibinfo{publisher}{Cambrdige University Press}, \bibinfo{address}{New
  York}, \bibinfo{year}{1990}).

\bibitem[{\citenamefont{Dray}(1985)}]{Dray:1984gy}
\bibinfo{author}{\bibfnamefont{T.}~\bibnamefont{Dray}}, \bibinfo{journal}{J.
  Math. Phys.} \textbf{\bibinfo{volume}{26}}, \bibinfo{pages}{1030}
  (\bibinfo{year}{1985}).

\bibitem[{\citenamefont{Martel}(2003)}]{Martel:2003th}
\bibinfo{author}{\bibfnamefont{K.}~\bibnamefont{Martel}},
  \emph{\bibinfo{title}{Particles and Black Holes: Time-Domain Integration of
  the Equations of Black-Hole Perturbation Theory}}, Ph.D. thesis,
  \bibinfo{school}{The University of Guelph} (\bibinfo{year}{2003}).

\bibitem[{\citenamefont{Brizuela} \emph{et~al.}(2006)\citenamefont{Brizuela,
  Martin-Garcia, and Marugan}}]{Brizuela:2006ne}
\bibinfo{author}{\bibfnamefont{D.}~\bibnamefont{Brizuela}},
  \bibinfo{author}{\bibfnamefont{J.~M.} \bibnamefont{Martin-Garcia}},
  \bibnamefont{and} \bibinfo{author}{\bibfnamefont{G.~A.~M.}
  \bibnamefont{Marugan}}  (\bibinfo{year}{2006}), \eprint{gr-qc/0607025}.

\bibitem[{\citenamefont{Messiah}(1962)}]{Messiah:1962am}
\bibinfo{author}{\bibfnamefont{A.}~\bibnamefont{Messiah}},
  \emph{\bibinfo{title}{Quantum Mechanics, Vol. 2}}
  (\bibinfo{publisher}{North-Holland}, \bibinfo{address}{Amsterdam},
  \bibinfo{year}{1962}).

\bibitem[{\citenamefont{Bishop}(1982)}]{Bishop:1982nt}
\bibinfo{author}{\bibfnamefont{N.~T.} \bibnamefont{Bishop}},
  \bibinfo{journal}{Gen. Rel. Grav.} \textbf{\bibinfo{volume}{14}},
  \bibinfo{pages}{717} (\bibinfo{year}{1982}).

\bibitem[{\citenamefont{Bishop}(1984)}]{Bishop:1984nt}
\bibinfo{author}{\bibfnamefont{N.~T.} \bibnamefont{Bishop}},
  \bibinfo{journal}{Gen. Rel. Grav.} \textbf{\bibinfo{volume}{16}},
  \bibinfo{pages}{589} (\bibinfo{year}{1984}).

\end{thebibliography}
\end{document}